\documentclass[a4paper,fleqn,usenatbib]{mnras} 
\usepackage[T1]{fontenc}
\usepackage{ae,aecompl}
\usepackage{graphicx}   
\usepackage{amsmath}    
\usepackage{amssymb}    

\defcitealias{Tremaine1993}{T93}

\title[How to design a planetary system]
{How to design a planetary system for different scattering outcomes: giant impact sweet spot, maximising exocomets,
scattered disks}
\author[M. C. Wyatt et al.]
  {M. C. Wyatt$^1$\thanks{Email: wyatt@ast.cam.ac.uk},
  A. Bonsor$^1$,
  A. P. Jackson$^2$,
  S. Marino$^1$,
  A. Shannon$^{1,3,4}$\\
  $^1$ Institute of Astronomy, University of Cambridge, Madingley Road,
  Cambridge CB3 0HA, UK \\
  $^2$ School of Earth \& Space Exploration, Arizona State University, 781 E Terrace Mall, Tempe, AZ 85287-6004, USA \\
  $^3$ Department of Astronomy \& Astrophysics, The Pennsylvania State University, State College, PA, USA \\
  $^4$ Center for Exoplanets and Habitable Worlds, The Pennsylvania State University, State College, PA, USA \\
}

\pubyear{2016}

\begin{document}
\label{firstpage}
\pagerange{\pageref{firstpage}--\pageref{lastpage}}
\maketitle

\begin{abstract}
This paper considers the dynamics of scattering of planetesimals or planetary embryos by a planet on
a circumstellar orbit.
We classify six regions in the planet's mass versus semimajor axis parameter space
according to the dominant outcome for scattered objects:
ejected, accreted, remaining, escaping, Oort Cloud, depleted Oort Cloud.
We use these outcomes to consider which planetary system architectures maximise the
observability of specific signatures,
given that signatures should be detected first around systems with optimal architectures
(if such systems exist in nature).
Giant impact debris is most readily detectable for $0.1-10M_\oplus$
planets at 1-5\,au, depending on detection method and spectral type.
While A stars have putative giant impact debris at $4-6$\,au consistent with this
sweet spot, that of FGK stars is typically $\ll1$\,au contrary to expectations;
an absence of $1-3$\,au giant impact debris could indicate a low frequency of
terrestrial planets there.
Three principles maximise cometary influx from exo-Kuiper belts:
a chain of closely separated planets interior to the belt, none of which is
a Jupiter-like ejector;
planet masses not increasing strongly with distance
(for a net inward torque on comets);
ongoing replenishment of comets, possibly by embedded low-mass planets.
A high Oort Cloud comet influx requires no ejector and architectures that maximise the
Oort Cloud population.
Cold debris disks are usually considered classical Kuiper belt analogues.
Here we consider the possibility of detecting scattered disk analogues, which could be betrayed by a
broad radial profile and lack of small grains, as well as spherical 100-1000\,au mini-Oort Clouds.
Some implications for escaping planets around young stars, detached planets akin
to Sedna, and the formation of super-Earths, are also discussed.
\end{abstract}

\begin{keywords}
  circumstellar matter --
  stars: planetary systems: formation.
\end{keywords}

\section{Introduction}
\label{s:intro}
While the dynamics of extrasolar planetary systems can appear complex,
consideration of how Keplerian orbits are perturbed using the disturbing
function shows that it is usually possible to consider dynamical
interactions as the sum of three distinct types of perturbation \citep[see][]{MurrayDermott1999}.
That is, resonant perturbations that act when the ratio of orbital periods are close
to a ratio of two integers, secular perturbations that act over long timescales
at all locations in the system, and short-period perturbations that usually average
to zero but become important when objects undergo close encounters (i.e., scattering).
For many populations of small bodies in planetary systems, and indeed the planetary
system itself, the dynamical evolution is dominated by one type of perturbation.
For example, since secular perturbations are unavoidable, these perturbations are
often dominant \citep{Wyatt1999, LeePeale2003}.
However, for many populations resonances are also important \citep{Fabrycky&Murray-Clay2010, Batygin2015},
even if the objects are not actually in resonance \citep[e.g., ][]{Rasio1992}.

Here we propose a framework within which to consider the outcomes of
scattering processes.
The pioneering work of \citet[][hereafter T93]{Tremaine1993}, itself based on previous work such as that
of \cite{Duncan1987}, showed how the ultimate fate of
any test particles encountering and subsequently being scattered by a planet
depends on the planet's mass and semimajor axis.
Thus \citetalias{Tremaine1993} showed that the planet mass versus semimajor axis
parameter space could be divided into regions with different outcomes.
That work focussed on the possibility of implanting comets in the Oort
Cloud, and that has also been the emphasis of subsequent work that extended this
analysis.
For example, \cite{BrasserDuncan2008} showed how additional constraints can be
placed in this parameter space that depend on the eccentricity of the planet's orbit
(which is ignored here in the first instance), and gave consideration to multiple
planet systems (which we will consider later).

However, the value of the division of parameter space in this way extends beyond the
formation of Oort Clouds.
Indeed this parameter space outlines the likely outcome for a particle undergoing
repeated encounters with that planet.
Many dynamical populations evolve in a manner that is dominated by multiple scatterings
with planets.
This applies to any objects that are on planet-crossing orbits, although
secular and resonant perturbations can also become important in this regime
\citep[e.g.,][]{Levison2003, Tamayo2014, Beust2014}.
Generally-speaking, the scattered population is made up of
objects that were either born on planet-crossing orbits, or those for which perturbing forces nudged
them onto planet-crossing orbits at a later date. 
In the Solar System the most obvious scattered population is the comets \citep[e.g.][]{Duncan1997}.
This includes both objects born in the vicinity of the planets that have been evolving
ever since through scattering, including placement in the Oort Cloud (e.g., the Long-Period comets),
and those born on quasi-stable orbits far from planets, but which have more recently
been perturbed onto planet-crossing orbits (e.g., the Jupiter-Family comets).
Other small body populations are also expected to follow this type of evolution.
For example, in the inner Solar System, debris that is born on planet-crossing orbits
includes that created in the giant impact which created the Earth's Moon \citep{Canup2001},
while debris that is perturbed onto planet-crossing orbits at a later date includes
the Near-Earth Asteroids \citep{Bottke2002}.

Analogy with the Solar System means that scattering processes may also apply to small
bodies such as asteroids and comets orbiting within extrasolar planetary systems.
Such small bodies are known to be present around many nearby stars from the detection
of circumstellar dust known as a debris disk that is created as these larger planetesimals are
destroyed \citep[e.g.,][]{Wyatt2008}.
If scattering processes are at play in these systems then the structure of their debris disks may
bear the imprint of those scattering processes.
Indeed, there are several systems for which it is proposed that debris has been seen following
a giant impact onto a planet which controls the subsequent evolution of that debris
\citep[e.g.,][]{Lisse2012}, and there is mounting evidence for the existence of exocomets
\citep[e.g.,][]{Beichman2005, Kiefer2014b, Boyajian2016}.
Furthermore, while most known debris disks are usually considered to be classical Kuiper belt
analogues, in that they are comprised of objects that orbit far enough from planets to
not undergo strong encounters, the possibility of strong scattering is brought home by the
discovery of the Fomalhaut system in which a planet-like object is seen to be on an orbit that
crosses the debris belt \citep{Kalas2013}.

It is also thought that planets themselves undergo epochs of intense scattering.
Such a rearrangement of the planets has been proposed in the Solar System to explain 
the moderate orbital eccentricities of the giant planets \citep[e.g.,][]{Tsiganis2005}.
The high eccentricities of the giant planets of extrasolar planetary systems has likewise
been proposed to originate in an epoch of planet-planet scattering \citep[e.g.,][]{Juric2008, Chatterjee2008}.
Many aspects of the scattering evolution discussed in this paper will also apply to populations
of larger bodies, and so the framework discussed herein can also be used to consider some aspects of the
dynamical evolution of, say, scattered planetary embryos.

In \S \ref{s:mpvap} we replicate the division of parameter space as presented
in \citetalias{Tremaine1993}, with only minor modifications, but give
equal consideration to outcomes other than the formation of
an Oort Cloud.
Then in \S \ref{s:applications} we show what the parameter space looks like for 4 specific cases
which are also used to corroborate the ability of the model to make predictions for
the outcome of numerical simulations of scattering processes in the literature.
In \S \ref{s:outcomes} we then use the parameter space division to consider
how to design planetary system architectures to maximise specific outcomes.
Whether such planetary system architectures exist in nature
is another matter, but throughout the paper we refer to observations
of extrasolar planets, Solar System minor planets, and extrasolar debris
disks to which this method may be applied.
Conclusions are given in \S \ref{s:conc}.

\section{Planet mass versus semimajor axis parameter space}
\label{s:mpvap}
\begin{figure*}
  \begin{center}
    \begin{tabular}{c}
      \includegraphics[width=2\columnwidth]{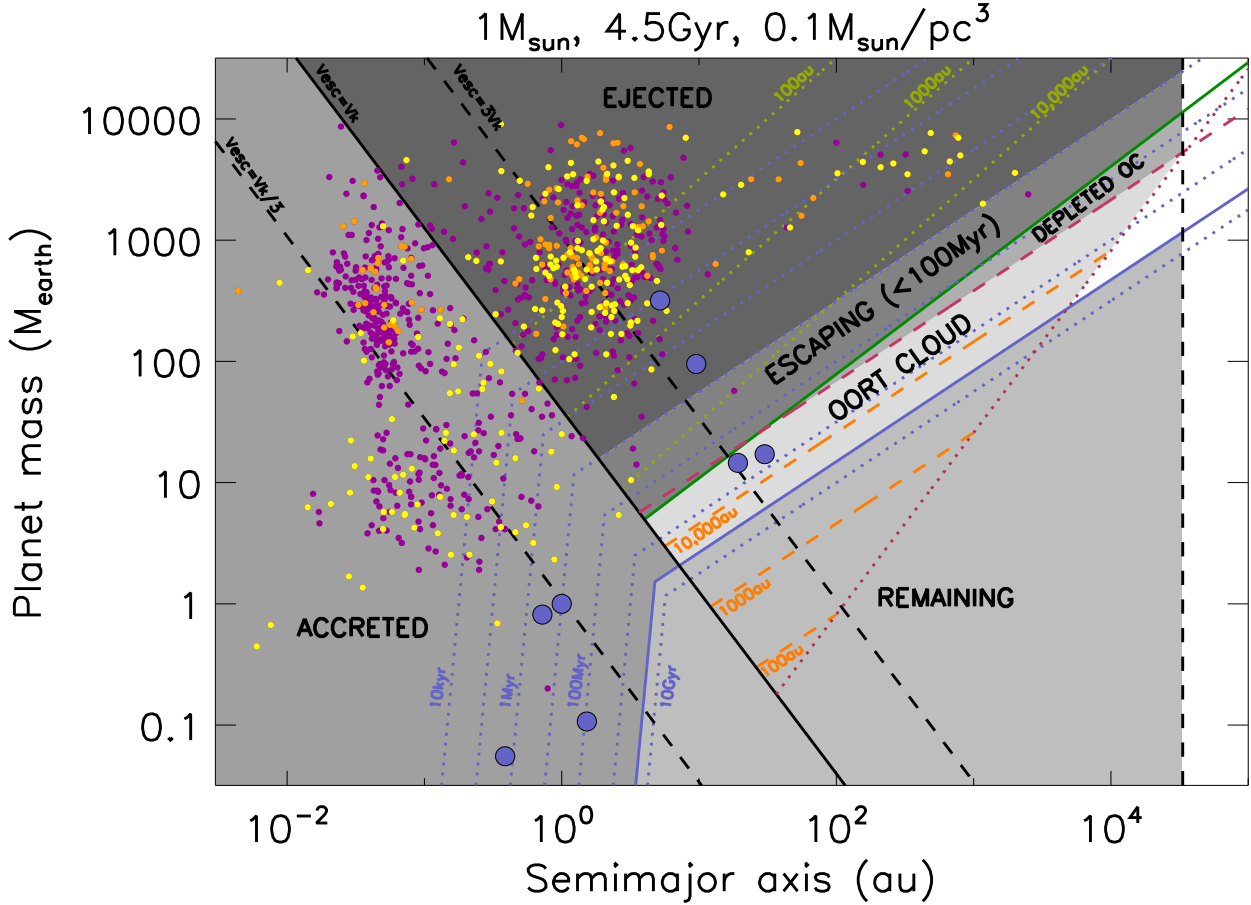}
    \end{tabular}
    \caption{Planet mass versus semimajor axis parameter space.
    The yellow, purple and orange dots show the known exoplanets found around $M_\star<0.6$,
    $0.6<M_\star<1.4$ and $M_\star>1.4$ stars, respectively;
    blue dots are the Solar System planets.
    The shading shows the dominant outcome of scattering interactions with a planet of
    given parameters (and density 1\,g\,cm$^{-3}$) orbiting a $1M_\odot$ star for 4.5\,Gyr of evolution
    in an environment with mass density 0.1\,$M_\odot$\,pc$^{-3}$.
    The timescales to achieve these outcomes are given by the blue dotted lines (eqs.~\ref{eq:tdiff}
    and \ref{eq:tacc});
    the solid blue line corresponds to 4.5\,Gyr.
    The semimajor axes at which planets eject particles are shown with yellow dashed
    lines (eq.~\ref{eq:aej}), and those at which particles are implanted in the Oort Cloud are shown with orange
    dashed lines (eq.~\ref{eq:af}).
    The black diagonal lines correspond to planets with a constant ratio of escape velocity
    to Keplerian velocity (eq.~\ref{eq:vescvk}).
    The red dashed line is that at which stellar encounters strip particles from the Oort Cloud (eq.~\ref{eq:thalf}),
    and the vertical black dashed line is the radius at which these encounters would have removed the
    planet over the system age (eq.~\ref{eq:ah}).
    }
   \label{fig:ss}
  \end{center}
\end{figure*}

Consider a planet of mass $M_{\rm p}$ on a circular orbit around a star of
mass $M_\star$ with a semimajor axis $a_{\rm p}$.
Throughout the paper we assume that $M_{\rm p}$ is in $M_\oplus$,
$M_\star$ is in $M_\odot$, and $a_{\rm p}$ is in au;
the units used in the paper are summarised in Table~\ref{tab:units}.
Fig.~\ref{fig:ss} shows the parameter space that is most important for
determining the outcome of scattering interactions with that planet, i.e.,
planet mass versus semimajor axis.
In \S \ref{ss:planets} we describe the populations of known planets shown
on the figure, which includes both Solar System and extrasolar planets,
as well as the debris populations (which are not shown).
The shading on Fig.~\ref{fig:ss} shows 6 different regions of parameter space that
are defined by the most likely outcome for planetesimals encountering a planet
in that region of parameter space (assuming it is the only planet in the system):
{\bf accreted} (planetesimal ends up colliding with the planet),
{\bf ejected} (planetesimal ends up being ejected from the system),
{\bf remaining} (planetesimal remains in the system),
{\bf escaping} (planetesimal will soon be ejected but is currently still undergoing scattering),
{\bf Oort cloud} (planetesimal ends up in the Oort Cloud),
{\bf depleted Oort Cloud} (planetesimal was put in the Oort Cloud but has subsequently been ejected).
This division is guided by lines that were derived in \citetalias{Tremaine1993}, which are
described in \S \ref{ss:vescvk}, \ref{ss:tdiff} and \ref{ss:ttide}.
More specifically, Fig.~\ref{fig:ss} shows how the parameter space is divided for
the Solar System, i.e., for planets orbiting a 4.5\,Gyr-old $1M_\odot$
star in a stellar environment with local mass density 0.1\,$M_\odot$\,pc$^{-3}$;
thus this figure is directly comparable with Fig.\,2 of \citetalias{Tremaine1993}, with mostly cosmetic
changes.
Note that this figure is not intended to show the only outcome for encounters with such
planets.
Rather the shading represents the expected dominant outcome, with the caveat that
the dominant outcome may also be influenced by the initial parameters of the
planetesimal's orbit as well as the other planets in the system.

\begin{table}
  \centering
  \caption{Units of parameters introduced in \S \ref{s:mpvap}.}
  \label{tab:units}
  \begin{tabular}{lll}
     \hline
     Parameter                & Symbol         & Units \\
     \hline
     Stellar luminosity       & $L_\star$      & $L_\odot$ \\
     Stellar mass             & $M_\star$      & $M_\odot$ \\
     Planet mass              & $M_{\rm p}$    & $M_\oplus$ \\
     Planet semimajor axis    & $a_{\rm p}$    & au \\
     Oort Cloud radius        & $a_{\rm f}$    & au \\
     Ejection semimajor axis  & $a_{\rm ej}$   & au \\
     Local mass density       & $\rho_0$       & 0.1\,$M_\odot$\,pc$^{-3}$ \\
     Planet density           & $\rho_{\rm p}$ & 1\,g\,cm$^{-3}$ \\
     Stellar age              & $t_\star$      & Gyr \\
     \hline
  \end{tabular}
\end{table}

\subsection{Planet and debris parameters}
\label{ss:planets}
One of the cosmetic improvements to Fig.~\ref{fig:ss} that was 
not available to \citetalias{Tremaine1993} is the addition of known exoplanets.
Here we took the known exoplanets from the NASA exoplanet archive
\footnote{http://exoplanetarchive.ipac.caltech.edu/} on 25 Feb 2016.
The planets are coloured by the mass of their host star (yellow for $M_\star < 0.6M_\odot$,
purple for $0.6M_\odot < M_\star < 1.4M_\odot$ and orange for $M_\star > 1.4M_\odot$).
The Solar System planets are also shown as the larger blue circles.
This does not necessarily include all known exoplanets, but serves to illustrate the main
features of the exoplanet population.

The main populations are \citep[see ][]{Udry2007}:
(i) the Hot Jupiters centred around $1M_{\rm jup}$, 0.03\,au,
which are found around $\sim 1$\% of stars;
(ii) the super-Earths centred around $10M_\oplus$, 0.03-1\,au,
which are found around 30-50\% of stars;
(iii) the eccentric Jupiters centred around a few $M_{\rm jup}$, 3\,au,
which are found around $\sim 5$\% of stars;
(iv) the long-period giants that are $\sim 10M_{\rm jup}$ at $>10$\,au,
which are found around a few $\%$ of stars \citep{Bowler2016}.
Detection biases mean that true terrestrial planets and Neptune analogues
are rare in the exoplanet population, so the ubiquity of such planets is
unknown at present.
Estimates of these populations can be made, however, either by extrapolation of
the super-Earth population \citep{Howard2010}, or from the small numbers
of micro-lensing detections \citep{Sumi2010}.

The last 20 years have also provided a significant increase in our understanding of the
populations of debris (i.e., the planetesimals which may be scattered by planets)
around nearby stars, and indeed our own Sun \citep[for reviews see ][]{Wyatt2008, Matthews2014pp6}.
This paper will summarise the main populations.
Approximately $20$\% of Sun-like stars, and a similar if not higher fraction of A stars, host
cold debris belts that are detected in the far-IR \citep[][Sibthorpe et al. in prep.]{Eiroa2013, Thureau2014}.
These all have large inner holes that are empty of dust and when imaged are often shown
to be radially confined to narrow rings, although broad disks covering a factor of
a few in radius are also known.
The inner holes have radii in the range 10-150\,au, with some obvious selection biases
towards larger disks in those that can be imaged.
Thus these are considered to be exo-Kuiper belts.
However, whether these are analogues to the classical Kuiper belt (i.e., objects born on
stable orbits) or to the scattered disk (i.e., objects undergoing scattering with planets),
is not much discussed.
The longevity of some disks \citep[such as HD~207129, ][]{Lohne2012}, and the narrowness
of others \citep[such as Fomalhaut, ][]{Kalas2005}, argue for a classical Kuiper belt interpretation
for these systems, while broad disks seen around young stars could have a scattered disk
interpretation.
The azimuthal structure seen toward several imaged disks is a strong clue to the dynamics
of these populations, and for some disks with clumpy structure this has been used to argue
for a population analogous to the resonant Kuiper belt objects that were trapped in
resonance with a migrating planet \citep{Wyatt2003, Dent2014}.

Less frequently the inner regions of planetary systems are also seen to host abundant debris.
In some cases the hot dust is the only debris component present in the system at detectable levels
\citep{Beichman2005}, but there are a number of debris disks with spectral energy distributions suggesting
two temperatures of debris \citep{Wyatt2005, Morales2009, Chen2014, Kennedy2014}.
The origin of this hot dust is a matter of considerable debate.
Possibilities include:
(i) These are analogues to the asteroid belt, although this is unlikely
for old systems with dust $\ll 3$\,au, since such belts would have been depleted by collisional
erosion \citep{Wyatt2007hotdust}.
(ii) This is dust released in a recent giant impact \citep{Rhee2008, Lisse2009, Jackson2012},
possibly similar to the Earth's Moon-forming collision.
(iii) This is dust dragged in from the outer Kuiper belt by Poynting-Robertson drag
\citep{vanLieshout2014,Mennesson2014,Kennedy2015prdrag}.
(iv) The dust is released from a comet-like population, either scattered in from an exo-Kuiper belt
\citep{Nesvorny2010, Bonsor2012nbody},
or from a population analogous to the long-period comets \citep{Beichman2005, Wyatt2010}.

Our ability to know the dynamics of the planetesimal populations in specific systems is hampered
by the fact that usually there is little information on the exoplanet system within which they
reside.
Nevertheless, a growing number of systems host planets and debris
\citep{Wyatt2012, Kalas2013, Kennedy2014, Moro-Martin2015}.
Perhaps the most famous planet plus debris system is HR8799 \citep{Marois2008, Su2009},
which will be considered in more detail in \S \ref{ss:younga}.

\subsection{Maximum kick: accretion vs ejection}
\label{ss:vescvk}
One of the most important lines on Fig.~\ref{fig:ss} is that for which the
planet's escape velocity $v_{\rm esc}$ is equal to its Keplerian velocity $v_{\rm k}$
\citep[e.g.,][]{Ford2008}.
Remembering the units given in Table~\ref{tab:units}, this can be found from the following
relation between planet and stellar properties and the ratio $v_{\rm esc}/v_{\rm k}$
\begin{equation}
  M_{\rm p} = 40 M_\star^{3/2} a_{\rm p}^{-3/2} \rho_{\rm p}^{-1/2} (v_{\rm esc}/v_{\rm k})^3,
  \label{eq:vescvk}
\end{equation}
where the planet's density $\rho_{\rm p}$ has units of 1\,g\,cm$^{-3}$, and is assumed to be
1\,g\,cm$^{-3}$ for figures in this paper.
The significance of this boundary is that the maximum velocity kick that a particle
can receive in a single encounter with the planet is $v_{\rm esc}/\sqrt{2}$, since larger kicks
require larger deflection angles that are only possible by approaching the planet with
a smaller impact parameter which would result in a collision.
Assuming an initially circular orbit, a kick of order the Keplerian velocity is sufficient to put
the particle on an unbound trajectory (if it is oriented in the right direction). 
Thus eq.~\ref{eq:vescvk} gives the approximate limit at which particles can be put on unbound trajectories
in a single encounter.
Since many more encounters are expected with deflection angles just below this limit than
those above (i.e., with impact parameters at larger distances), one can expect that the
further to the right of this line a planet is, the more
kicks an object receives that can eject it from the system before it has encounter that is close
enough to collide with the planet;
i.e., collisions with the planet become unlikely as an outcome compared to ejection of the object.
Conversely, the further to the left of this line a planet is, the more likely an object is
to collide with the planet before it receives sufficient kicks to increase its eccentricity so that
its orbit evolves in a cometary diffusion regime.

This argument is phrased slightly differently to that presented in \citetalias{Tremaine1993}, which considered the
collisional lifetime for objects in the cometary diffusion regime,
but eq.~14 of \citetalias{Tremaine1993} has the same scaling and is identical to eq.~\ref{eq:vescvk} within a
factor of 2.
Also shown on Fig.~\ref{fig:ss} for reference is the planets for which $v_{\rm esc}/v_{\rm k}$
is 1/3 and 3.

Thus the conclusion is that, in the absence of other considerations, the eventual fate of objects
to the left of the line is accretion onto the planet, while that of objects to the right of that
line is ejection from the system.
This is broadly in agreement with, for example, the simulations of \cite{Raymond2010} which showed that Jupiter-mass planets
are more likely to lead to ejections whilst lower mass planets are more likely to collide (see their Fig.~4).
The multi-planet N-body simulations of \cite{Veras2016} also exhibit a clear difference between the fate of 
scattered planets that started in different regions of Fig.~\ref{fig:ss};
e.g., 70/82 of their high mass planets (Jupiters and Saturns that lie in the ejection regime) that
leave the simulations are ejected (see their Table A12), whereas of their Uranus and
Neptune mass planets (that start in or close to the accretion regime) that leave the simulations,
8/33 are ejected and 9/33 collide with one another (see their Table A13).

\subsection{Timescale: remaining vs lost}
\label{ss:tdiff}
The discussion in \S \ref{ss:vescvk} for the eventual fate of scattered particles does not account for the timescale
for the ejection and accretion outcomes to occur.
Thus another important line on Fig.~\ref{fig:ss} is that for which the timescale for these outcomes is
the age of the system (which is the solid blue line).

For the region to the right of the line described by eq.~\ref{eq:vescvk}, i.e. for particles
for which the eventual outcome is ejection, the outcome timescale is taken to be the cometary diffusion
time as empirically derived in \citetalias{Tremaine1993} (their eq. 3), and later derived analytically in
\citet{BrasserDuncan2008} (see their Appendix A).
Thus the planet parameters for which ejection occurs on a timescale $t_\star$ in Gyr are given by
\begin{equation}
  M_{\rm p} = M_\star^{3/4} a_{\rm p}^{3/4} t_\star^{-1/2}. \label{eq:tdiff}
\end{equation}
Fig.~\ref{fig:ss} shows with dashed blue lines the planets for which ejection
occurs on different timescales, as given in the annotation, and in solid blue line
that for which ejection occurs on a timescale of the assumed age of the system $t_\star = 4.5$\,Gyr.

For the region to the left of the line described by eq.~\ref{eq:vescvk}, i.e. for particles
for which the eventual outcome is accretion, the outcome timescale is taken to be the collision time
under the assumption that the relative velocities at which the particles encounter the planet
is of order the escape velocity of the planet.
More specifically, we consider an outcome timescale that might be expected from debris released in a
giant impact, and so use a relative velocity distribution scaled by the planet's escape velocity
to that expected for ejecta released in the Moon-forming collision, further assuming an axisymmetric spatial
distribution \citep[see][]{Jackson2012}.
This results in a gravitational focussing factor is the same for all planets \citep{Wyatt2016},
and means that the planet parameters for which accretion occurs on a timescale $t_\star$ are
\cite[using eq. 11 of ][ with $\Delta v = v_{\rm esc}$]{Jackson2012} given by
\begin{equation}
  M_{\rm p} = 10^{-6} M_\star^{-3} a_{\rm p}^{12} \rho_{\rm p}^{5/2} t_\star^{-3}, \label{eq:tacc}
\end{equation}
and similar lines to those for the ejection outcome are plotted on Fig.~\ref{fig:ss}.
For example, eq.~\ref{eq:tacc} shows that debris from giant impacts involving the Earth, or those
involving Mercury, is reaccreted onto those planets on timescales of $\sim 10$\,Myr and
$\sim 0.6$\,Myr, respectively (for the nominal assumption of $\rho_{\rm p}=1$\,g\,cm$^{-3}$, and a
factor of a few longer for their actual densities).

While the above assumptions mean that the timescale given by eq.~\ref{eq:tacc} is most applicable to
giant impact debris, a relative velocity that is comparable to the planet's escape velocity may also be a
reasonable estimate for debris that is being stirred by the planet.
However, stirring by other planets could set a higher relative velocity for the debris in which case
the lines on Fig.~\ref{fig:ss} would have a shallower dependence on $a_{\rm p}$.

Since the two lines described by eqs.~\ref{eq:tdiff} and \ref{eq:tacc} are not equal at the boundary
given by eq.~\ref{eq:vescvk}, this would result in the appearance of a discontinuity at that line
(which is avoided on Fig.~\ref{fig:ss} by showing the lines up to the point where they intersect).
Clearly the approximations used to delineate the different outcomes, and to quantify the timescales
for those outcomes, break down close to the boundaries
(e.g., eq.~11 of \citet{Jackson2012} used an expansion applicable only for small $\Delta v / v_{\rm k}$
and furthermore assumed a toroidal distribution of debris).
This emphasises the point that Fig.~\ref{fig:ss}
should only be used as a guide to indicate the expected outcome and its timescale, and that more
detailed numerical simulations are needed, in particular to assess the outcomes near the boundaries.
In any case, the assumption here is that objects below the solid blue lines have not had
sufficient time to achieve their eventual fate described in \S \ref{ss:vescvk},
and so this region is labelled as remaining.

\subsection{Tide: Oort cloud vs ejected}
\label{ss:ttide}
\citetalias{Tremaine1993} show how the eventual outcome in the region that would have been
considered to result in ejection by the reasoning above can instead result in the
particles being deposited in the Oort Cloud.
This is because as the objects undergo cometary diffusion, i.e., keeping their pericentres
close to the planet but receiving kicks which increase their semimajor axes,
they would be expected to be ejected after they reach a semimajor axis $a_{\rm ej}$ (in au)
at which the individual kicks they receive when encountering the planet are sufficient to unbind them
from the system.
This means that planets of
different masses eject particles after they reach different distances as indicated
with the yellow dotted lines on Fig.~\ref{fig:ss} given by (see eq. 8 of \citetalias{Tremaine1993})
\begin{equation}
  M_{\rm p} = 3 \times 10^4 M_\star a_{\rm p} a_{\rm ej}^{-1}.\label{eq:aej}
\end{equation}

However, the diffusion takes place on a finite timescale that depends on the semimajor axis
that the particle has reached, and there are additional perturbations to the particle's
orbit from the Galactic tide and from stellar encounters, the timescales for which also depend
on semimajor axis \citep[see][]{Heisler1986}.
Since both tides and encounters can act to raise the particle's pericentre this would stop the cometary diffusion
and freeze the particle's semimajor axis at whatever value it has reached at that time, thus depositing it in the Oort Cloud
\citep[][]{Duncan1987}.
In Appendix~\ref{a:enc} we show that, unless the object being scattered is orbiting a massive star
(see eq.~\ref{eq:enc}), the relevant
timescales for Galactic tides are shorter than those of stellar encounters, and so the latter can be neglected
in the following analysis.
Thus the semimajor axis at which this freezing occurs, $a_{\rm f}$ (in au), is the location at which the tidal timescale
equals that for cometary diffusion, and the orange dashed lines on Fig.~\ref{fig:ss} show 
the planets that result in freezing at different Oort Cloud sizes given by (rearranging eq. 7 of \citetalias{Tremaine1993})
\begin{equation}
  M_{\rm p} = 0.8 \times 10^{-3} M_\star^{1/2} a_{\rm p}^{3/4} a_{\rm f}^{3/4} \rho_0^{1/2},\label{eq:af}
\end{equation}
where $\rho_0$ is the local mass density in units of that near the Sun of 0.1\,$M_\odot$\,pc$^{-3}$ \citep{Holmberg2000}.

Comparing the lines given by eqs.~\ref{eq:aej} and \ref{eq:af} it is clear that there is
a line on Fig.~\ref{fig:ss} above which a particle
is ejected before it can reach the semimajor axis at which it would be implanted
in the Oort Cloud.
This is shown in green on Fig.~\ref{fig:ss} and corresponds to (see eq. 9 of \citetalias{Tremaine1993})
\begin{equation}
  M_{\rm p} = 1.5 M_\star^{5/7} a_{\rm p}^{6/7} \rho_0^{2/7}.
  \label{eq:mejoc}
\end{equation}
Ejection is the outcome above this line, and the yellow dotted lines show the semimajor axis
at which ejection occurs (eq.~\ref{eq:aej}).
While the outcome for particles encountering all planets above the green line should be {\it ejected},
we further subdivide this ejection outcome to include an {\it escaping} region to emphasise that
the timescale for ejection can be relatively long in this region, and so it is possible to see particles
in the process of being ejected.
The motivation for this area of parameter space will become clear in \S \ref{ss:younga}.
Below the green line the eventual outcome would be for particles to be implanted in the
Oort Cloud, which would be at a radius given by the orange dashed lines (eq.~\ref{eq:af}), but noting that the
timescale to reach this outcome could be longer than the age of the system for low mass
planets.

Two other lines are shown on Fig.~\ref{fig:ss} relating to the Oort Cloud outcome.
One is that for which the half-life for objects in the Oort Cloud due to perturbations
from passing stars is equal to the age of the system, which is shown as the
red dashed line given by (see eq.~16 of \citetalias{Tremaine1993})
\begin{equation}
  M_{\rm p} = 7 M_\star^{3/4} a_{\rm p}^{3/4} \rho_0^{-1/4} t_\star^{-3/4}.\label{eq:thalf}
\end{equation}
Above this line, objects may be implanted in the Oort Cloud (if the planet is not massive
enough to eject the particles before they reach this location), but they would be subsequently
removed by the passage of nearby stars;
i.e., the ultimate fate of particles encountering planets in the depleted Oort Cloud region is
ejection, though some fraction may remain in a depleted Oort Cloud
\footnote{Note that eq.~\ref{eq:thalf} becomes inaccurate for Oort Clouds with semimajor axes
approaching the tidal radius of the star \citep[see Fig. 7 of][]{Weinberg1987};
i.e., large radius Oort Clouds may be more readily depleted than assumed.}.
The other line is the red dotted line, which is that for which the semimajor axis at which the
planet would implant objects in the Oort Cloud is equal to that of the planet itself
(see eq.~13 of \citetalias{Tremaine1993})
\begin{equation}
  M_{\rm p} = 0.8 \times 10^{-3} M_\star^{1/2} a_{\rm p}^{3/2} \rho_0^{1/2}.\label{eq:afapl}
\end{equation}
Clearly a planet to the right of this line could not form an Oort Cloud.

Finally, planets are not considered if they are either outside the tidal radius of the
star (see eq.~10 of \citetalias{Tremaine1993}), which is those beyond
\begin{equation}
  a_{\rm p} = 1.9 \times 10^5 M_\star^{1/3} \rho_0^{-1/3}, \label{eq:at}
\end{equation}
or if they have a have a half-life due to perturbations from passing stars that is
shorter than the system age (see eq.~15 of \citetalias{Tremaine1993}), which is those beyond
\begin{equation}
  a_{\rm p} = 1.5 \times 10^{5} M_\star \rho_0^{-1} t_\star^{-1}. \label{eq:ah}
\end{equation}
Equations \ref{eq:at} and \ref{eq:ah} are shown with vertical black dotted and dashed
lines respectively on the figures (when they fall within the plotted range).

\subsection{Applicability to scattered planets}
\label{ss:app}
The division of parameter space described in this section applies specifically to the scattering
of test particles, i.e., those with insufficient mass to affect the orbit of the planet doing the
scattering.
However, this only restricts the applicability to the scattering of objects that are an
order of magnitude or so less in mass than the planet, which means that it can also apply to
scattered planets (so long as they are small in mass in comparison with the planet doing the
scattering).

To assess this applicability more quantitatively, consider two planets of mass $M_1$ and $M_2$ on
circular coplanar orbits at a distance $a$ that undergo scattering leaving
$M_1$ on an orbit with an apocentre at $a$ and eccentricity $e_1$, and $M_2$ on an orbit with a
pericentre at $a$ and an eccentricity $e_2 \approx 1$.
Conservation of angular momentum shows that
\begin{equation}
  \frac{M_2}{M_1}=\frac{1-\sqrt{1-e_1}}{\sqrt{2}-1},
  \label{eq:m2m1}
\end{equation} 
which means that $M_2/M_1 < 2.4$, and so the planet that is scattered out cannot be significantly more
massive than that which did the scattering, but can be comparable in mass.
This means, for example, that Jupiter mass planets at large distance would require planets of Jupiter
mass or greater orbiting close-in if the former are to have been scattered out, though the close-in planet
need not continue to exist after the scattering, since it could have collided with the star. 

Equation~(\ref{eq:m2m1}) can also be used to estimate the eccentricity imparted to the inner planet in
this process, since for small $e_1$ this reduces to $M_2/M_1 \approx 1.2e_1$.
Thus a rule of thumb is that the maximum eccentricity gained by $M_1$ is of order $M_2/M_1$ (unless $M_2$
is ejected in a single encounter that placed it significantly above escape velocity from the star's
gravitational potential).
This explains the magnitude of the eccentricity imparted to Jupiter on scattering Neptune
and Uranus into the Kuiper belt in the instability proposed by \cite{Tsiganis2005},
since $e_{\rm Jup} \sim M_{\rm Nep}/M_{\rm Jup}$.

\section{Applications to specific system parameters}
\label{s:applications}
Here we consider what the planet mass versus semimajor axis parameter space looks like for 4 different
system parameters.
The primary aims of this section are to illustrate how this parameter space can be used to arrive at
conclusions about the dynamics of a particular system, to corroborate the success of this framework
at predicting outcomes by comparison with numerical simulations in the literature, and to introduce
some potential outcomes to be explored further in \S \ref{s:outcomes}.

\subsection{Current Solar System}
\label{ss:ss}
Having outlined the method for dividing the planet mass - semimajor axis parameter space
into regions in which planets have different outcomes in \S \ref{s:mpvap}, let us first
consider the implications of the resulting parameter space for a system with parameters
similar to that of the Solar System.
That does not mean a system with planets like those in the Solar System (although such a system
will be considered), rather a system orbiting a $1M_\odot$ star with an age of 4.5\,Gyr in
a local mass density of 0.1\,$M_\odot$\,pc$^{-3}$ (and assuming planet densities of 1\,g\,cm$^{-3}$),
as plotted in Fig.~\ref{fig:ss}.

{\it Oort Cloud: }
The conclusions that can be reached from this figure about the formation of the Oort Cloud
are well known.
For example, \citetalias{Tremaine1993} showed that the parameter space in which an Oort Cloud forms
is quite restricted, and that those Oort Clouds that do form have a narrow range of semimajor axes $\sim 10,000$\,au.
While this parameter space is inhabited by Uranus and Neptune in the Solar System, which should thus readily
supply objects to the Oort Cloud \citep[even accounting for the possibility that these planets may have
started closer to the Sun;][]{Tsiganis2005}, it could be that Oort Clouds are relatively rare.
Many simulations have confirmed these predictions regarding the ability of planets to implant objects
in the Oort Cloud \citep[e.g., ][]{Dones2004}, while also showing further subtleties such as 
the ability of Jupiter and Saturn to place a small fraction of the objects they scatter
into the Oort Cloud even if ejection is the predominant outcome in such encounters \citep{Brasser2008}.
The timescale predicted for the scattering process to occur is also born out in numerical simulations.
For example, compare the prediction of Fig.~\ref{fig:ss} that it should take 0.1-1\,Gyr for
Uranus and Neptune to implant material in the Oort Cloud with Fig. 13 of \cite{Dones2015}.
The radius at which the Oort Cloud forms in the simulations also agrees with that predicted
of $\sim 10,000$\,au, with some studies including differentiation between inner
and outer Oort cloud \citep[e.g., ][]{Lewis2013, Brasser2015}.
Inspection of Fig.~\ref{fig:ss} shows that Oort Clouds could form at smaller orbital radii,
but that such an outcome requires both low mass planets and a large system age;
e.g., if Neptune and Uranus were each Earth mass then the Oort Cloud would be at $\sim 1000$\,au,
but would take $\sim 20$\,Gyr to form.
Another way to achieve a small Oort Cloud on a shorter timescale is to place the planetary system
in a dense stellar environment (as discussed in \S \ref{ss:youngss}).

{\it Ejected: }
As noted in \cite{Brasser2008}, many of the exoplanets known at that time
will end up ejecting most of the material they encounter.
The timescale for these planets to eject material is usually relatively rapid, typically $\ll 10$\,Myr.
That Jupiter is an efficient ejector of comets is common knowledge given a basic understanding of cometary
dynamics.
That planets across a wide range of masses and semimajor axes far from their star also eject nearby
material is also recognised by those studying planet formation \citep{Goldreich2004} and those studying
cometary evolution (see e.g., the simulations of \citet{Higuchi2006} that confirm
ejection as the dominant outcome for Jupiter mass planets at 1-30\,au and
0.1-10 Jupiter mass planets at 5\,au).
However, the region of parameter space of planets for which ejection is the most
likely outcome should be more widely acknowledged, since
the presence of such planets in a system has a significant
effect on the dynamics of scattered material.
Any orbit that crosses such planets has a high probability of being ejected from the system, and
so while material may pass back and forth between an orbit interior and exterior to a planet as
it undergoes multiple scattering events, it is unlikely for material to pass from an orbit that
is entirely interior to such a planet to one entirely outside the planet
(and likewise from an exterior to interior orbit), unless there is a force acting on this material
that changes the orbit faster than the scattering timescale.
Thus any system known to have an eccentric Jupiter planet or a long-period giant planet (i.e., 5-10\% of stars)
has an efficient ejector, and ejectors could be much more common given that there is a
large region of parameter space in which it cannot yet be known whether a system has an
ejector (e.g, Saturn-mass planets at 10-30\,au).
We return to this in \S \ref{ss:comet}, since it has implications for comet-like populations.
It is also the case that for circumbinary planets, the secondary star of the system would lie in the
ejected regime, explaining why planets that are placed onto orbits that cross the stellar region
end up being ejected from the system rather than impacting one of the stars \citep{Smullen2016}.

{\it Accreted: }
More exoplanets are now known in the region in which the dominant outcome is accretion onto the planet;
these are the Hot Jupiter and super-Earth populations discussed in \S \ref{ss:planets}.
Any material being scattered by such planets will end up being accreted onto the planets on
a relatively short timescale, which among other consequences facilitates continued growth of
these planets.
That material could consist of left-over debris from the planet formation process \citep[e.g., the
late veneer thought to have been accreted by the Earth;][]{Schlichting2012}, or debris that
finds its way into this region at a later epoch \citep[such as the Near Earth Asteroids,][]{Bottke2000}), 
and could include planetary embryos (e.g., \S \ref{ss:app}).
Thus, extra embryos do not tend to escape this region, but are accreted onto existing planets
\citep[e.g.,][]{Ford2008, Petrovich2014}.
Even if collisions between the embryos and the planets release a large mass of debris, this
debris would eventually be reaccreted onto the planets, as would any
escaping planetary moon.
This is confirmed in numerical simulations of the dynamical evolution of giant impact debris
which find that debris from the Earth's Moon-forming impact reaccretes onto the Earth on a
timescale of $\sim 15$\,Myr \citep{Jackson2012}, and that it takes
$\sim 0.3$\,Myr for debris released from a $18M_\oplus$ planet at 0.63\,au to reaccrete onto that
planet \citep{Wyatt2016}.
These timescales agree well with those predicted by eq.~\ref{eq:tacc} given that
this calculation does not account for the non-axisymmetric geometry or multi-planet interactions
involved in the simulations being compared to.
The eventual fate of giant impact debris is discussed further in \S \ref{ss:gi}.
The only way for mass that finds itself in this region to avoid ending up on a planet is for
it to end up on the star, or to get ground into dust that is removed by radiation pressure.
This principle may go some way to explaining why these inner
regions commonly retain a large mass in planets.
Again, this is well known from simulations of terrestrial planet formation \citep[e.g., ][]{Chambers2001},
but the region of parameter space to which this outcome applies as a fundamental principle
should be more widely acknowledged, since it implies that planets interact differently with circumstellar
material inside and outside the $v_{\rm esc}=v_{\rm k}$ line.
Indeed, it is suggestive that the known exoplanet populations appear to be separated by
this line, although the absence of $\sim 100M_\oplus$ planets in the $\sim 0.3$\,au
region is more likely explained by the rapid growth and migration of such planets in 
interactions with the gas disk \citep{Ida2004, Mordasini2009}.
Some aspects of super-Earth formation are discussed further in \S \ref{sss:se}.

{\it Remaining: }
The conclusion that planets in the inner regions ($\ll 5$\,au) accrete everything they
encounter only applies to planets above a certain mass, since low mass planets would not
have had enough time to accrete all of the material they encounter.
For a 4.5\,Gyr system this implies that primordial debris may be able to persist near
Mars-mass planets at 4\,au, or near more massive planets further out.
The same applies in the outer regions;
e.g., at the 700\,au distance of the putative planet nine in the Solar System \citep{Batygin2016},
even a $10M_\oplus$ planet would not have ejected planetesimals in its vicinity over 4.5\,Gyr.
However, particularly at small separations (but also further out), mutual collisions amongst
the debris need to be considered, since these would deplete this population at a rate which
depends on the total mass of debris and the size of the largest object in the debris population
as well as the level of stirring \citep{Wyatt2007collisionalcascade, Heng2010}.
Thus the outcomes shown on Fig.~\ref{fig:ss} are strictly only those that apply to
low mass debris populations, and it should be noted that for sufficiently massive debris populations
it may be possible to remove some of the debris expected to be {\it remaining} within the given
timescale by collisional grinding;
collisional grinding may also allow debris to avoid the fate of being accreted onto the planet
\citep[e.g., ][]{Jackson2012}.
This is considered further in \S \ref{ss:gi} in application to giant impact debris.

\begin{figure*}
  \begin{center}
    \begin{tabular}{c}
      \includegraphics[width=2\columnwidth]{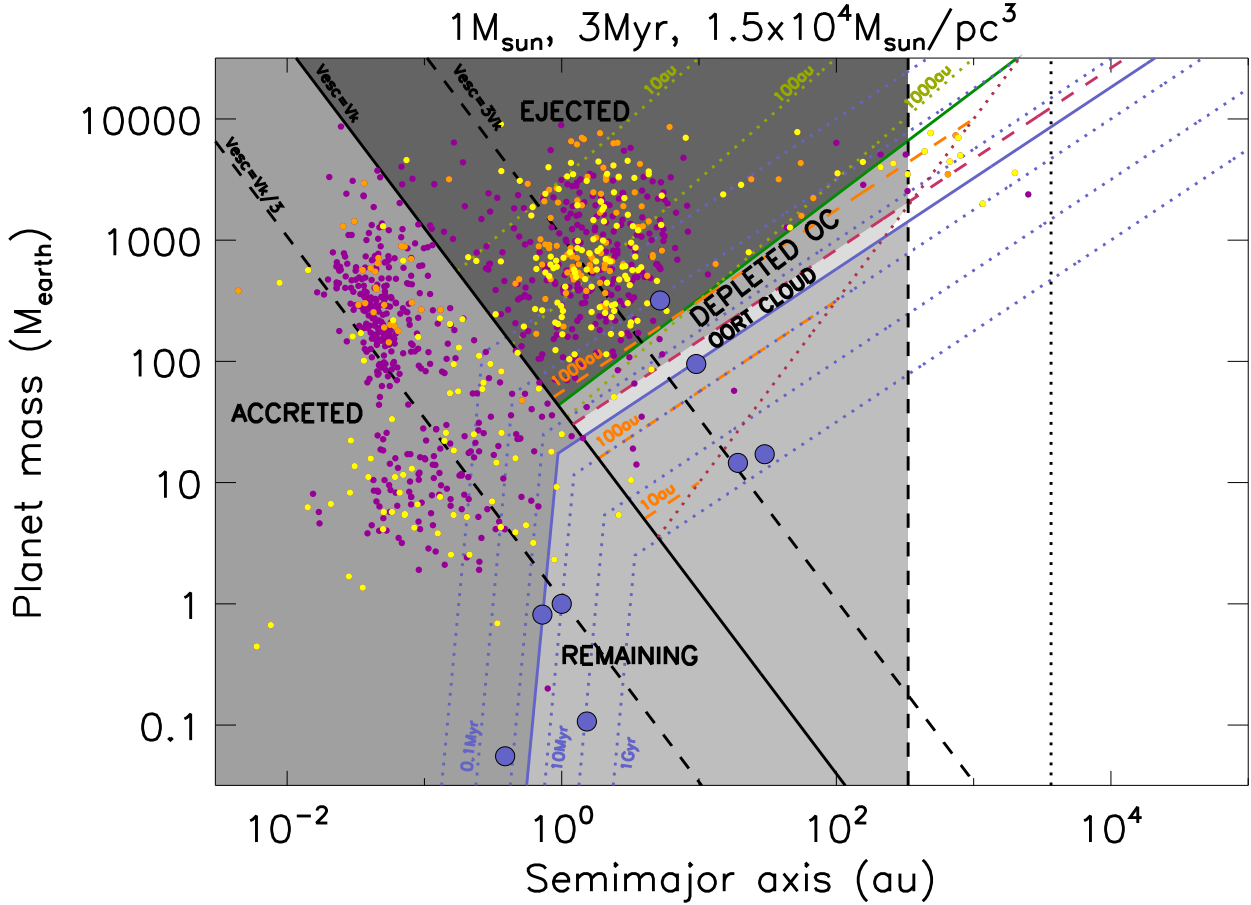}
    \end{tabular}
    \caption{As for Fig.~\ref{fig:ss}, but for a system orbiting a $1M_\odot$ star for 3\,Myr of evolution
    in an environment with mass density $1.5 \times 10^4$\,$M_\odot$\,pc$^{-3}$.
    The vertical black dotted line is the tidal radius of the star for this local mass density (eq.~\ref{eq:at}).
    }
   \label{fig:youngss}
  \end{center}
\end{figure*}

\subsection{Young solar system formed in a cluster}
\label{ss:youngss}

Fig.~\ref{fig:youngss} aims to recreate the conditions in the simulations of \cite{Brasser2006}
which explored the consequence of the Sun spending the first 3\,Myr of its life in a stellar cluster
with mean density $1.5 \times 10^4$\,$M_\odot$\,pc$^{-3}$, evidence for which may be present in
the isotopic composition of minor Solar System bodies \citep{Adams2010}.
For this calculation we used the equations of \S \ref{ss:ttide} assuming $\rho_0$ as the mean density.
While the tide from a stellar cluster acts on a slightly different timescale from that of a galactic disk,
its dependence on the various parameters scale in the same way, and the change in the pre-factor in
eq.~\ref{eq:af} is $<30$\% and so is ignored for the purposes of this plot given the much larger uncertainty
in the mean density.
Indeed, considerations of the birth environment of the Sun suggest that a mass density that is an order of
magnitude lower may be more appropriate \citep{Adams2010}, and more recent simulations have shown the need to
include the effect of gas in the cluster \citep[e.g., ][]{Brasser2007}.
Nevertheless, this plot illustrates the different possible outcomes that may be
achieved by placing the planetary system in a dense environment.
Apart from the younger age, which means that the planets have to be considerably more massive to
have induced their respective outcome by 3\,Myr, the main consequence of this scenario
is that the higher mass density increases the importance of tides.
This changes both the types of planets which can implant material in the Oort Cloud (eq.~\ref{eq:mejoc})
and the orbital radius of that Oort Cloud (eq.~\ref{eq:af}).
The region of parameter space in which planets can cause Oort Clouds is still relatively small, but 
includes Saturn for the parameters chosen here.
Such a scenario is often invoked to explain the origin
of detached Kuiper belt objects like Sedna on wide orbits at 100-1000\,au \citep[e.g., ][]{Kaib2008},
and indeed the putative planet nine \citep{Batygin2016}.
The simulations show that, for suitable cluster conditions, it is plausible that such objects can have
formed in the inner regions of the Solar System and have been scattered out by interactions with
the planets, whereupon they were detached from the planetary system by tides \citep{Brasser2006}.
Some nearby stars likely were also born in dense clusters and so could have analogous populations of
detached objects orbiting relatively close to their star, a possibility which is discussed in
\S \ref{sss:minioort} and \S \ref{sss:exosedna}.
It is also possible that some of the Solar System's Oort Cloud was captured by the Sun's gravitational field
after these objects were ejected following formation in the circumstellar disks of other stars in the
Sun's birth cluster \citep{Levison2010}.
Thus another consequence of the scenario in which a star forms in a dense cluster is that some of
the planetesimals that escape the system (i.e., those interacting with planets in the
Ejected or Depleted Oort Cloud regions of Fig.~\ref{fig:youngss}) could end up in the Oort Clouds of other stars in the cluster.

\subsection{Young A stars like HR8799}
\label{ss:younga}

\begin{figure*}
  \begin{center}
    \begin{tabular}{c}
      \includegraphics[width=2\columnwidth]{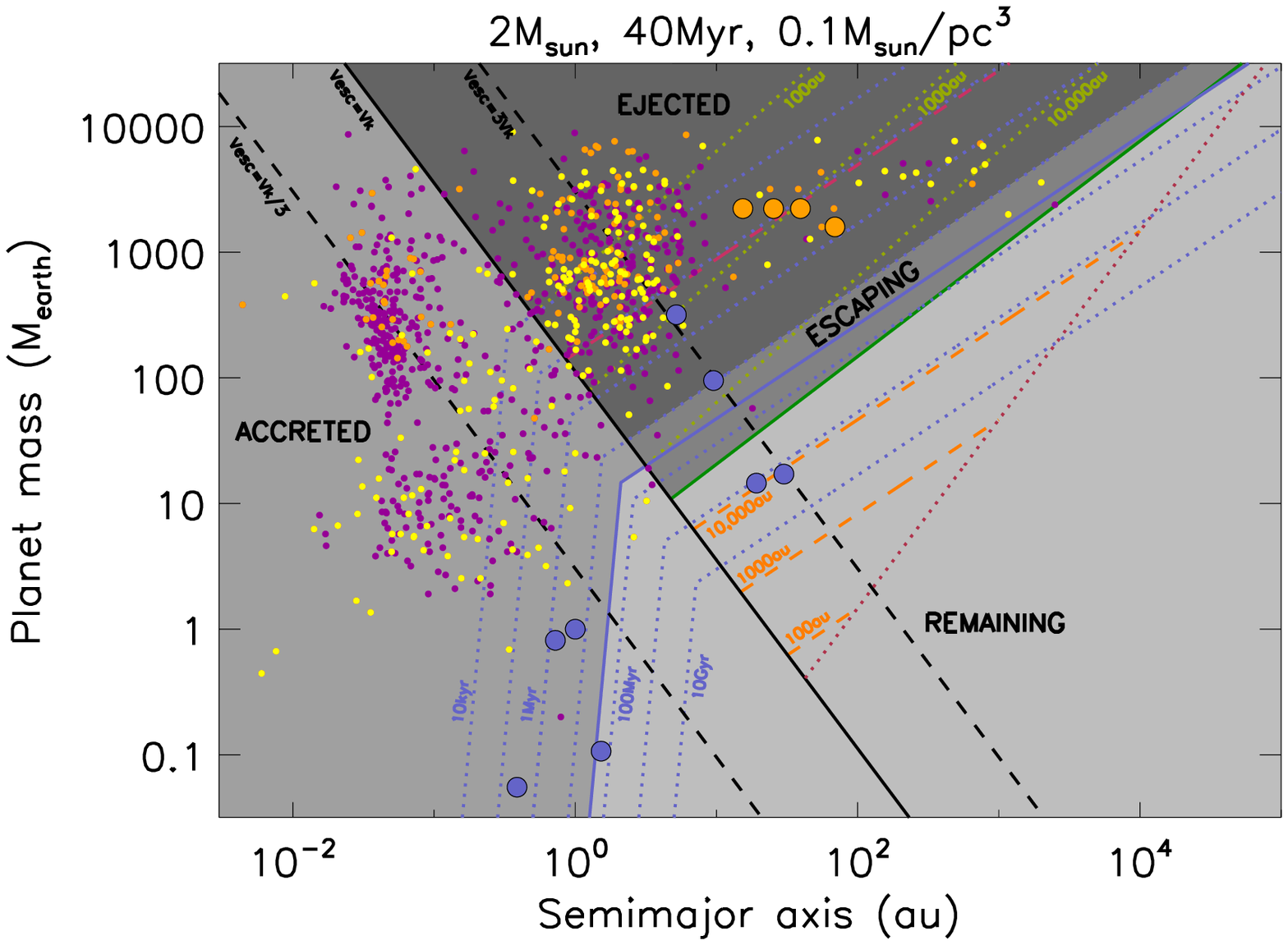}
    \end{tabular}
    \caption{As for Fig.~\ref{fig:ss}, but for a system orbiting a
      $2M_\odot$ star for 40\,Myr of evolution in an environment with
      mass density 0.1\,$M_\odot$\,pc$^{-3}$. The four HR8799
      planets are shown as large orange dots at the locations of
      \protect\cite{Gozdziewski2014} and masses of
      \protect\cite{Konopacky2016}.}
   \label{fig:younga}
  \end{center}
\end{figure*}

Fig.~\ref{fig:younga} considers the parameter space of a nearby young
A star, that is, a 40\,Myr-old $2M_\odot$ star in an environment
with local mass density 0.1\,$M_\odot$\,pc$^{-3}$ and local
stellar density 0.045\,$M_\odot$\,pc$^{-3}$.
Note that stellar encounters dominate over Galactic tides for such a high
mass star, which has been accounted for by increasing the effective local mass density 
by a factor 1.6 (see Appendix~\ref{a:enc}), though this is of little
consequence since there is no Oort Cloud region for these parameters.
This is meant to be appropriate for systems like those currently being surveyed by direct imaging to search
for planets orbiting young stars in nearby moving groups.
Such surveys have been successful at discovering both directly imaged planets and debris disks,
often in the same system.
This type of system is epitomised by HR8799, which is of comparable mass and age to those
plotted, with four long-period giant planets of mass $3-9M_{\rm Jup}$ imaged orbiting 12-60\,au
\citep{Marois2010}, and a debris disk extending both exterior \citep[$>90$\,au; ][]{Matthews2014hr8799, Booth2016}
and interior \citep[$<10$\,au; ][]{Su2009} to the planets.

{\it Oort Cloud: }
Most of the differences between Figs.~\ref{fig:ss} and \ref{fig:younga} arise from the
difference in system age; the difference in the stellar mass is not so important.
One of the first things to note, as pointed out by \citetalias{Tremaine1993}, is that young stars such as
that plotted have not had time to form an Oort Cloud.
That is, there is no Oort Cloud (or indeed depleted Oort Cloud) parameter space
because the solid blue line lies entirely above the solid green line.
Stars this massive can still form Oort Clouds by the end of their main sequence lifetime (if they
have suitable planets), and these would be at $\sim 10,000$\,au like that in the Solar System.
However, \cite{Jura2011} suggested that the Oort Clouds of A stars are on average
less massive than that in the Solar System, based on the pollution signature 
of their white dwarf descendants.
While their Oort Clouds could also have been depleted in the star's post-main sequence evolution \citep{Veras2011},
if confirmed this could set constraints on the prevalence of planets in the relevant Oort Cloud region
on these figures.
As noted by \citetalias{Tremaine1993}, for stars that are massive enough ($>7M_\odot$), their main sequence
lifetime is not long enough for an Oort Cloud to form, although such stars also
predominantly form in clusters which may aid the rapid formation of a close-in Oort Cloud
(e.g., \S \ref{ss:youngss}).

{\it Escaping: }
Otherwise the parameter space looks similar in so far as the existence of ejected, accreted
and remaining regimes.
However, the star's youth implies the existence of a population of escaping bodies which was
not discussed in \S \ref{ss:ss}.
These are the objects which are still being scattered by planets in the cometary diffusion regime.
While their ultimate fate is ejection from the system, many persist on highly eccentric
orbits since the timescale for that outcome is comparable to the age of the system.
Such objects could be planets scattered from the planetary region
or comet-like debris in a population analogous to the Kuiper belt's Scattered disk.
\cite{Veras2009} predicted a population of escaping planets that may be detectable around young stars
scattered out during planetary system instabilities, while the Scattered disk in the Solar System
is (to some extent) a remnant of an escaping population which would have been much more massive
at earlier times (e.g., eq.~1 of \citet{Booth2009} shows that the characteristic timescale for
Uranus and Neptune to deplete the Scattered disk is $\sim 280$\,Myr).
These populations are discussed further in \ref{sss:sd} and \S \ref{sss:escape}.

{\it Ejected: }
Similar to the conclusions in \S \ref{ss:ss}, the population of known eccentric Jupiters
and long-period giant planets would put material encountering them onto unbound orbits
very rapidly.
For example, the blue dotted lines 
show that the HR8799 planets have diffusion times that are $\ll 1$\,Myr,
and so would have long since removed any nearby planets and depleted any scattered disk by ejection.
This means that the outer debris disk in HR8799 cannot be comprised of material currently being scattered
by the known planets;
rather this debris has only managed to survive this long because it is not encountering those
planets, and so is a population more analogous to the classical Kuiper belt.
Similar reasoning shows that the planet-like object Fomalhaut-b found 130\,au from its 440\,Myr-old
A star host Fomalhaut \citep{Kalas2013} cannot be Jupiter in mass unless it was put on this orbit
very recently.
This was shown in numerical simulations \citep{Beust2014, Tamayo2014}, but is
implied from Fig.~\ref{fig:younga}, since the diffusion time is
$\sim 40$\,Myr at that distance and so the timescale for disruption of the narrow
debris ring it traverses must be much shorter than this (assuming that the orbit of the planet brings
it close enough to the debris for scattering to ensue), which in turn is clearly shorter
than the age of the system.
If instead Fomalhaut-b is a low mass scattered disk object \citep[e.g., closer to the proposal of][]{Lawler2015},
the planet which scattered it onto such an eccentric orbit can be predicted to lie close to the
line for which ejection takes 440\,Myr (unless the object was only scattered recently);
e.g., eq.~\ref{eq:tdiff} shows that such a planet at 32\,au \citep[the mean of the distribution of
possible values for the pericentre of Fomalhaut-b's orbit,][]{Kalas2013}
would be $\sim 30M_\oplus$.

{\it Remaining: }
The above reasoning does not preclude that the debris beyond 90\,au from HR8799 is in fact
still interacting with unseen planets, since planets in that region that are in the {\it remaining} regime
would not have had time yet to eject the debris in their vicinity.
Fig.~\ref{fig:younga} shows that the diffusion time for planets in this region
can be longer than the age of the system even for planets up to Saturn in mass.
The existence of such a planet could help to explain why the debris distribution
extends from $\sim 145$\,au to beyond 400\,au \citep{Booth2016}, since a planet
orbiting close to the inner edge could have excited eccentricities in the debris population
creating an exterior scattered disk, an idea which is explored further in \S \ref{sss:sd}.
Such a planet would need to be massive enough to stir the disk over the system age, but not so massive
that it has ejected the majority of the debris, implying a roughly Saturn mass planet,
although a comparison of the debris distribution with numerical simulations is needed for a
more accurate determination.
Note, however, that there is no requirement to invoke such a planet in this system, 
since the disk's breadth may alternatively be explained by an initially broad Kuiper belt.

{\it Exocomets: }
Infrared observations indicate the presence of dust at $\sim 9$\,au which is interior to the known
planets of HR8799 \citep{Su2009}.
This is far enough from the star (and the planets) that it is compatible with an origin in the steady state
grinding of a planetesimal belt at that location analogous to the Solar System's asteroid belt
\citep{Wyatt2007hotdust, Contro2015}.
In this paper we consider an alternative explanation, which is that the hot dust is fed by comets
scattered into the inner regions from the planetesimal belt beyond 90\,au.
While all four of the known HR8799 planets lie deep in the {\it ejected} region,
seemingly presenting a formidable barrier for any exocomets to cross,
comets must still undergo many scatterings before ejection, some of which will have
passed them inward.
Thus at any given time there should be a population of objects residing in
the inner regions that were scattered in from the outer disk.
However, the short timescale to achieve the ultimate fate of ejection ($\sim 0.1$\,Myr)
means that HR8799's comet population is expected to be relatively small (at least compared to what it would
be if the planets were lower in mass), and so may be unlikely to be the
origin of its hot dust.
Nevertheless, other systems may have architectures that are more suited to replenishing
hot dust from exocometary populations, which are discussed further in \S \ref{ss:comet}.

\subsection{Planets around low-mass stars}
\label{ss:bd}

\begin{figure*}
  \begin{center}
    \begin{tabular}{c}
      \includegraphics[width=2\columnwidth]{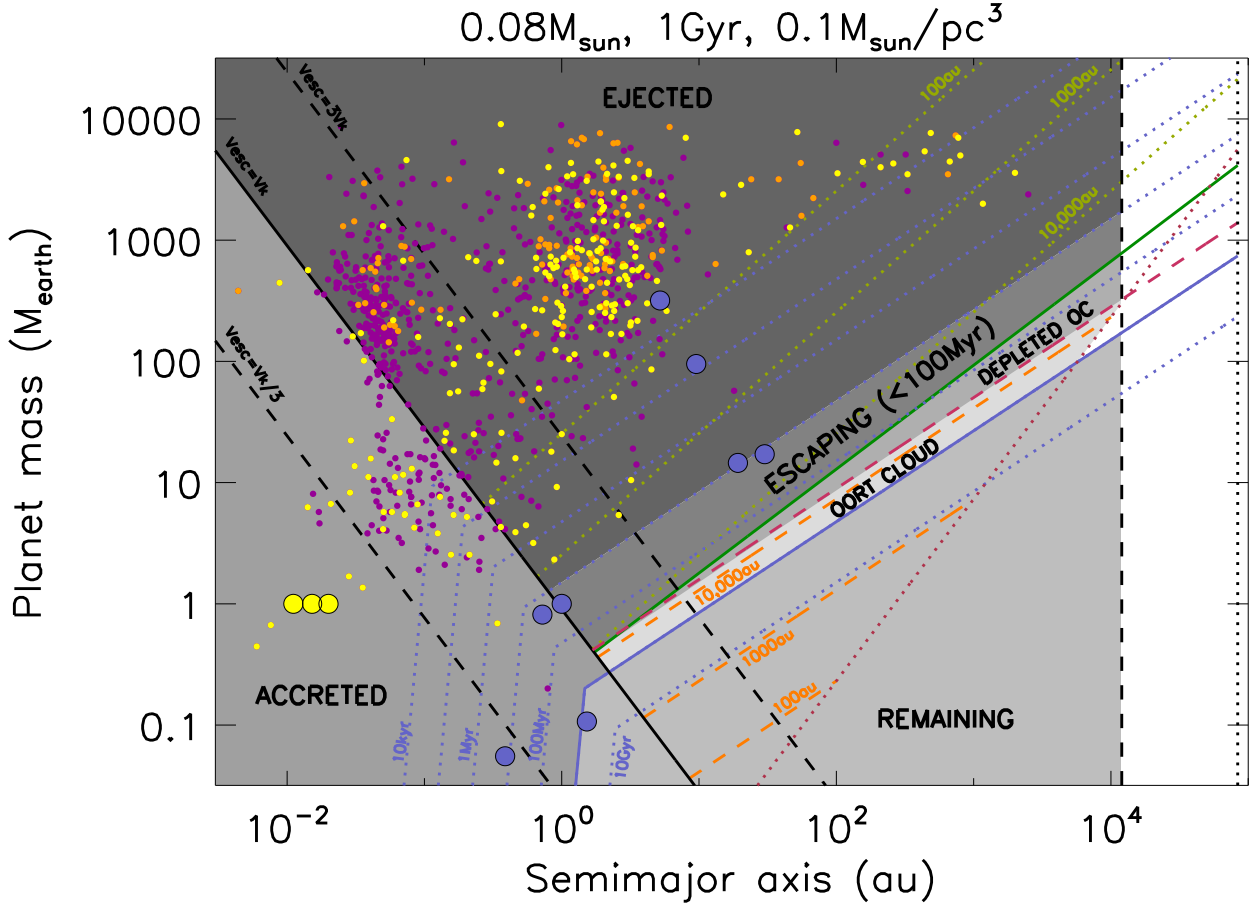}
    \end{tabular}
    \caption{As for Fig.~\ref{fig:ss}, but for a system orbiting a
      $0.08M_\odot$ star for 1\,Gyr of evolution in an environment with
      mass density 0.1\,$M_\odot$\,pc$^{-3}$.
      The three planets found around TRAPPIST-1 are shown as large yellow
      dots \protect\citep{Gillon2016}.
      }
   \label{fig:bd}
  \end{center}
\end{figure*}

The discovery of 3 Earth-sized planets orbiting the M8 brown dwarf TRAPPIST-1 emphasises that planetary
systems are present around stars of all masses \citep{Gillon2016}.
While the stellar mass made little difference to the scattering outcomes for the stellar parameters
considered in Figs.~\ref{fig:ss} and \ref{fig:younga}, the mass of TRAPPIST-1 is just $0.08M_\odot$,
which causes its scattering outcomes shown in Fig.~\ref{fig:bd} to be shifted substantially relative
to their location on Fig.~\ref{fig:ss}.
While the three known TRAPPIST-1 planets would be expected to accrete most of the material they encounter, this
is true for a smaller region of parameter space than for higher mass stars
\citep[although note that this is also a heavily populated part of parameter space;][]{Winn2015}. 
The most notable consequence of the star's low mass is the relative ease with which planets can eject objects
which encounter them.
The larger ejected regime makes it harder for a low-mass star to build up a planetary
core that is capable of runaway accretion of gas \citep[see also][]{Payne2007}, and means that
even Earth-like planets can present a barrier that prevents comets from reaching the inner regions of a system.
Oort Clouds can still form, slightly within 10,000\,au, and could be fed by scattering from planets
as low in mass as the Earth at 5\,au.

\section{How to maximise desired outcomes}
\label{s:outcomes}
The interpretion of observations of planets or debris around nearby
stars is usually hampered by the fact that we have only incomplete
(if any) information about the rest of the planetary system.
This leads to the necessity to consider the dynamics in a range of
hypothetical systems to see if a plausible explanation for the
observations can be found.
Naturally this requires consideration of an impossibly wide range of parameter space,
and there is often no guarantee that any given plausible explanation is a unique explanation. 
Here we consider a number of different possible scattering outcomes that may
be (or may have been) observed.
On the basis that the first systems detected with a given outcome are likely to be those
which nature has provided the most favourable planetary systems for achieving
that outcome, here we focus on determining the architectures of the planetary
system that would maximise the chance of observing the different outcomes.
For the most part we are not concerned with how such a system might form, so this
does not mean that such systems are a plausible outcome of planet formation processes
for which other considerations are involved.
However, if the required outcome cannot be reproduced with the most favourable
planetary system architecture imaginable, then it is likely that the proposed
mechanism cannot be invoked to explain the observation under consideration.
We also consider what constraints might be placed on a system's planets
based on the planetary system architectures which cannot produce a given outcome.

\subsection{Giant Impact Debris}
\label{ss:gi}
The final stage in the formation of the terrestrial planets is thought to have been characterised
by multiple giant impacts as the large number of embryos formed is whittled down by merging collisions
to the few terrestrial planets seen today \citep{Chambers2001, Kenyon2006}.
Debris released in such impacts may persist in the system at levels that are detectable due to the dust
created in its mutual collisions for 10s of Myr \citep{Jackson2012, Genda2015}.
While the collisional evolution of giant impact debris must be accounted for when considering
its detectability \citep[e.g., ][]{Wyatt2016}, Fig.~\ref{fig:ss} can already be used to come to some
conclusions about the types of terrestrial planet which are most favourable for producing detectable debris.
For example, this shows that the planets that make debris that can persist in the face of reaccretion
onto the planet for $>10$\,Myr are found at $>1$\,au (see eq.~\ref{eq:tacc}), with relatively little dependence on planet mass
(because while higher mass planets have a larger collision cross-section, their higher escape velocity
means that their debris extends across a larger volume).
However, the planets cannot be too far from the star, because planets at large distances have an escape velocity that
is higher than their Keplerian orbital velocity (e.g., $\gg 7$\,au, eq.~\ref{eq:vescvk}),
which means that most of the debris that is created is quickly placed onto unbound orbits.
Furthermore, the planets cannot be too low in mass, since larger quantities of debris are likely to be created
in collisions with more massive planets \citep[with the caveat that if the planet has an atmosphere then
this might prevent the escape of debris in impacts,][]{Inamdar2016}, which is thus likely to be
more readily detectable.
These considerations suggest that there is a sweet spot in the planet mass - semimajor axis parameter space
at which giant impact debris is most likely to be detected (in that it will be both bright and long-lived).
This would be expected to be for planets of a few $M_\oplus$ at a few au.
The sweet spot is quantified in \S \ref{sss:ss}  and \S \ref{sss:varpar}, and its implications considered in \S \ref{sss:ssimp}.

\subsubsection{Characterising the sweet spot}
\label{sss:ss}

\begin{figure*}
  \begin{center}
    \begin{tabular}{cc}
      \hspace{-0.0in} \includegraphics[width=1.0\columnwidth]{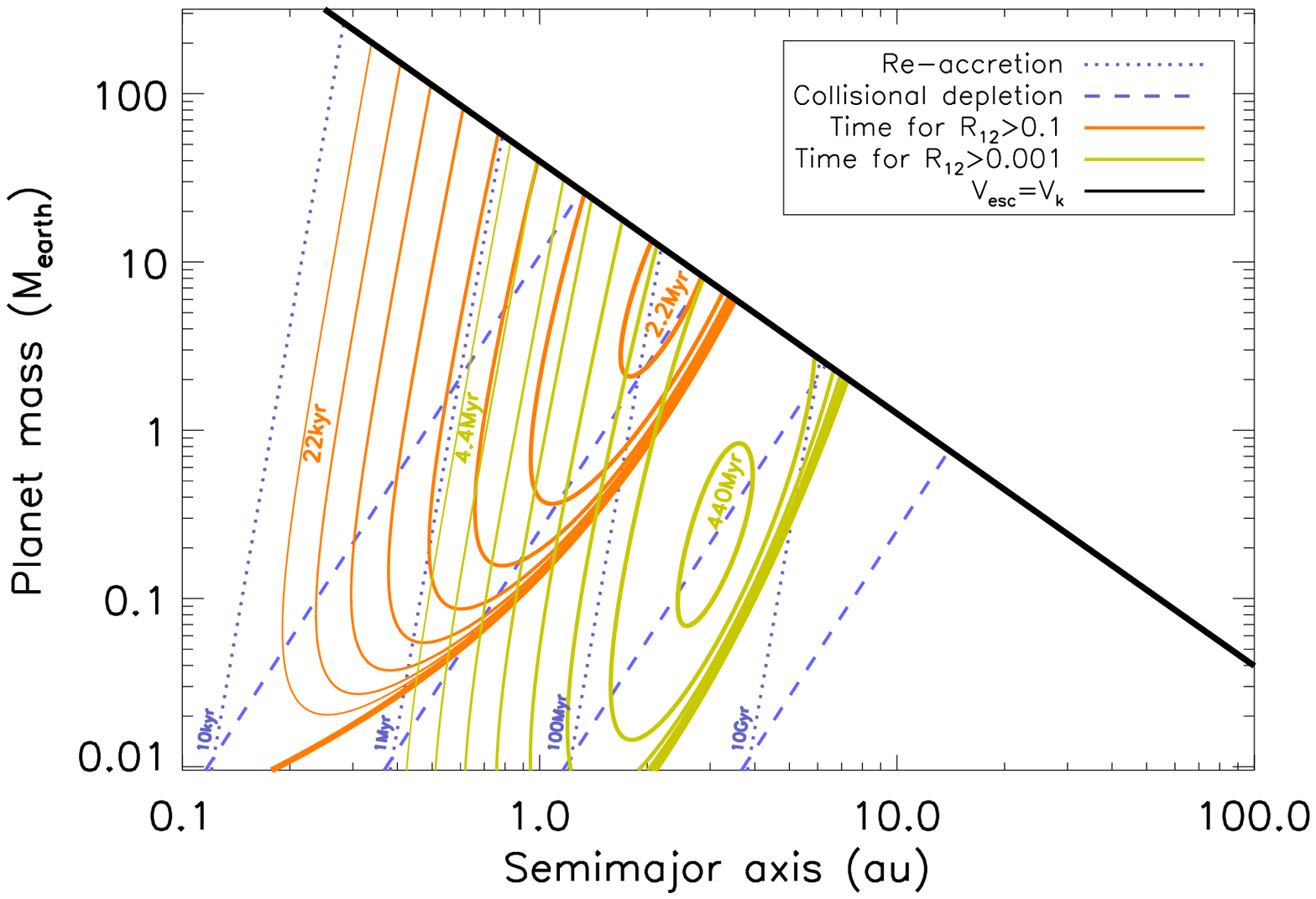} &
      \hspace{-0.0in} \includegraphics[width=1.0\columnwidth]{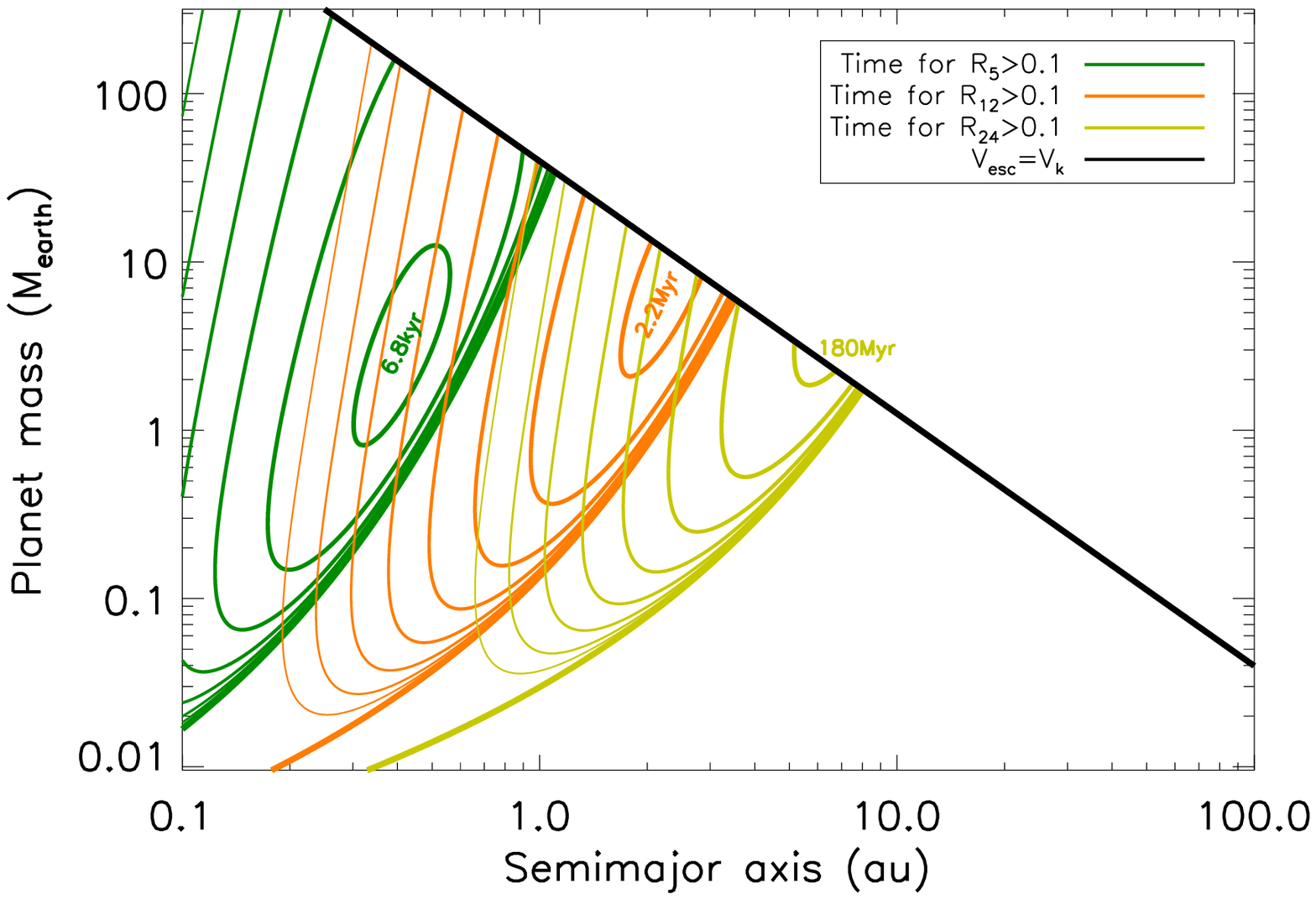} \\ 
      \hspace{-0.0in} \includegraphics[width=1.0\columnwidth]{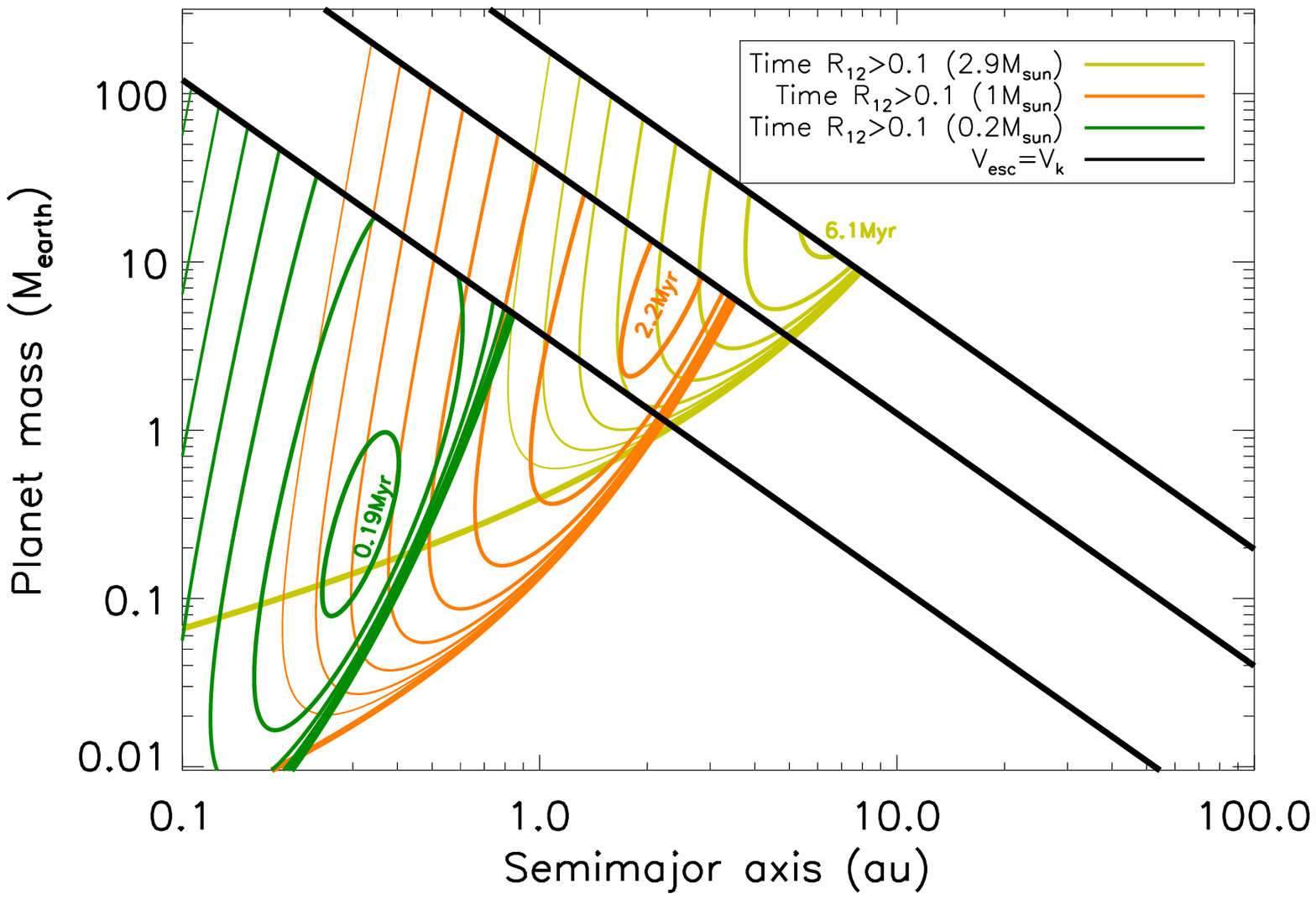} &
      \hspace{-0.0in} \includegraphics[width=1.0\columnwidth]{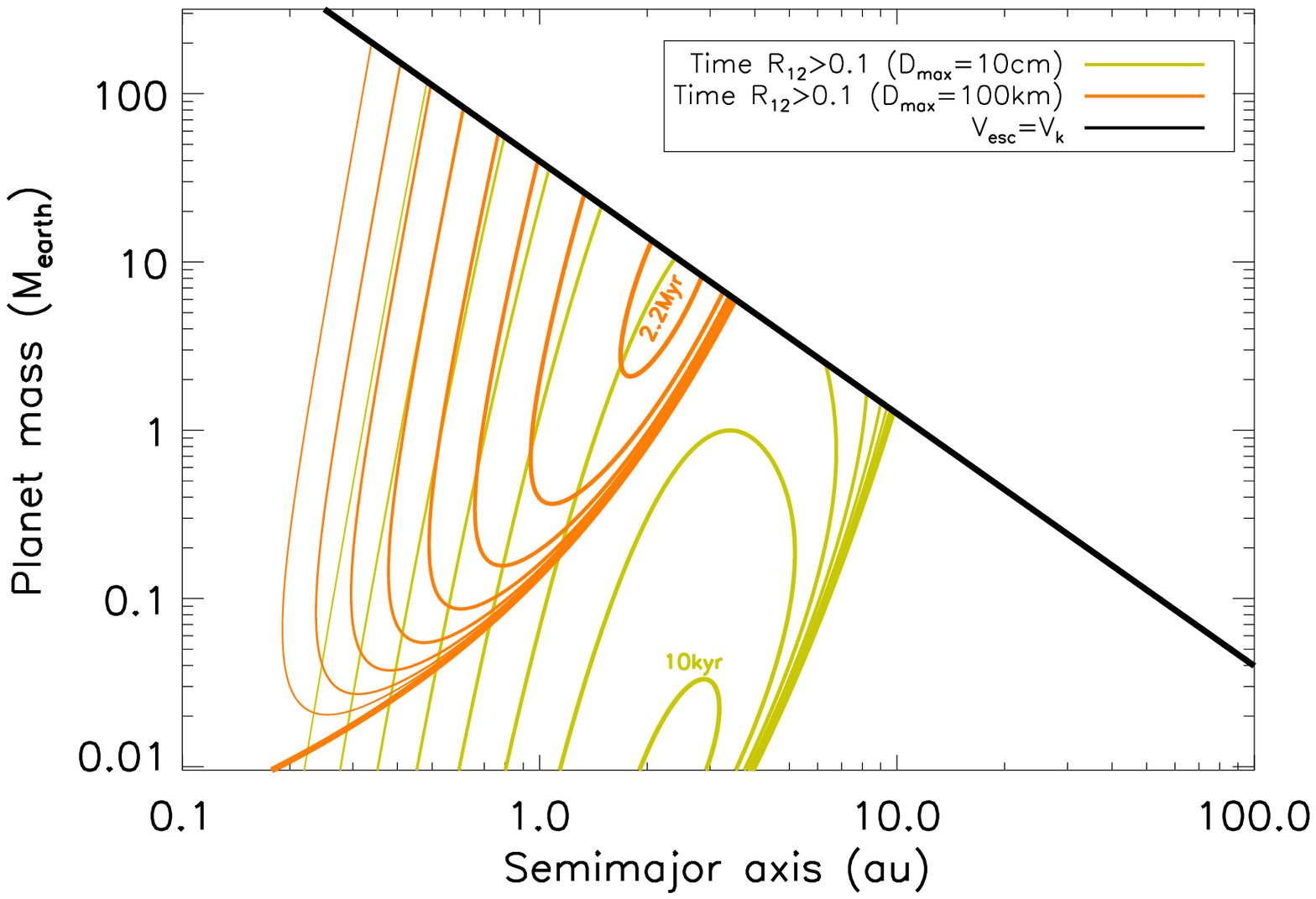}
    \end{tabular}
    \caption{Planet mass versus semimajor axis parameter space showing timescales relevant for
    the evolution of giant impact debris originating from such planets.
    Unless stated otherwise it is assumed that the planet is orbiting a Sun-like
    star ($1M_\odot$, $1L_\odot$, 5780\,K), and the escaping debris has $f_{\rm esc}=0.05$,
    $D_{\rm max}=100$\,km, $Q_{\rm D}^\star=10^5$\,J\,kg$^{-1}$.
    Debris is not considered above the black line for which the planet's
    escape velocity equals its Keplerian velocity.
    The blue dotted lines on the top left panel show the timescale for reaccretion onto the
    planet (eq.~\ref{eq:tacc}), while that for depletion in mutual collisions is shown with blue dashed lines
    (eq.~\ref{eq:mptc});
    the annotation is placed where these two timescales are equal.
    The other lines are contours on which the timescale is the same for the thermal
    emission from the debris to remain above a given fractional excess level at a given
    wavelength, which for most panels means a fractional excess of 0.1 at a wavelength of 12\,$\mu$m.
    The contours for each set of parameters are shown with a different colour,
    as indicated in the legend for each panel.
    In each case the maximum time is indicated on the figure in the same colour as the
    contours, with the contours drawn at logarithmic intervals of 0.01, 0.019, 0.036, 0.069,
    0.13, 0.25, 0.47 and 0.9 times this maximum time.
    The thick coloured lines show the planet mass below which the debris is never detectable
    at the relevant level (eq.~\ref{eq:rlambda}).
    The top left panel shows the effect of changing $R_{\lambda}$, the top right panel the effect of
    changing $\lambda$, the bottom left panel the effect of changing stellar mass (assuming a
    $2.9M_\odot$, $54L_\odot$, 9500\,K star and a $0.21M_\odot$, $0.011L_\odot$, 3250\,K star),
    and the bottom right panel the effect of changing $D_{\rm max}$, as noted in the legend.
    }
   \label{fig:gi}
  \end{center}
\end{figure*}

\begin{table}
  \centering
  \caption{Summary of parameters introduced in \S \ref{ss:gi}.}
  \label{tab:units2}
  \begin{tabular}{lll}
     \hline
     Parameter                                 & Symbol             & Units \\
     \hline
     Stellar temperature                       & $T_\star$          & K \\
     Fraction of mass escaping as debris       & $f_{\rm esc}$      & dimensionless \\
     Initial mass of debris                    & $M_{\rm d0}$       & $M_\oplus$ \\
     Instantaneous mass of debris              & $M_{\rm d}$        & $M_\oplus$ \\
     Diameter of largest debris fragment       & $D_{\rm max}$      & km \\
     Debris temperature                        & $T$                & K \\
     Observing wavelength                      & $\lambda$          & $\mu$m \\
     Planck function                           & $B_\nu$            & Jy\,sr$^{-1}$ \\
     Fractional excess                         & $R_{\lambda}$      & dimensionless \\
     Dispersal threshold                       & $Q_{\rm D}^\star$  & J\,kg$^{-1}$ \\
     Initial collisional lifetime of debris    & $t_{\rm c0}$       & Gyr \\
     Instantaneous collisional lifetime        & $t_{\rm c}$        & Gyr \\
     Reaccretion timescale                     & $t_{\rm acc}$      & Gyr \\
     Fraction of debris mass reaccreted        & $f_{\rm acc}$      & dimensionless \\
     Ratio of $t_{\rm acc}/t_{\rm c0}$         & $\eta$             & dimensionless \\
     Time debris mass is above $M_{\rm d}$     & $t_{(>M_{\rm d})}$ & Gyr \\
     Time fractional excess is above $R_{12}$  & $t_{(>R_{12})}$    & Gyr \\
     \hline
  \end{tabular}
\end{table}

To quantify this sweet spot including a consideration of the effect
of collisions amongst the debris population, Fig.~\ref{fig:gi} focusses on the planets for which
giant impact debris is expected to be potentially detectable.
For this figure it was assumed that a giant impact puts into
circumstellar orbit a fraction $f_{\rm esc}=0.05$ of the mass of the planet, so that the initial
mass of debris
\begin{equation}
  M_{\rm d0} = f_{\rm esc}M_{\rm p};
  \label{eq:md0}
\end{equation}
see Table~\ref{tab:units2} for a summary of the additional parameters used in this section.
This debris is assumed to have a power law size distribution with index $-3.5$ (i.e., $dn/dD \propto D^{-3.5}$)
extending from a diameter of $D_{\rm max}=100$\,km down to the radiation pressure blow-out limit
\citep[see eq.~14 of ][]{Wyatt2008}, 
where the debris is assumed to have the same density as the planet ($\rho_{\rm p}$).

Further assuming that the dust acts like a black body, this means that the debris has a
temperature
\begin{equation}
  T = 278.3L_\star^{1/4}a_{\rm p}^{-1/2},
  \label{eq:tdust}
\end{equation}
and starts with a fractional luminosity that is given in eq.~15 of \cite{Wyatt2008}, from
which the thermal emission at a wavelength $\lambda$ can be derived using eq.~10
of that paper.
The thick coloured lines on Fig.~\ref{fig:gi} show the planets for which their
giant impact debris starts out with thermal emission at the given level of fractional
excess $R_{\lambda}$ (i.e., the disk flux divided by the stellar
flux at a wavelength $\lambda$), which are given by
\begin{equation}
  M_{\rm p} = 2.4 \times 10^{10} T_\star^{-4}L_\star^{3/2}M_\star^{-1/2}
              \left[ \frac{B_\nu(\lambda,T_\star)}{B_\nu(\lambda,T)} \right]
              D_{\rm max}^{1/2}\rho_{\rm p}^{-1/2}f_{\rm esc}^{-1}R_{\lambda},
  \label{eq:rlambda}
\end{equation}
where $L_\star$ has units of solar luminosity, $B_\nu$ is the Planck function at
the given wavelength and temperature, and $T_\star$ is the temperature of the stellar emission.
Planets must be above these lines for their giant impact-generated debris to be
detectable at the given level.

Assuming that the debris has a dispersal threshold $Q_{\rm D}^\star=10^5$\,J\,kg$^{-1}$ that is
independent of size, the timescale for mutual collisions amongst the debris population to deplete the
debris is given by eq.~16 of \cite{Wyatt2008} with the further assumption that the width of the torus and its stirring
are set by the planet's escape velocity.
This gives a collisional lifetime of
\begin{equation}
  t_{\rm c} = 6.4 \times 10^{-12} M_\star^{-1}
              a_{\rm p}^4 M_{\rm p}^{-2/9} \rho_{\rm p}^{-1/9}
              D_{\rm max} {Q_{\rm D}^\star}^{5/6} M_{\rm d}^{-1}
  \label{eq:tc}
\end{equation}
in Gyr, where $M_{\rm d}$ is the total mass of debris at that time in $M_\oplus$.
This collision lifetime has a very strong dependence on distance from the star,
which goes some way to explaining why there are so few
Vulcanoid asteroids in the Solar System \citep{Steffl2013}, since these are otherwise
long-term dynamically stable \citep{Evans1999}, but would have been eroded by mutual
collisions over the age of the Solar System \citep{Stern2000}.

Equation~\ref{eq:tc} can be rearranged to find that the planet mass that results in giant impact debris that starts out
with a collisional lifetime $t_{\rm c0}$ in Gyr is
\begin{equation}
  M_{\rm p} = 6.9 \times 10^{-10} M_\star^{-9/11}
              a_{\rm p}^{36/11} \rho_{\rm p}^{-1/11}
              D_{\rm max}^{9/11} {Q_{\rm D}^\star}^{15/22} f_{\rm esc}^{-9/11} t_{\rm c0}^{-9/11},
  \label{eq:mptc}
\end{equation}
which is plotted in the top left panel of Fig.~\ref{fig:gi} with dashed blue lines, along with the dotted blue lines
which give the planets that result in a given reaccretion timescale (i.e., the same as those on Fig.~\ref{fig:ss}).

The eventual fate of the mass of debris depends on $\eta=t_{\rm acc}/t_{\rm c0}$, which is the ratio of
the timescale on which the debris is reaccreted onto the planet to that on which its initial mass is ground into
dust (which is subsequently removed by radiation pressure).
Rearranging eq.~\ref{eq:tacc} shows that reaccretion onto the planet takes place on a
timescale
\begin{equation}
  t_{\rm acc} = 0.01 M_\star^{-1} a_{\rm p}^4 \rho_{\rm p}^{5/6} M_{\rm p}^{-1/3},
  \label{eq:tacc2}
\end{equation}
while $t_{\rm c0}$ is calculated using eq.~\ref{eq:tc} with $M_{\rm d}=M_{\rm d0}$,
giving
\begin{equation}
  \eta = 1.6 \times 10^9 \rho_{\rm p}^{17/18} M_{\rm p}^{8/9} D_{\rm max}^{-1} {Q_{\rm D}^\star}^{-5/6} f_{\rm esc}.
\end{equation}
With these two loss mechanisms the rate of debris mass loss is
\begin{equation}
  \dot{M}_{\rm d} = -M_{\rm d}/t_{\rm acc} - M_{\rm d}/t_{\rm c} =
                    -M_{\rm d}/t_{\rm acc} - M_{\rm d}^2/(M_{\rm d0}t_{\rm c0}), 
  \label{eq:mddot}
\end{equation}
which can be solved to give
\begin{equation}
  M_{\rm d} = M_{\rm d0}[\exp{(t/t_{\rm acc})} + \eta(\exp{(t/t_{\rm acc})}-1)]^{-1},
  \label{eq:mm0}
\end{equation}
as well as showing that the fraction of mass that is eventually accreted is
\begin{equation}
  f_{\rm acc} = \eta^{-1}\ln{(1+\eta)}.
  \label{eq:facc}
\end{equation}

Equation~\ref{eq:mm0} can be rearranged to give the time for which the debris has a mass above a given level
\begin{equation}
  t_{(>M_{\rm d})} = t_{\rm acc} \ln{\left( \frac{\eta + M_{\rm d0}/M_{\rm d}}{\eta+1} \right) }.
  \label{eq:tgm}
\end{equation}
This is used in Fig.~\ref{fig:gi} to work out, for each planet mass and semimajor axis, the length of
time giant impact debris from that planet would have a fractional excess above a given level $R_{\lambda}$
(i.e., $t_{(>R_{\lambda})}$), by using eq.~\ref{eq:md0} for $M_{\rm d0}$, and using for $M_{\rm d}$ the mass
required to give this fractional excess (which is $f_{\rm esc}$ times the right hand side of eq.~\ref{eq:rlambda}).
Lines of constant $t_{(>R_{\lambda})}$ are shown on Fig.~\ref{fig:gi} as the thinner solid lines with 
the different colours corresponding to different combinations of model parameters or detection thresholds.
This is one way to visualise the sweet spot, since for a uniform frequency of giant impacts per logarithmic
bin of planet mass and semimajor axis, debris would be expected to be seen at a higher incidence
in regions of highest $t_{(>R_{\lambda})}$.

For Fig.~\ref{fig:gi} we have not considered debris above the $v_{\rm esc}=v_{\rm k}$ line, which is shown in
black.
However, it is in principle possible to take this into account by working out the fraction of debris that is placed
onto bound orbits in this regime, and to use as the lifetime of this debris that from cometary diffusion.

\subsubsection{Parameter dependence of sweet spot}
\label{sss:varpar}
The shape of the sweet spot in Fig.~\ref{fig:gi} is close to that predicted in \S \ref{ss:gi}.
Indeed, the lowest contours (i.e., those for which the debris does not last long above the detectable
level) are bounded at the top right and bottom right by eqs.~\ref{eq:vescvk} and \ref{eq:rlambda}, respectively,
while the left hand edge is near the corresponding isochrone for reaccretion onto the planet (i.e., the dotted
line given by eq.~\ref{eq:tacc}).
It may seem counter-intuitive that the timescale for the debris to remain above the given $R_{12}$
level can be set by the reaccretion timescale in this regime, since the initial collisional depletion timescale
(i.e., the dashed line given by eq.~\ref{eq:mptc}) is shorter than that for reaccretion.
However, the collision timescale $t_{\rm c}$ gets longer as the debris mass is depleted leading to that
mass dropping inversely with age, which is much slower than the exponential depletion in mass caused by
reaccretion (albeit on a longer timescale).
Thus the relevant timescale depends on whether a large or small fraction of the initial debris mass needs to be
removed to drop below the detection threshold (in which case reaccretion or collisional depletion, respectively, are
dominant).
This means that the contours at larger semimajor axes do not follow the reaccretion timescale, but instead
reach a maximum time for the debris to remain above the detection threshold
that is set by collisional depletion.
It is interesting to note that it is not necessarily the most massive planets for which the debris is
detectable the longest (i.e., the maximum time does not occur on the $v_{\rm esc}=v_{\rm k}$ line).
This arises because collisional depletion is faster for debris from higher mass planets 
(see eq.~\ref{eq:tc}), which is because mutual collisions amongst the debris occur at higher velocities
(as well as the debris being dispersed into a larger volume).

With this understanding of the origin of the sweet spot, the dependence of its shape on the different
parameters can also be readily understood.
For example, the top left panel of Fig.~\ref{fig:gi} shows two values of $R_{12}$.
The $R_{12}>0.1$ threshold is achieved for tens of thousands of stars by photometric instruments
such as WISE \citep[e.g.,][]{Kennedy2013}, while the $R_{12}>10^{-3}$ threshold
is the goal of cutting edge nulling interferometry techniques that may be achieved on bright
stars \citep[e.g.][]{Defrere2016}.
In the former case we find that the sweet spot is for planets more massive and further from the star
than the Earth, for which the debris remains detectable for $1-2$\,Myr.
For the lower detection threshold the debris from lower mass (i.e., Mars-mass) planets becomes
accessible, with planets at $\sim 3$\,au having debris that remains detectable for 200-400\,Myr.
This can be understood from the lower detection threshold given by eq.~\ref{eq:rlambda}, which means that
lower masses of debris are detectable, which allows the debris from planets at larger orbital radii
to be detectable, whereupon the longer evolutionary timescales mean that it can remain so for longer
periods.

For a similar reason there is also a dependence on the wavelength of observation, which
sets the lower envelope of the sweet spot through eq.~\ref{eq:rlambda}, and in particular through the
ratio of Planck functions in the square brackets.
This ratio is minimised when the wavelength is long enough for the dust to be emitting in the Rayleigh-Jeans
limit, or equivalently for debris that is close enough to the star for this to be the case, at which point
the ratio scales $\propto T_\star/T \propto T_\star L_\star^{-1/4} a_{\rm p}^{1/2}$.
This is true for the closest planets at $\sim 0.1$\,au on the top left panel of Fig.~\ref{fig:gi}.
Thus it can be expected that the $R_{24}>0.1$ line would be similar to that of $R_{12}$ at small radii
(where it would scale $\propto a_{\rm p}^{1/2}$),
but would depart from this scaling and turn up towards higher planet masses at orbital radii that are further out
than the turn-up for $R_{12}$.
This is confirmed in the top right panel of Fig.~\ref{fig:gi}, and means that giant impacts should be more
readily detected at longer wavelengths, with the caveat that observations at longer wavelengths do
not necessarily have the same sensitivity to fractional excess, and/or it may be harder to distinguish
giant impact debris from steady state grinding of exo-Kuiper belts at longer wavelengths.
Conversely at shorter wavelengths like 5\,$\mu$m, detected impact debris would be expected to be closer to
the star, and to be seen relatively infrequently given the shorter duration of detectability.

The properties of the star also affect the location of the sweet spot. 
The lifetimes have a dependence on the stellar mass through eqs.~\ref{eq:vescvk}, \ref{eq:tacc} and \ref{eq:mptc}. 
However, the strongest effect on the sweet spot is again through the lower detection threshold
limit in eq.~\ref{eq:rlambda}, which scales $\propto T_\star^{-3} L_\star^{5/4} M_\star^{-1/2}$ for the
Rayleigh-Jeans limit discussed above.
This means that the detection threshold for giant impacts around an A0V star in this limit is an order of magnitude
higher than those plotted on the top left panel of Fig.~\ref{fig:gi}.
However, the bottom left panel of Fig.~\ref{fig:gi} shows that this does not necessarily mean that giant
impact debris is less readily detected around higher mass stars, because the debris is also hotter around a
higher luminosity star which causes the turn-up in the lower limit discussed in the previous paragraph
to occur at larger radii.
Overall the bottom left panel of Fig.~\ref{fig:gi} shows
that giant impact debris around higher mass stars is seen out to larger radii,
where it lasts slightly longer and requires a higher mass planet progenitor than a Sun-like star.
Conversely, giant impact debris is detectable from planets orbiting lower mass stars, even for Mars-mass planets,
but only if they are close-in ($<0.5$\,au).

Some of the parameters in the calculation are quite uncertain, such as the largest planetesimal size
$D_{\rm max}$ and the dispersal threshold $Q_{\rm D}^\star$.
It must also be recognised that the dispersal threshold is known to be dependent on
planetesimal size which results in a debris size distribution that is more involved than
that assumed here \citep[e.g.,][]{Wyatt2011}. 
Nevertheless, the assumptions used here provide a self-consistent and transparent model
that also provides a reasonable approximation to the evolution of mass and debris luminosity.
This means that the dispersal threshold is some kind of average of the size distribution
and will not be discussed further except to note that more accurate calculations can 
be done to explore this issue, but will not affect the qualitative results presented
here.

The largest planetesimal size is, however, an important parameter. 
For example, if most of the debris mass was vaporised and subsequently condensed into 10\,cm-sized
grains \citep{Johnson2012}, this would result in $D_{\rm max}$ being reduced by six orders of
magnitude.
The lower envelope of the sweet spot would be reduced by 3 orders of magnitude (eq.~\ref{eq:rlambda})
making the aftermath of impacts involving small mass planets detectable (see bottom right panel of
Fig.~\ref{fig:gi}).
However, the collisional lifetime would also be reduced by six orders of magnitude, meaning that
the debris would be short-lived at detectable levels, though again would be most readily detected
at large orbital radii where collisional depletion times are longest.
Conversely, if most of the debris mass was placed into larger objects, the giant impacts would need
to involve higher mass planets to create detectable debris, but that debris would be detectable for
longer.

\subsubsection{Implications of the sweet spot}
\label{sss:ssimp}

\begin{table*}
  \centering
  \caption{Proposed giant impact debris around $\leq 120$\,Myr stars with excess emission detected at 12\,$\mu$m.
  Parameters are taken from the literature, except for stellar mass which is that appropriate for the spectral type.}
  \label{tab:gieg}
  \begin{tabular}{lllllll}
     \hline
     Star           & Sp. Type  & $L_\star$     & $M_\star$     & $T_\star$ & Dust location  & Reference \\
     \hline
     $\eta$ Tel     & A0V       & $22L_\odot$   & $2.9M_\odot$  & 9506\,K   & 4\,au          & \citet{Smith2009etatel} \\
     HD172555       & A5V       & $9.5L_\odot$  & $2.0M_\odot$  & 8000\,K   & 5.8\,au        & \citet{Lisse2009,Smith2012} \\
     EF Cha         & A9        & $10L_\odot$   & $1.7M_\odot$  & 7400\,K   & 4.3\,au        & \citet{Rhee2007} \\
     HD113766       & F3/F5     & $4.4L_\odot$  & $1.4M_\odot$  & 5878\,K   & 1.8\,au        & \citet{Lisse2008} \\
     HD15407A       & F5V       & $3.9L_\odot$  & $1.4M_\odot$  & 6500\,K   & 0.6\,au        & \citet{Melis2010,Fujiwara2012} \\
     HD23514        & F6V       & $2.8L_\odot$  & $1.3M_\odot$  & 6400\,K   & 0.25\,au       & \citet{Rhee2008} \\
     ID8            & G6V       & $0.8L_\odot$  & $0.9M_\odot$  & 5500\,K   & 0.33\,au       & \citet{Meng2014} \\
     TYC8241 2652 1 & K2        & $0.7L_\odot$  & $0.7M_\odot$  & 4950\,K   & 0.4\,au        & \citet{Melis2012} \\
     \hline
  \end{tabular}
\end{table*}

\begin{figure}
  \begin{center}
    \begin{tabular}{c}
      \hspace{-0.0in} \includegraphics[width=1.0\columnwidth]{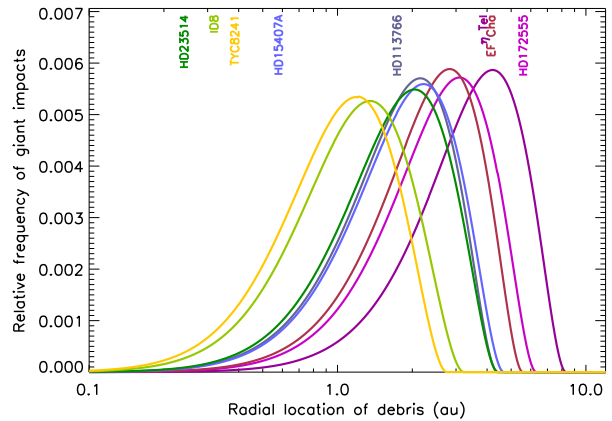}
    \end{tabular}
    \caption{Distribution of radial locations of giant impact debris expected for the 8 stars
    in Table~\ref{tab:gieg}; the number of expected detections scales with the area under the curve.
    Giant impacts are assumed to occur with equal frequency per logarithmic
    bin of planet mass and semimajor axis, and it is assumed that the debris is characterised by $f_{\rm esc}=0.05$,
    $D_{\rm max}=100$\,km, $Q_{\rm D}^\star=10^5$\,J\,kg$^{-1}$, and is detected if $R_{12}>0.1$.
    The different colours indicate the different stars, as shown in the annotation which is placed at the
    observed radial location of debris for this star.
    }
   \label{fig:gieg}
  \end{center}
\end{figure}

Having outlined the region in which giant impact debris lasts longest, the implication is that
this should be where the first giant impact debris is detected.
If not, this could imply that such planets do not exist, or that they do not suffer giant impacts if
they do.
One of the clearest examples of a star with giant impact debris is $\sim 20$\,Myr-old HD172555,
with dust at 5.8\,au from this A5V star \citep[see Table~\ref{tab:gieg};][]{Lisse2009, Smith2012}.
A giant impact origin for the dust is inferred from its silica composition and abundance of sub-micron sized
grains.
Taking the model at face value, given that the parameters of this star are closest to that of the
$2.9M_\odot$ star in the bottom left panel of Fig.~\ref{fig:gi}, the detection of debris
at this radial location is consistent with expectations.
In fact, Fig.~\ref{fig:gieg} shows that the dust is slightly further out than the nominal model would predict for this star,
but this can be accounted for by reducing the largest debris fragment size (see bottom right panel of Fig.~\ref{fig:gi}),
or by recognising that there is a contribution to the dust luminosity from grains expected to have been removed by
radiation pressure.
Thus, if this is giant impact debris, we predict the presence of a $3-10M_\oplus$ planet
orbiting coincident with the debris at 5-6\,au (though the parent planet could be less massive depending on
model parameters).
Several other young A-type stars have also been suggested to have giant impact debris at a similar
location (see Table~\ref{tab:gieg}, e.g., 4.3\,au for EF Cha, \citet{Rhee2007}; 4\,au for $\eta$ Tel,
\citet{Smith2009etatel}), again consistent with their expected location (Fig.~\ref{fig:gieg}).
This could mean that planets in this region of parameter space are relatively common around A-type stars.
Such planets are absent in the known exoplanet population (see, e.g., Fig.~\ref{fig:ss}), but this is
because they are below the detection threshold of radial velocity and transit surveys.

There are also several examples of proposed giant impact debris found much closer to the
star (e.g., HD23514 at 0.25\,au, \citet{Rhee2008}; TYC8241 2652 1 at 0.4\,au, \citet{Melis2012};
ID8 at 0.33\,au, \citet{Meng2014}).
It is notable that the stars with close-in debris are of later spectral type (F-, G- and K-type stars)
than those mentioned above with debris at larger distance.
Indeed, Table~\ref{tab:gieg} shows a significant dependence of the radial location of the debris
on spectral type.
This trend was anticipated from the bottom left panel of Fig.~\ref{fig:gi}. 
However, since these stars are closest to the Sun in their properties, and the nominal model predicts that
giant impact debris from planets at 1-3\,au would be much longer-lived than that at $\ll 1$\,au,
it is still surprising that more distant debris is not more prevalent in the population of known
giant impact debris (see Fig.~\ref{fig:gieg}).
There are three possible explanations for this:
(i) the model parameters are wrong;
(ii) the population in Table~\ref{tab:gieg} is biased;
(iii) there a few giant impacts occuring with planets in the 1-3\,au region.

(i) Increasing the largest debris fragment size is one way to preferentially detect debris at smaller radii
(see bottom right panel of Fig.~\ref{fig:gi}), but it is only possible to favour debris at a few tenths of
an au with unrealistic ($>1000$\,km) debris sizes.
Favouring such small radii could be achieved by modifying the debris size distribution in other ways,
to push up the detection threshold set by eq.~\ref{eq:rlambda},
but the upper envelope set by eq.~\ref{eq:vescvk} means this would inevitably require the debris to have been
released from $\sim 100M_\oplus$ planets, which seems unlikely if such planets are primarily gaseous \citep[e.g.,][]{Rogers2015}.
It may also be possible to preferentially increase the duration of detectability of close-in debris by
including additional physics in the model.
For example, one aspect that could increase the duration of detectability of hot giant impact debris
is its vaporisation (and subsequent recondensation).
The abundance of gas and optical depth effects during this early phase could make such debris persist
at detectable levels much longer than predicted here, and such effects may be particularly relevant closer
to the star (Jackson et al., in prep.).

(ii) Table~\ref{tab:gieg} does not include all possible examples of giant impact debris, which should be
a subset of the 12\,$\mu$m excess candidates found in surveys that searched all nearby stars
\citep[e.g.,][]{Kennedy2013, Cotten2016}.
The problem is that interpretation of such a set of candidates requires careful consideration
of whether the excesses come from giant impacts, protoplanetary disks, or from a more distant
exo-asteroid belt. 
We have concentrated here on the systems claimed in the literature as giant impact debris, but it
could be that the more ambiguous interpretation of colder debris at $>1$\,au has biased against
including them in Table~\ref{tab:gieg}.
To assess this we considered the \citet{Kennedy2013} sample of 12\,$\mu$m excesses, excluding those thought to
be protoplanetary or transition disks, those not confirmed to be $<120$\,Myr, and those of earlier
spectral type than F3.
This left 7 candidates, including HD113766 and HD15407 that are noted in Table~\ref{tab:gieg},
and five F3V-F8 stars with dust in the range $1.3-2.7$\,au (HD115371, HD103703, HD106389, HD108857, HD22680).
If these are confirmed to have giant impact debris, then given the expected distribution of dust locations
for stars with similar stellar properties shown in Fig.~\ref{fig:gieg}, it is likely that this distribution
is consistent with giant impacts that occur with equal frequency in logarithmic bins of
planet mass and semimajor axis.
Note that the $\sim$Myr duration of detectability predicted by the model is also consistent with
that estimated from giant impact debris from super-Earths in section 8 of \citet{Wyatt2016} from the
fraction of 10-100\,Myr stars with 12\,$\mu$m excesses, which was in the range 0.1-10\,Myr.
Given the implications discussed in (iii), a thorough analysis of potential giant impact debris
to assess its origin is warranted.

(iii) The most interesting possibility is that the lack of giant impacts 1-3\,au arises from a lack
of planets in this region.
Like the A stars, the other detection techniques have yet to fully characterise the frequency of
planets in this region.
It is well known that super-Earth planets $\ll 1$\,au are common, but the population further out
can only be assessed from extrapolation of the population of planets closer in \citep[see][]{Winn2015}.
Thus searches for giant impact debris potentially provide a way of assessing the ubiquity of planets in a
region of interest to those studying habitable planets.
While we cannot say that terrestrial planets cannot be commonly present at 1-3\,au from
Sun-like stars, if they were inferred to be less common than closer-in super-Earths, this would be in contradiction
to some extrapolations of that population \citep[e.g.,][]{traub2012}.
For now, it is worth emphasising that the underlying planet population should be imprinted in observations of the
debris from giant impacts.

Another point to note from Table~\ref{tab:gieg} is the absence of giant impact debris around late-type stars,
such as M stars, since the bottom left panel of Fig.~\ref{fig:gi} shows that even relatively low
mass planets can result in giant impact debris that is at a detectable level.
Moreover, transit studies have shown us that planets are relatively common in exactly this region
\citep{Winn2015}.
Since the duration of detectability is only an order of magnitude lower than for Sun-like stars, in
a region known to be abundant with planets, it would appear surprising that giant impact debris
has yet to be detected around an M star.
The explanation could be that the low luminosity of M stars means that they are inherently faint, and so
it is not possible to detect debris down to levels of $R_{12}=0.1$ except for the nearest stars.
Alternatively these planetary systems are born both stable and without extra embryos, such that giant
impacts do not occur on the main sequence.

The above discussion has focussed on young stars (within a few 100\,Myr) and the possibility that
embryos formed near the planet in question are the origin of giant impacts.
However, another source of impactors is the exocomet population discussed in more detail in \S \ref{ss:comet}
which means that Gyr-old stars can also exhibit the giant impact phenomenon,
and indeed the \citet{Kennedy2013} sample of 12\,$\mu$m excess star includes old stars such as BD+20307.
\citet{Wetherill1994} pointed out that increasing the comet scattering rate in the Solar
System also increases the expected size of the largest impactor to have hit the Earth, from a
25\,km body at the current epoch to one the size of Ceres in their simulation B.
This points to the possibility that an increased comet population, as well as releasing dust through
sublimation, mutual collisions and disintegration, could result in a giant impact that releases large
quantities of dust into the inner regions of a system.
\citet{Lisse2012} suggested such a mechanism for the origin of hot dust in the
$\eta$ Corvi system, which is inferred to be at $\sim 3$\,au from this 1.4\,Gyr-old F2V star.
The spatial distribution of the hot dust provides one method to test this scenario,
because the geometry of collisional debris results in an asymmetry in the dust production
location that is enhanced at the point of impact for several 1000 orbits \citep{Jackson2014}.
Observations of $\eta$ Corvi are consistent with such an asymmetry, since while the dust temperature
puts it at 3\,au from the star \citep{Lisse2012}, a significant fraction of dust is seen at a closer
projected separation of $\sim 0.7$\,au \citep{Defrere2015, Kennedy2015lbti}.
If this is the correct interpretation then we have some constraints on the planet that was impacted,
since a planet at 3\,au which does not eject comets that approach it, but does
accrete them, would be $0.1-10M_\oplus$.
The number of known two-temperature debris disks is growing \citep{Morales2011, Chen2014}.
Like $\eta$ Corvi, the warmer inner components of these disks can often be confirmed to be spatially
separated from the outer cooler components \citep{Kennedy2014}, which may provide a source of
impacts onto planets in the inner regions and so an explanation for the presence of the warm dust.

\subsection{Exocomets}
\label{ss:comet}
The flux of comets in the inner reaches of planetary systems is of particular interest
because of its implications for the habitability of planets in the habitable
zone (e.g., due to the delivery of water to the planets, and catastrophic impacts) and as a
possible explanation for the Earth's Late Veneer \citep[e.g.,][]{Morbidelli2015}.
These comets may also replenish the dust disks known as exozodi that are seen within a few au
of several stars from their excess near-IR to mid-IR emission \citep{Absil2006, Smith2009etacorvi, Absil2013,
Ertel2014, Mennesson2014}.
If they reach close enough to the star, or are favourably aligned to our line-of-sight, such
comets may also show up in transit as they pass in front of the star \citep[e.g.,][]{Jura2005, Kiefer2014b,
Vanderburg2015, Boyajian2016}.

In the Solar System there are two main families of comets. 
The Ecliptic comets (such as the Jupiter-Family comets) originate in the Kuiper belt,
but following some perturbation they are dislodged from their initially stable orbit
and start undergoing encounters with Neptune.
Some of these scatterings pass the comets inward through the Centaur region where
they come under the gravitational influence of scattering by the closer in planets.
Ultimately, most are ejected by Jupiter.
However, some make it to the inner Solar System where they
appear as comets and disintegrate to replenish the dust in the zodiacal cloud \citep{Nesvorny2010};
a small fraction find a dynamical path to remain there as Encke-type comets \citep[e.g., ][]{Levison2006}.
The long-period comets originate in the Oort Cloud, and arrive in the inner Solar System once Galactic
tides have sufficiently reduced the pericentres of their orbits.
Planetary architectures that maximise the extrasolar analogues to these two comet families
are discussed in \S \ref{sss:jfc} and \ref{sss:lpc}, respectively.

\subsubsection{High cometary flux from exo-Kuiper belt}
\label{sss:jfc}
The question of how to maximise the cometary flux scattered in from an exo-Kuiper belt has been
studied by several authors.
For example, \cite{Bonsor2012analytic} showed how constraints on the Tisserand parameter give an indication
of the types of chains of planets required to scatter comets in, which must be sufficiently tightly
packed for scattered comets to reach a given proximity to the star or to start being scattered
by the next planet further down the chain (see, e.g., their eq.~3).
This was followed up in \cite{Bonsor2012nbody} by numerical simulations which quantified the comet influx
from chains of planets, the conclusion of which was that comet influx is necessarily rather small.
\cite{Bonsor2013, Bonsor2014} considered ways to increase the comet influx by allowing the outer planet in
the chain to be scattered into, or to migrate into, the Kuiper belt to replenish the cometary population.
Consideration of Fig.~\ref{fig:ss} shows that there are 3 main requirements in the design of
a planetary system to maximise the comet influx.
In addition to requiring a chain of closely separated planets:

{\it (1) No ejector: }
The most important requirement was raised in \S \ref{ss:ss}, which is that if there is
a planet in the chain that is in the {\it ejected} region, then the comets may never make it into
the inner system (or if they do they do not remain there for very long), but instead are ejected from the system.
This explains why \cite{Raymond2014} concluded that the presence of a Jupiter-mass planet
at $>15$\,au is inconsistent with a scattering origin for the exozodiacal dust of Vega.
It also explains why the \cite{Horner2009} simulations in which they considered how the
impact rate on the Earth from Centaurs (i.e., objects scattered in from the Kuiper belt)
changes as the mass of Jupiter is changed.
They found that the impact rate is maximised when the mass of Jupiter is close to that of Saturn.
The increase in impact rate as Jupiter's mass is decreased is readily understandable from the arguments
above, because Jupiter is in the {\it ejected} region, and so decreasing its mass reduces its ability
to eject comets (or rather they survive longer and so a greater fraction reach the inner
Solar System).
The decrease in comet flux as Jupiter's mass is reduced below Saturn mass arises
because Saturn is also in the {\it ejected} region, and so can eject all of the comets
that pass by the radius of its orbit, but only so long as there is no massive planet interior
to its orbit that scatters comets away from Saturn's orbit on a shorter timescale than that
which Saturn ejects them.
The conclusion that a large exocomet population is incompatible with a system with an ejector
planet seems contradictory to the abundance of exocomets seen toward the star $\beta$ Pic
as FEBs \citep[Falling Evaporating Bodies; e.g.,][]{Kiefer2014b}, since this system is known to host
the $9M_{\rm jup}$ planet $\beta$ Pic-b orbiting at 9\,au \citep{Lagrange2010}.
However, the favoured model for the dynamical origin of the FEBs is in a mean motion resonance
that is interior to $\beta$ Pic-b \citep{Beust2000};
that is the exocomets are not required to have crossed the planet's orbit in this system.
The edge-on orientation and youth of the $\beta$ Pic system may also play a role in the detectability
of this phenomenon.

{\it (2) Inward torque: }
Another requirement on the chain of planets
is that the comets have to be passed inward.
If not, they will end up undergoing the cometary diffusion described by \citetalias{Tremaine1993} that either
implants them in the Oort Cloud or ejects them.
The resulting requirement on the planets is beyond the scope of this paper, but a
general comment is that during the early stages of scattering the particles tend to be
scattered both interior and exterior to the planet.
The requirement for an interior planet to start dominating the scattering process is thus likely
to be when the timescale for scattering by that inner planet is shorter than the timescale for
scattering by the planet in question.
Since these timescales will scale with orbital period, the expectation would be that the
tipping point would be close to flat in mass with distance (hence why the \citet{Horner2009}
simulations found that Jupiter should be comparable in mass to Saturn for its ability
to scatter comets inward to be comparable with the ability of Saturn to eject comets).
However, the exact mass distribution required to pass comets inward also depends on the separation
between the planets.
\cite{Bonsor2013} and \cite{Raymond2014} show that decreasing the planet separation increases
the rate at which particles are scattered inwards, although they also found that there are specific
configurations related to resonances that can increase the efficiency of inward scattering;
if too closely packed the planetary system may also become unstable \citep[e.g., ][]{Faber2007}.
Nevertheless, it is clear that an inward flow requires that planet mass does not increase too
strongly with radius.
Indeed, \cite{Raymond2014} show that chains of planets with decreasing planet mass as a funcion of distance
scatter planetesimals inward more efficiently than those with increasing mass, though note that all of
their simulations included a planet in the ejection region.

{\it (3) Replenishment: }
Finally, another requirement is for the population of comets undergoing scattering to be
replenished.
The problem is that objects on unstable orbits tend to undergo scattering and are
consequently removed relatively quickly from the system, on a timescale $t_{\rm sca}$,
while objects on stable orbits can remain unperturbed over Gyr timescales.
Defining $M_0$ to be the initial mass of the comet belt, another way of stating the above
problem is that it is only possible to get comet mass influx rates approaching $M_0/t_{\star}$
early on in the evolution (i.e., for $t_{\star} \ll t_{\rm sca}$), as shown for example in the simulations
of \citet{Bonsor2012nbody}.
This is why \citet{Bonsor2014} invoked outward migration of the outer planet, because
as long as this can be sustained over $t_{\star}$ timescales, the resulting migration of
its unstable resonance overlap region causes objects on previously long-term stable orbits
to end up in an unstable region where they could undergo scattering with the planet.
Alternatively \cite{Bonsor2013} invoked a dynamical instability which was triggered 
late on, starting at a time $t_{\star} \gg t_{\rm sca}$,
which allowed comet influx rates of $M_0/t_{\rm sca}$ that are significantly in excess
of $M_0/t_{\star}$, but only for short periods of time.
Another mechanism proposed by Faramaz et al. (in prep.) involves long-timescale diffusion
in resonances \citep[e.g., ][]{Murray1997} which can allow large influx rates on Gyr
timescales, but requires the majority of the belt to be near resonances to achieve influx
rates of $M_0/t_{\star}$.
Here we propose that a planet (or planets) embedded within a planetesimal belt could place those
planetesimals onto orbits that cross an interior planetary system where they may undergo
further scattering.
This is similar to the suggestion that dwarf planets embedded in the Kuiper belt
excite eccentricities in the Kuiper belt population which could place them
on unstable orbits over a timescale comparable to the age of the Solar System
\citep{Munoz-Gutierrez2015}.
It is also similar to the embedded planets scenario proposed in \S \ref{ss:younga} to explain
the broad disk of HR8799, since as well as scattering debris out, some will also get scattered
in where it can interact with an inner planetary system potentially ending up on comet-like
orbits.
For a system of given age, the maximum rate of comet influx ($M_0/t_{\star}$) is likely to
arise for an embedded planet which has a timescale for depleting the disk of order the age of the system
(since more massive planets would have depleted the disk long ago, while less massive planets
would scatter material in too slowly).
For HR8799-like parameters, this would argue for a planet at 100\,au that is around Saturn-mass, similar
to the planet required for the broad debris disk, though noting (as in \S \ref{ss:younga}) that HR8799
does not satisfy requirement (1) and so is not expected to have a maximised exocomet population.

The three requirements proposed above should be tested against numerical simulations, of which there are
several in the literature.
For example, \cite{Bonsor2012nbody} did simulations of chains of planets between 5-50\,au with masses of
either Jupiter, Saturn or Neptune.
In all systems most particles are scattered out, which is expected from requirement (1) because all systems have
planets in the {\it ejected} region.
They also found that while putting planets closer together increases the inward scattering
rate, and that having high mass planets does this faster, the overall fraction
of the belt scattered in (referring to particles reaching $<1$\,au) is both similar and
a small fraction of $M_0/t_{\star}$ (see their Fig.~9).
This can be understood, as lower mass planets may result in a larger fraction of the scattered
material ending up at $<1$\,au, but a higher mass planet may destabilise a larger fraction of the
planetesimal belt.
Regardless, these simulations are not optimised for inward scattering, because they include a
planet in the {\it ejected} region, and there is a finite source of planetesimals in the scattering
region (requirement (3)).
A closer example of simulations predicted to be optimised for inward scattering is
given in \cite{Wetherill1994};
these simulations were not full N-body, but instead used a Monte-Carlo scattering approach
based on the \"{O}pik-Arnold method.
That paper gave arguments similar to those above about the importance of
reducing Jupiter and Saturn's mass to stop the leak of particles being ejected,
also pointing out that this allows Earth to capture more comets into orbits interior
to Jupiter.
Their simulation B which decreased Jupiter and Saturn to $15M_\oplus$, found that
this results in a factor 100-1000 increase in the comet influx.
Fig.~2 of \cite{Raymond2014} also showed an increased scattering rate as planet mass,
in a system of 5 equal mass planets, is decreased from $100M_\oplus$ to $10M_\oplus$, though
the scattering rate decreased as planet mass was decreased further to $5M_\oplus$.
Decreasing planet mass further in the Wetherill (1994) simulations did end up with a higher
comet flux (their simulation D which had 24 planets of $0.2-5M_\oplus$ in the 3.5-10\,au region),
but the character of the simulation had changed;
these planets ended up on highly eccentric orbits, and Earth also grew by a factor of 8,
which is readily understood to arise because these planets are in the
{\it accreted} regime, and were placed too close to prevent instability and
scattering amongst the planets.
Thus for now the extent to which comet influx can be increased by reducing planet masses
below $10M_\oplus$ is unclear from simulations in the literature, and deserves further
attention.


\subsubsection{High long-period cometary flux}
\label{sss:lpc}
The requirements on planetary architecture to maximise the long-period comet flux
are similar to those for Jupiter-family comets in \S \ref{sss:jfc}.
Requirement (1) for an absence of planets in the {\it ejected} region
still stands, since once Galactic tides have perturbed Oort Cloud comets into
the planetary system they still need to pass the planets and can be ejected before reaching
the innermost parts of the system.
In numerical simulations which considered the effect of changing the mass of Jupiter,
\cite{Horner2010} concluded that the rate of comets passed in from the Oort Cloud
is reduced by the presence of a Jupiter-like ejector planet.
Requirement (2) from \S \ref{sss:jfc} does not apply to comets torqued in from the Oort Cloud,
for which tidal perturbations provide the inward torque.
The replenishment requirement (3) from \S \ref{sss:jfc} still applies to this scenario,
but is perhaps most clearly rephrased as a requirement to maximise the mass implanted in the
Oort Cloud (although the time taken for objects to reach the Oort Cloud and that to perturb them back to
the inner system may argue for a specific configuration to explain replenishment at a particular
stellar age).
One way of achieving this is by placing the planets in a chain in which they all lie in the
Oort Cloud region on Fig.~\ref{fig:ss}, since all planets acting individually would act to scatter
planetesimals in their vicinity toward the Oort Cloud.
For example, placing all planets along the blue dotted line corresponding to a diffusion
timescale of 1\,Gyr on Fig.~\ref{fig:ss} may be suitable.
However, the constraints on the inner planets in the chain may in fact be less stringent,
since as long as the planet masses increase sufficiently with orbital radius, the
comets would be passed outward along the chain, rather than inward, with the consequence
that it is the outermost planet in the chain that is important for understanding the
eventual outcome.
Just as late dynamical instabilities in the planetary system can enhance the exo-Kuiper belt
comet flux, so can stellar flybys also in the short term enhance the long-period comet flux
\citep{Hills1981, Fouchard2011}, thus allowing high long-period comet influx rates in systems
with planetary system architectures that result in low mass Oort Clouds.

To test these proposed requirements, consider the simulations of \cite{Lewis2013}
for the formation of the Oort Cloud in systems similar to the Solar System, but reducing
the masses of various planets down to either Saturn or Neptune mass.
By the reasoning above, the resulting Oort Cloud from their simulations should not be too
different to the current Oort Cloud, because all simulations had Uranus and Neptune at their
current masses, both of which Fig.~\ref{fig:ss} shows are able to scatter objects into the Oort Cloud.
Although a lower efficiency would be expected for simulations with more
massive inner planets.
Their Fig. 1 shows that this is indeed the case, also confirming the $\sim 1$\,Gyr
timescale we predict for the creation of the Oort Cloud as well as its depletion by
passing stars.
Table 1 of \cite{Lewis2013} also gives the rate of long period comets reaching $<2$\,au which were
very similar for their simulations, although the rate was highest for the simulations in which all 4 giant
planets had Neptune mass.
This configuration is closest to the optimal configuration we predict, but it may be possible to
increase this rate further, since a Neptune-mass Jupiter and Saturn would still be in the {\it ejected}
region.
That is, these planets would have ejected material which could have ended up in the Oort Cloud
had their masses been reduced to $\sim 5M_\oplus$ (or lower).

\subsection{Debris in outer regions}
\label{ss:outerdd}
Debris disks are most often detected by their far-IR emission which originates in dust at $\gg 10$\,au
from the star. 
While this is most commonly interpreted as dust created in the collisional cascade of classical Kuiper
belt analogues, this section explores how it might be possible to determine if the emission instead
had a dominant scattering component, either a scattered disk (\S \ref{sss:sd}) or an Oort Cloud
(\S \ref{sss:minioort}).

\subsubsection{Scattered Disk}
\label{sss:sd}
The possibility of observing debris that is being scattered by a planet,
such as the component of the Kuiper belt that
extends beyond Neptune that has orbits with pericentre close to Neptune,
was already introduced in \S \ref{ss:younga}.
This was used as a possible explanation for the broad debris disk
of HR8799, and the breadth of the disk is one observational
manifestation of the scattering process.
Simulations show that scattering processes cause an extended scattered disk to have a surface
density distribution that falls off $\propto r^{-3.5}$ \citep{Duncan1987}.
However, the observed profile of a scattered disk might be flatter than this due
to collisional evolution which preferentially erodes the inner parts of the
distribution \citep{Wyatt2010}. 
Nevertheless, a scattered disk would be brightest at the inner edge, and
detailed collisional modelling can be used to determine the expected radial
profile. 

Another potentially observable characteristic of such a disk is an absence of small dust
grains.
This is because most collisions occur at pericentre which means that
even large grains (i.e., even those with a radiation pressure coefficient $\beta<0.01$)
can be put on unbound orbits by radiation pressure 
resulting in the size distribution being cut off at a size above that expected
for a cascade of planetesimals on low eccentricity ($e<0.3$) orbits for which the
cut-off is at $\beta \approx 0.5$ \citep{Wyatt2010}.
This provides a means to test whether a debris disk originates in such a population, since a lack of
small grains affects the temperature of the emission at a given radial location, and so may be in
evidence as an unexpectedly low temperature either in the main ring, or in the debris disk's halo
\citep[e.g., ][]{Matthews2014hr8799}.
Such an interpretation is complicated, however, by the fact that a lack of small grains in the
size distribution can also arise from a very low level of stirring \citep[e.g., $e \ll 0.01$,][]{Thebault2008}.
While this only affects grains small enough to be put on elliptical orbits by radiation
pressure (i.e., those with $0.1<\beta<0.5$), and the halo is still comprised of only small grains,
constraints on the energy available in a collision to create new surface area are another
reason why small grains could be under-abundant in systems with low levels of stirring
\citep{Krijt2014, Thebault2016}.

If a star is concluded to have a scattered disk component (e.g., from the temperature
and radial distribution of its debris disk emission), the properties of the scattering
planet can be inferred from Fig.~\ref{fig:ss} given the stellar age, mass and the radius
of the scattered disk's inner edge.
A lack of small grains has been inferred for the outer 150\,au ring of $\eta$ Corvi
\citep{Duchene2014}, and also for other disks imaged by Herschel \citep{Pawellek2014, Pawellek2015}.
Application of this interpretation to $\eta$ Corvi shows that at $\sim 1.4$\,Gyr this
could still have a significant scattered disk population, of objects on their way to
being implanted in the Oort Cloud, for a planet that is relatively low mass (i.e.,
with a scattering timescale of $\sim 10$\,Gyr, and so $1-10M_\oplus$).
A more massive planet could also be the origin of the scattered disk, as long as it
was put onto an orbit in the belt more recently.
While it has been proposed that this system could be currently observed at an epoch
analogous to that of the Solar System's Late Heavy Bombardment \citep{Gomes2005},
as an explanation that expands on that discussed in \S \ref{sss:ssimp} 
for the unusually high quantities of hot dust in the system
\citep{Lisse2012}, this may be ruled out by high resolution imaging of the
structure of the 150\,au belt (Marino et al., in prep.).
Note, however, that the low temperature of the outer belt of $\eta$ Corvi may also be explained
by the composition of the debris, rather than by its dynamics \citep[see discussion in][]{Duchene2014},
and so it is premature to make claims of embedded planets in this system without more
detailed analysis of the dust distribution.

A less extreme version of the scattered disk with its $r^{-3.5}$ profile can
be made by decreasing the planet mass (or by looking at a younger star),
since in this case the planetesimals may not yet have reached very high eccentricity
orbits.
Instead the debris disk will appear broader than a typical disk, i.e. the possibility discussed
in \S \ref{ss:younga}.
Examples of broad disks other than HR8799 \citep{Booth2016} include
61 Vir \citep{Wyatt2012}, $\gamma$ Tri \citep{Booth2013}, and $\gamma$ Dor \citep{Broekhoven-Fiene2013}.
It may be hard to distinguish between a belt that was initially broad with low eccentricities,
and one which started narrower but was broadened by interactions with a planet.
However, numerical simulations of this interaction might show the surface
density distribution has certain characteristics that can be compared with observations.
As one example, Fig.~1 of \cite{Booth2009} shows the surface density distribution in the
Nice model of \cite{Gomes2005} at snapshots of before the instability (when the
mass is concentrated in the belt), during the instability (when the belt is broad),
and after the instability (when the belt is still broad but much depleted).

\subsubsection{Mini-Oort Clouds}
\label{sss:minioort}
Consider the scenario described in \S \ref{ss:youngss} for the formation of Sedna-like objects applied
to a larger population of detached objects.
These objects would form a mini-Oort Cloud, i.e., one with a radius of $\sim 1000$\,au.
Since the orbital planes of this population would be randomised, they would form a spherical shell
around the star, which if dense enough would collide to result in a collisional cascade and so dust
\citep[e.g., ][]{Howe2014} which should emit at infrared-radio wavelengths and could be confused with a
debris disk that would be interpreted as a Kuiper belt analogue.
The detached disk population is not thought to be that significant in the Solar System, somewhere
between $0.01-5M_\oplus$ depending on the size distribution \citep{Brown2004, Schwamb2009,
Trujillo2014}.
However, extrasolar mini-Oort Clouds could be enhanced relative to ours, since the presence of
Jupiter would have ejected many objects being scattered by Saturn before they would have reached our
mini-Oort Cloud, and so systems without a Jupiter-like ejector would have enhanced versions of our
detached disk population.
The debris disk of Vega was originally interpreted as a spherical shell \citep{Aumann1984},
which might have been a reasonable explanation for the symmetrical dust distribution seen on the
sky \citep{Sibthorpe2010}.
However, this star is being viewed pole-on \citep{Aufdenberg2006}, and so the observed spherical symmetry of
the dust distribution is more likely explained as a face-on viewing geometry of a planar debris disk
that is aligned with the stellar equator.
Thus there is no convincing evidence that (mini-)Oort Clouds have been detected yet.

The existence of mini-Oort Clouds has already been proposed in the literature using a different formation
mechanism, i.e., planet-planet scattering \citep{Raymond2013}.
This is not an outcome that was included on Fig.~\ref{fig:youngss} because it likely results from
a multi-planet interaction which causes the scattering planet to be moved
far enough from the debris that further scattering interactions are prevented.
The mechanism proposed here is different, since continued scattering by the planet is prevented
by the tidal interaction of the debris with nearby stars (in the same way our Oort Cloud formed) rather than by
perturbations to the scattering planet.
It is also relevant that \citet{Kaib2011} found that the inner edge of the Oort Cloud can be strongly affected
by radial migration of the host star through the Galaxy, and that encounters with field stars can be efficient at
implanting objects on Sedna-like orbits if the host star spends significant time in dense environments.

\subsection{Exoplanet populations}
\label{ss:plpop}
This section combines some thoughts on how scattering processes might be evident in the exoplanet
populations.
There is no point in repeating well known conclusions (e.g., about the origin of the eccentric Jupiters),
but there are prospects for discovering new scattered planets through direct imaging
(\S \ref{sss:escape} and \S \ref{sss:exosedna}), or indeed for discovering those that have since been
ejected (\S \ref{sss:ffp}),
and scattering processes may have direct relevance to the formation of super-Earth planets (\S \ref{sss:se}).

\subsubsection{Escaping planets}
\label{sss:escape}
\S \ref{ss:younga} already introduced the possibility that young stars may contain a population
of planets that are in the process of being ejected from their systems in multiple scattering
events off other close-in planets.
However, \S \ref{ss:app} also pointed out one potential limitation on this population
which is that scattered planets cannot be significantly more massive than the planet that
scattered them.
This implies that the known long-period giant planets, all of which are
more massive than a few Jupiter mass (the detection threshold of current
instrumentation), are unlikely to be this population of escaping planets \citep[see also][]{Bryan2016}.
Even if their systems do contain a several Jupiter mass planet orbiting closer in that is
capable of scattering the planets to this distance, the escaping planet would have been put onto an
unbound orbit on a timescale much shorter than the system age (multi-planet interactions
not-withstanding).

The planetary system architecture which maximises the population of escaping planets
can be derived from Fig.~\ref{fig:younga};
the scattering planets should all lie in the {\it escaping} shaded region
(i.e., Neptune-mass planets at 5\,au or Saturn-mass planets at 30-100\,au),
with no adjacent planets in the {\it ejected} region.
As noted above, this restricts the scattered planets to be of similar or lower mass,
which means they are too faint for detection with current instruments unless surrounded
by large quantities of dust \citep[e.g., ][]{Kennedy2011}.
If such escaping planets are discovered, then Fig.~\ref{fig:younga} provides a framework
within which to consider what additional planets may be present in the system (e.g., see discussion
in \S \ref{ss:younga} about the origin of Fomalhaut-b's orbit), 
albeit with caveats for possible multi-planet interactions.

A direct comparison of these predictions with the results of numerical simulations of planet-planet
scattering \citep[e.g., ][]{Chatterjee2008, Juric2008, Veras2009, Raymond2010} is hard, because most simulations involved an initial population of
planets with a range of masses drawn from a distribution.
Consequently, a range of outcomes were found, not necessarily optimised to maximising the
scattered population.
However such simulations did demonstrate that such a population exists around young
stars.
For example, \cite{Veras2009} placed 6 planets at 3-7\,au in the mass range 3-$3\times 10^4M_\oplus$
and found that most were ejected, but that some were still escaping at 10s of Myr (see their Fig.~1).
This is expected from Fig.~\ref{fig:younga}, since the assumed distribution means that one of the
planets surely lies in the {\it ejected} region.
While the few $>170M_\oplus$ planets in their escaping population (see their Fig.~2) seem to contradict the
prediction that the escaping planets cannot be massive, these likely originate from systems which also
host massive inner planets, and the escaping planets may have been prevented from rapid ejection (which is
the fate that Fig.~\ref{fig:younga} would otherwise predict) by encounters with other planets in the
system.
An escaping planet population is also seen in the simulations of \cite{Raymond2010} which,
in agreement with the predictions of Fig.~\ref{fig:ss}, showed that a larger
fraction of 3-planet systems with planets in (or close to) the escaping region
(i.e., Neptune-mass planets) spend more time in a transitional ejection phase
with pericentre larger than 5\,au.

\subsubsection{Exo-Sednas}
\label{sss:exosedna}
As an addendum to the scenario discussed in \S \ref{sss:escape}, it is possible that the
escaping planets never escaped, but were pulled away from the inner planetary system by
tidal forces which implanted them in the Oort Cloud at a distance appropriate for the
cluster environment it was in.
This could thus result in planets at an intermediate $\sim 1000$\,au distance 
\citep[similar to the putative planet nine in the Solar System;][]{Batygin2016}
which would be stable against further perturbations once the cluster has dissipated.
Future surveys may find such planets on stable orbits and question whether
they formed in situ, were captured from an orbit around a nearby star \citep{Mustill2016},
or were scattered out (as proposed here).
One test of the latter hypothesis would be to search for the scattering planet, the properties
of which can be predicted from Fig.~\ref{fig:youngss} (i.e., they would be expected
to lie in the Oort Cloud region on that figure) with the additional constraints of eq.~\ref{eq:m2m1}
(i.e., that they would be more massive than the scattered planet), albeit with the caveat that
multiple scattering planets may complicate the predictions.

The existence of such planets could also have implications for the inner planetary system.
That is, if such planets are massive enough, and detached at a close enough separation from the
star, these could have caused subsequent disruption to the inner planetary system.
This is because, in the same way that the orbits of Oort Cloud comets are isotropic, tidal forces
would have randomised the detached planet's orbital plane relative to that of its progenitor planetary
system, and so the Oort Cloud planet could thus induce Kozai
oscillations and so excite a large eccentricity in the inner planetary system.
Indeed, \cite{Martin2016} invoked this mechanism as a way for a circumbinary planet to
influence the orbit of its host stars. 
Such oscillations take time however, and a multi-planet system may be stable against such
perturbations, even if their timescale is shorter than the age of the system.

\subsubsection{Escaped Planets}
\label{sss:ffp}
A further addendum to the scenario discussed in \S \ref{sss:escape} is the possibility that planets
might be detected after having been ejected from the system. 
Indeed, Jupiter mass interstellar (or free floating) planets are possibly more common than main sequence stars
\citep{Sumi2011}, and it is possible that such planets formed in a circumstellar disk, but were since ejected
in scattering interactions with other planets in the system.
This would imply that either planets capable of ejecting Jupiter mass planets (i.e., those of comparable or greater
mass, \S \ref{ss:app}) are common, or the number of planets ejected per ejector is high, or that the interstellar
planets have an origin in a different mechanism.
The first of these possibilities can be assessed observationally, and current estimates would place the fraction
of systems with Jupiter mass planets at closer to 10\% than 100\% \citep{Winn2015}, though this cannot be claimed with any confidence
since the full range of parameter space has yet to be explored (in particular the occurrence rate of Jupiter
mass planets at large orbital radii is poorly constrained).

\subsubsection{Super-Earth Formation}
\label{sss:se}
Consider a system in which multiple embryos form within a few au.
Fig.~\ref{fig:ss} shows that, if these are nudged onto crossing orbits so that scattering ensues,
then the result will be that the embryos collide and coalesce, until they are sufficiently 
separated.
There have been many papers on this process such that it is unnecessary to repeat here
\citep[see, e.g., ][]{Chambers2001, Petrovich2014}.
However, one point to make is that the same applies to any solid mass that makes its way into the inner region.
Thus, if we assume that planets in the outer region are scattering planetesimals, and potentially passing them
inward onto comet-like orbits, then if those planetesimals encroach into the growing super-Earth
region, they will be accreted onto the super-Earth.
The resulting exchange of angular momentum would make the super-Earth planet move out;
how far depends on how much mass is accreted.
Since the angular momentum of the planet scales as $\sim \mu m_{\rm p}a_{\rm p}^{1/2}$
(where $\mu=GM_\star$),
and that gained from accreting a small mass $dm$ at the pericentre of a high eccentricity
orbit is $\mu dm \sqrt{2a_{\rm p}}$, the super-Earth planet would grow by this mechanism keeping
$m_{\rm p}^{2(\sqrt{2}-1)}a_{\rm p}^{-1}$ constant, i.e., along a track on Fig.~\ref{fig:ss}
of $M_{\rm p} \propto a_{\rm p}^{1.2}$.
Growth by this mechanism is fundamentally limited, however, by the rate at which mass
can be scattered into the super-Earth's feeding zone, a topic discussed in \S \ref{ss:comet}.
It is thus perhaps likely that scattering of planetesimals is an inefficient method of mass transfer.

It is also worth considering the implications of Fig.~\ref{fig:ss} for a model in which
a super-Earth forms further out then migrates in \citep[e.g., ][]{Alibert2006}.
In this case the planet transitions from a region in which the material it encounters is destined
for ejection to one which it starts to accrete everything it encounters.
Indeed, the simulations of \cite{Payne2009} showed that a large fraction of planetesimals
are not accreted onto the migrating planet in this process, but instead end up in a broad scattered disk
extending beyond where the planet started.
Thus we suggest from Fig.~\ref{fig:ss} that super-Earths might grow more efficiently in
this way by evolving up a track on that figure that keeps them below the $v_{\rm esc}=v_{\rm k}$ line.

More generally we can note that, since the $v_{\rm esc}=v_{\rm k}$ line applies to the accretion of
solid material, but not that of gas, the $v_{\rm esc}/v_{\rm k}=1$ line should represent the maximum
core mass for an object formed at that location.
This explains why it is impossible to form Uranus and Neptune through collisional growth at their current
locations \citep{Levison2001}.
The discovery of solid planets at large distance thus may provide a challenge for planet
formation models, which could be resolved if such planets form closer to the star and then migrate
outward, or if growth occurs through a mechanism such as pebble accretion \citep[since gas drag can prevent
the pebbles being ejected in the way discussed here;][]{Levison2015}.

\section{Conclusion}
\label{s:conc}
This paper considered the dynamical outcome for an object orbiting a star that is being scattered through close encounters
with a planet. 
It was shown that, assuming a single low-eccentricity planet system, that outcome
is divided into six main regions that are described in \S \ref{s:mpvap}, and depend only on
the mass and semimajor axis of the planet:
accreted, ejected, remaining, escaping, Oort Cloud, depleted Oort Cloud. 
While this division was known from the previous work of \citetalias{Tremaine1993}, this
paper gives equal emphasis to all outcomes and considers the implications for the
various components of extrasolar planetary systems that are much better known now.
It also emphasises the importance of the ratio of the planet's escape velocity to its Keplerian
velocity in determining the outcome.
After considering a few example systems and comparison with dynamical simulations
in the literature (\S \ref{s:applications}), the paper focusses on the types of planetary sytem architectures
that favour specific outcomes (\S \ref{s:outcomes}).

The outcome for scattering by terrestrial and super-Earth planets is accretion onto the
planet, although mutual collisions with other objects undergoing scattering and
eventual grinding into dust is another loss mechanism.
Debris released in a giant impact involving the planet is a typical origin of a population
undergoing such scattering.
It was shown that the planet's mass and semimajor axis have a strong effect on the duration of
detectability of giant impact debris which peaks at a specific planet mass and
semimajor axis, the exact value of which depends on the detection threshold and wavelength
of observation, as well as the spectral type of the star.
Whereas the examples of giant impact debris proposed in the literature around A-type stars are
found at radial locations compatible with expectations (i.e., at 4-6\,au, implying progenitor
planets of $3-10M_\oplus$), those found around Sun-like stars are found at $\ll 1$\,au much closer
to the star than expected.
This could indicate an absence of terrestrial planets beyond 1\,au around Sun-like stars,
emphasising the potential of giant impact debris searches to constrain the frequency of
habitable planets.
However, for now we cannot rule out that giant impacts occur with equal frequency per logarithmic
bin in planet mass and semimajor axis.

While the framework considered in this paper only applies to single planet systems, scattering in a
multiple planet system can also be inferred by considering how each planet acts in isolation. 
This paper specifically considers which planetary system architectures favour the production of exocomets,
identifying three principles that maximise exocomets being scattered in from an exo-Kuiper belt.
Significant exocomet populations require a chain of closely spaced planets in which none of the
planets is massive enough to favour ejection of objects that encounter it.
Planet masses should not increase with distance from the star, so that comets are passed in rather
than out.
Constant replenishment of the comet population is also required, which we suggest could be facilitated
by a low-mass planet embedded in the exo-Kuiper belt.
Similar principles apply to exocomets that arrive in the inner regions of the system from an
exo-Oort Cloud.
In this case an absence of ejecting planets is also required, and the planet masses
should be appropriate to maximise the amount of material that ends up in the exo-Oort Cloud.

Extrasolar debris disks are usually interpreted by analogy with the Solar System's classical Kuiper
belt (i.e., objects on stable low eccentricity orbits).
Here we suggest the possibility that debris disks may have a significant
scattered disk component (i.e., of objects with high eccentricities currently undergoing
scattering with a planet).
Such scattered disks could be inferred from the radial breadth of the debris disk and from
a lack of small grains;
e.g., the broad disk of HR8799 could be caused by an embedded Saturn-mass planet.
We also proposed that mini-Oort Clouds could result from a planetary system that was born in
a dense cluster.

With many direct imaging campaigns currently searching for exoplanets at large distance from their
host star, this paper also considered the possibility of observing planets that are in the process
of being ejected through interactions with an inner planetary system, or of seeing detached
planets akin to Sedna in the Solar System.
The framework presented in this paper readily provides predictions for the scattering planets
that can be used in the interpretation of any such detections; 
e.g., we showed how a $30M_\oplus$ planet at 32\,au could explain the origin of the high eccentricity
of the Fomalhaut-b orbit.

The value of the framework presented in this paper is in its simplicity, and 
as such its limitations should also be born in mind.
In particular, we did not consider the possibility that the planets are on eccentric orbits,
which may aid scattering \citep[e.g., ][]{Frewen2014}, and lead to secular evolution of the
planetary system which could be important \citep[e.g.,][]{Beust2014, Pearce2014, Read2016}.
Also, the planet mass versus semimajor axis diagram as presented only claims to predict the
dominant outcome.
Other outcomes are also possible but at lower probability.
Furthermore, only limited consideration was given to the dynamics of multi-planet systems.
There is no substitute for N-body simulations which would provide a more definitive
answer as to the outcome of scattering in a specific system.
However, the framework presented herein provides a useful tool to interpret N-body simulations,
which can also be used to devise planetary system architectures for specific outcomes,
even if the predictions still need to be tested with more detailed simulations.

\appendix
\section{Stellar Encounters}
\label{a:enc}
In \S \ref{ss:ttide} the timescale for nearby stars to modify the orbit of a scattered
object was calculated assuming that this was dominated by Galactic tides.
Here we consider how including perturbations from stellar encounters would have
changed (if at all) any of the resulting conclusions.
For the case of the Solar System, \citet{Heisler1986} concluded that the timescales for
stellar encounters to modify a comet's orbit scale in the same way as those for Galactic tides,
but are longer and so can be ignored.
However, this conclusion may not apply to the broader range of system parameters considered in this
paper, and moreover when calculating the perturbations from stellar encounters it is important 
to note that the way these scale with system parameters depends on the rate of stellar encounters.

\begin{table}
  \centering
  \caption{Units of parameters introduced in Appendix~\ref{a:enc}.}
  \label{tab:unitsapp}
  \begin{tabular}{lll}
     \hline
     Parameter                           & Symbol         & Units \\
     \hline
     Specific angular momentum           & $J$            & au$^2$\,yr$^{-1}$ \\
     Local stellar mass density          & $\rho_{\rm s}$ & 0.045\,$M_\odot$\,pc$^{-3}$ \\
     Velocity dispersion of nearby stars & $\sigma$       & 20\,km\,s$^{-1}$ \\
     Impulsive/diffusive boundary        & $a_{\rm imp}$  & au \\
     \hline
  \end{tabular}
\end{table}

If the rate of stellar encounters is low enough that the resulting change in a comet's
orbit is dominated by the single strongest stellar encounter, rather
than by the cumulative effect of many weaker encounters
(i.e. that the perturbation is impulsive rather than diffusive),
this leads to a mean square change in the comet's specific angular momentum per orbit of
\citep[eq.~37 of][]{Heisler1986}
\begin{equation}
  \left< \Delta J^2 \right> = 8 \times 10^{-30} \rho_{\rm s}^2 M_\star a^7
  \label{eq:ht37}
\end{equation}
in au$^4$\,yr$^{-2}$.
In eq.~\ref{eq:ht37}, $\rho_{\rm s}$ is the local stellar mass density of nearby stars in units of
0.045\,$M_\odot$\,pc$^{-3}$, which is that appropriate for stars near the Sun \citep{Bahcall1980,Holmberg2000},
and assumes the stellar mass distribution has the same shape as that near the Sun;
see Table~\ref{tab:unitsapp} for a summary of the units of parameters introduced in this Appendix.

This assumption breaks down when the comet is far enough from the star that the smallest impact
parameter expected over the comet's orbital period is inside the comet's orbit;
i.e., when the comet's semimajor axis $a>a_{\rm imp}$, where \citep[see eq.~35 of][]{Heisler1986}
\begin{equation}
  a_{\rm imp} \approx 3.5 \times 10^4 M_\star^{1/7} \rho_{\rm s}^{-2/7} \sigma^{-2/7},
  \label{eq:aimp}
\end{equation}
and $\sigma$ is the velocity dispersion of nearby stars in units of 20\,km\,s$^{-1}$
\citep[the value appropriate for stars near the Sun;][]{Heisler1986}.
In this regime the change per orbit is instead given by \cite[eq.~33 of][]{Heisler1986}
\begin{equation}
  \left< \Delta J^2 \right> = 3.5 \times 10^{-13} \rho_{\rm s} M_\star^{3/2} a^{7/2} \sigma^{-1}.
  \label{eq:ht33}
\end{equation}

Comparing eqs.~\ref{eq:ht37} and \ref{eq:ht33} shows that for
orbits at $a \gg a_{\rm imp}$ the diffusive perturbations from stellar
encounters are much weaker than the impulsive approximation would have predicted.
While this comparison also seems to imply that stellar encounters are stronger
by a factor of $\sim 5$ for orbits at the boundary between these regimes
(i.e., at $a=a_{\rm imp}$) than would have been assumed
by calculating their effect using the impulsive approximation,
this factor is close to unity and independent of other parameters.
Thus we consider it more realistic that there is a smooth transition between the two regimes
at a semimajor axis $\sim 1.6$ times further out than given by eq.~\ref{eq:aimp},
and that using the impulsive approximation will never underestimate the effect of stellar
encounters, though it will overestimate it at $a\gg a_{\rm imp}$.

The mean square change in the comet's specific angular momentum due to Galactic tides can
also be calculated \citep[eq.~20 of][]{Heisler1986}
\begin{equation}
  \left< \Delta J^2 \right> = 1.2 \times 10^{-29} \rho_0^2 M_\star^{-1} a^7.
  \label{eq:ht20}
\end{equation}
This means that the ratio of the perturbation (to a comet's angular momentum squared) from
stellar encounters to that from Galactic tides is a factor
$0.65 (\rho_{\rm s} M_\star / \rho_0)^2$ for $a<1.6a_{\rm imp}$,
and lower than this for comets orbiting at larger semimajor axes.
As such we conclude that Galactic tides dominate over stellar encounters (which can thus be ignored)
as long as
\begin{equation}
  M_\star < 1.2 \rho_0 / \rho_{\rm s}.
  \label{eq:enc}
\end{equation}
Equation~\ref{eq:enc} is satisfied for most of the systems considered in this paper except that in
\S \ref{ss:younga}.

The analysis presented in this paper (which assumed Galactic tides dominate) can also be readily
modified to account for stellar encounters.
The simplest way to account for impulsive stellar encounters for systems which do not satisfy
eq.~\ref{eq:enc} is to replace all instances of $\rho_0$ in the equations with
$0.81 \rho_{\rm s} M_\star$.
However, if such an analysis concludes that objects are placed by stellar encounters in an Oort
Cloud at $a>1.6a_{\rm imp}$, then this calculation would have overestimated the effect of stellar encounters
which should instead have been considered in the diffusive regime.
Replacing $a_{\rm f}$ in eq.~\ref{eq:af} with $1.6a_{\rm imp}$ (from eq.~\ref{eq:aimp})
shows that this applies to Oort Clouds formed by planets that are more massive than
\begin{equation}
  M_{\rm p} = 2.9 M_\star^{17/28} a_{\rm p}^{3/4} \rho_0^{1/2} \rho_{\rm s}^{-3/14} \sigma^{-3/14}.
  \label{eq:mpimpf}
\end{equation}
Since planets more massive than eq.~\ref{eq:aej} would still eject objects before
placing them in the Oort Cloud, this means that only a narrow region of parameter 
space of planets beyond 
\begin{equation}
  a_{\rm p} = 800 M_\star^{-1} \rho_0^2 \rho_{\rm s}^{-2} \sigma^{-2}
\end{equation}
is potentially affected, though stellar encounters or Galactic tides may still implant objects in an
Oort Cloud from scattering by planets in this region.

\section*{Acknowledgments}
MCW, AB and AS acknowledge the support of the European Union through
ERC grant number 279973.
APJ acknowledges support from NASA grant NNX16AI31G.
AS is partially supported by funding from the Center for Exoplanets and Habitable Worlds.
The Center for Exoplanets and Habitable Worlds is supported by the Pennsylvania State University,
the Eberly College of Science, and the Pennsylvania Space Grant Consortium.

\bibliographystyle{mnras}
\bibliography{refs.bib} 

\begin{thebibliography}{}
\makeatletter
\relax
\def\mn@urlcharsother{\let\do\@makeother \do\$\do\&\do\#\do\^\do\_\do\%\do\~}
\def\mn@doi{\begingroup\mn@urlcharsother \@ifnextchar [ {\mn@doi@}
  {\mn@doi@[]}}
\def\mn@doi@[#1]#2{\def\@tempa{#1}\ifx\@tempa\@empty \href
  {http://dx.doi.org/#2} {doi:#2}\else \href {http://dx.doi.org/#2} {#1}\fi
  \endgroup}
\def\mn@eprint#1#2{\mn@eprint@#1:#2::\@nil}
\def\mn@eprint@arXiv#1{\href {http://arxiv.org/abs/#1} {{\tt arXiv:#1}}}
\def\mn@eprint@dblp#1{\href {http://dblp.uni-trier.de/rec/bibtex/#1.xml}
  {dblp:#1}}
\def\mn@eprint@#1:#2:#3:#4\@nil{\def\@tempa {#1}\def\@tempb {#2}\def\@tempc
  {#3}\ifx \@tempc \@empty \let \@tempc \@tempb \let \@tempb \@tempa \fi \ifx
  \@tempb \@empty \def\@tempb {arXiv}\fi \@ifundefined
  {mn@eprint@\@tempb}{\@tempb:\@tempc}{\expandafter \expandafter \csname
  mn@eprint@\@tempb\endcsname \expandafter{\@tempc}}}

\bibitem[\protect\citeauthoryear{{Absil} et~al.,}{{Absil}
  et~al.}{2006}]{Absil2006}
{Absil} O.,  et~al., 2006, \mn@doi [\aap] {10.1051/0004-6361:20054522}, \href
  {http://adsabs.harvard.edu/abs/2006A%26A...452..237A} {452, 237}

\bibitem[\protect\citeauthoryear{{Absil} et~al.,}{{Absil}
  et~al.}{2013}]{Absil2013}
{Absil} O.,  et~al., 2013, \mn@doi [\aap] {10.1051/0004-6361/201321673}, \href
  {http://adsabs.harvard.edu/abs/2013A%26A...555A.104A} {555, A104}

\bibitem[\protect\citeauthoryear{{Adams}}{{Adams}}{2010}]{Adams2010}
{Adams} F.~C.,  2010, \mn@doi [\araa] {10.1146/annurev-astro-081309-130830},
  \href {http://adsabs.harvard.edu/abs/2010ARA%26A..48...47A} {48, 47}

\bibitem[\protect\citeauthoryear{{Alibert} et~al.,}{{Alibert}
  et~al.}{2006}]{Alibert2006}
{Alibert} Y.,  et~al., 2006, \mn@doi [\aap] {10.1051/0004-6361:20065697}, \href
  {http://adsabs.harvard.edu/abs/2006A%26A...455L..25A} {455, L25}

\bibitem[\protect\citeauthoryear{{Aufdenberg} et~al.,}{{Aufdenberg}
  et~al.}{2006}]{Aufdenberg2006}
{Aufdenberg} J.~P.,  et~al., 2006, \mn@doi [\apj] {10.1086/504149}, \href
  {http://adsabs.harvard.edu/abs/2006ApJ...645..664A} {645, 664}

\bibitem[\protect\citeauthoryear{{Aumann} et~al.,}{{Aumann}
  et~al.}{1984}]{Aumann1984}
{Aumann} H.~H.,  et~al., 1984, \mn@doi [\apjl] {10.1086/184214}, \href
  {http://adsabs.harvard.edu/abs/1984ApJ...278L..23A} {278, L23}

\bibitem[\protect\citeauthoryear{{Bahcall} \& {Soneira}}{{Bahcall} \&
  {Soneira}}{1980}]{Bahcall1980}
{Bahcall} J.~N.,  {Soneira} R.~M.,  1980, \mn@doi [\apjs] {10.1086/190685},
  \href {http://adsabs.harvard.edu/abs/1980ApJS...44...73B} {44, 73}

\bibitem[\protect\citeauthoryear{{Batygin}}{{Batygin}}{2015}]{Batygin2015}
{Batygin} K.,  2015, \mn@doi [\mnras] {10.1093/mnras/stv1063}, \href
  {http://adsabs.harvard.edu/abs/2015MNRAS.451.2589B} {451, 2589}

\bibitem[\protect\citeauthoryear{{Batygin} \& {Brown}}{{Batygin} \&
  {Brown}}{2016}]{Batygin2016}
{Batygin} K.,  {Brown} M.~E.,  2016, \mn@doi [\aj]
  {10.3847/0004-6256/151/2/22}, \href
  {http://adsabs.harvard.edu/abs/2016AJ....151...22B} {151, 22}

\bibitem[\protect\citeauthoryear{{Beichman} et~al.,}{{Beichman}
  et~al.}{2005}]{Beichman2005}
{Beichman} C.~A.,  et~al., 2005, \mn@doi [\apj] {10.1086/430059}, \href
  {http://adsabs.harvard.edu/abs/2005ApJ...626.1061B} {626, 1061}

\bibitem[\protect\citeauthoryear{{Beust} \& {Morbidelli}}{{Beust} \&
  {Morbidelli}}{2000}]{Beust2000}
{Beust} H.,  {Morbidelli} A.,  2000, \mn@doi [\icarus]
  {10.1006/icar.1999.6238}, \href
  {http://adsabs.harvard.edu/abs/2000Icar..143..170B} {143, 170}

\bibitem[\protect\citeauthoryear{{Beust} et~al.,}{{Beust}
  et~al.}{2014}]{Beust2014}
{Beust} H.,  et~al., 2014, \mn@doi [\aap] {10.1051/0004-6361/201322229}, \href
  {http://adsabs.harvard.edu/abs/2014A%26A...561A..43B} {561, A43}

\bibitem[\protect\citeauthoryear{{Bonsor} \& {Wyatt}}{{Bonsor} \&
  {Wyatt}}{2012}]{Bonsor2012analytic}
{Bonsor} A.,  {Wyatt} M.~C.,  2012, \mn@doi [\mnras]
  {10.1111/j.1365-2966.2011.20156.x}, \href
  {http://adsabs.harvard.edu/abs/2012MNRAS.420.2990B} {420, 2990}

\bibitem[\protect\citeauthoryear{{Bonsor}, {Augereau}  \&
  {Th{\'e}bault}}{{Bonsor} et~al.}{2012}]{Bonsor2012nbody}
{Bonsor} A.,  {Augereau} J.-C.,   {Th{\'e}bault} P.,  2012, \mn@doi [\aap]
  {10.1051/0004-6361/201220005}, \href
  {http://adsabs.harvard.edu/abs/2012A%26A...548A.104B} {548, A104}

\bibitem[\protect\citeauthoryear{{Bonsor}, {Raymond}  \& {Augereau}}{{Bonsor}
  et~al.}{2013}]{Bonsor2013}
{Bonsor} A.,  {Raymond} S.~N.,   {Augereau} J.-C.,  2013, \mn@doi [\mnras]
  {10.1093/mnras/stt933}, \href
  {http://adsabs.harvard.edu/abs/2013MNRAS.433.2938B} {433, 2938}

\bibitem[\protect\citeauthoryear{{Bonsor}, {Raymond}, {Augereau}  \&
  {Ormel}}{{Bonsor} et~al.}{2014}]{Bonsor2014}
{Bonsor} A.,  {Raymond} S.~N.,  {Augereau} J.-C.,   {Ormel} C.~W.,  2014,
  \mn@doi [\mnras] {10.1093/mnras/stu721}, \href
  {http://adsabs.harvard.edu/abs/2014MNRAS.441.2380B} {441, 2380}

\bibitem[\protect\citeauthoryear{{Booth}, {Wyatt}, {Morbidelli},
  {Moro-Mart{\'{\i}}n}  \& {Levison}}{{Booth} et~al.}{2009}]{Booth2009}
{Booth} M.,  {Wyatt} M.~C.,  {Morbidelli} A.,  {Moro-Mart{\'{\i}}n} A.,
  {Levison} H.~F.,  2009, \mn@doi [\mnras] {10.1111/j.1365-2966.2009.15286.x},
  \href {http://adsabs.harvard.edu/abs/2009MNRAS.399..385B} {399, 385}

\bibitem[\protect\citeauthoryear{{Booth} et~al.,}{{Booth}
  et~al.}{2013}]{Booth2013}
{Booth} M.,  et~al., 2013, \mn@doi [\mnras] {10.1093/mnras/sts117}, \href
  {http://adsabs.harvard.edu/abs/2013MNRAS.428.1263B} {428, 1263}

\bibitem[\protect\citeauthoryear{{Booth} et~al.,}{{Booth}
  et~al.}{2016}]{Booth2016}
{Booth} M.,  et~al., 2016, \mn@doi [\mnras] {10.1093/mnrasl/slw040}, \href
  {http://adsabs.harvard.edu/abs/2016MNRAS.tmpL..24B} {}

\bibitem[\protect\citeauthoryear{{Bottke}, {Jedicke}, {Morbidelli}, {Petit}  \&
  {Gladman}}{{Bottke} et~al.}{2000}]{Bottke2000}
{Bottke} W.~F.,  {Jedicke} R.,  {Morbidelli} A.,  {Petit} J.-M.,   {Gladman}
  B.,  2000, \mn@doi [Science] {10.1126/science.288.5474.2190}, \href
  {http://adsabs.harvard.edu/abs/2000Sci...288.2190B} {288, 2190}

\bibitem[\protect\citeauthoryear{{Bottke}, {Morbidelli}, {Jedicke}, {Petit},
  {Levison}, {Michel}  \& {Metcalfe}}{{Bottke} et~al.}{2002}]{Bottke2002}
{Bottke} W.~F.,  {Morbidelli} A.,  {Jedicke} R.,  {Petit} J.-M.,  {Levison}
  H.~F.,  {Michel} P.,   {Metcalfe} T.~S.,  2002, \mn@doi [\icarus]
  {10.1006/icar.2001.6788}, \href
  {http://adsabs.harvard.edu/abs/2002Icar..156..399B} {156, 399}

\bibitem[\protect\citeauthoryear{{Bowler}}{{Bowler}}{2016}]{Bowler2016}
{Bowler} B.~P.,  2016, preprint, \href
  {http://adsabs.harvard.edu/abs/2016arXiv160502731B} {} (\mn@eprint {arXiv}
  {1605.02731})

\bibitem[\protect\citeauthoryear{{Boyajian} et~al.,}{{Boyajian}
  et~al.}{2016}]{Boyajian2016}
{Boyajian} T.~S.,  et~al., 2016, \mn@doi [\mnras] {10.1093/mnras/stw218}, \href
  {http://adsabs.harvard.edu/abs/2016MNRAS.457.3988B} {457, 3988}

\bibitem[\protect\citeauthoryear{{Brasser} \& {Duncan}}{{Brasser} \&
  {Duncan}}{2008}]{BrasserDuncan2008}
{Brasser} R.,  {Duncan} M.~J.,  2008, \mn@doi [Celestial Mechanics and
  Dynamical Astronomy] {10.1007/s10569-007-9100-y}, \href
  {http://adsabs.harvard.edu/abs/2008CeMDA.100....1B} {100, 1}

\bibitem[\protect\citeauthoryear{{Brasser} \& {Schwamb}}{{Brasser} \&
  {Schwamb}}{2015}]{Brasser2015}
{Brasser} R.,  {Schwamb} M.~E.,  2015, \mn@doi [\mnras]
  {10.1093/mnras/stu2374}, \href
  {http://adsabs.harvard.edu/abs/2015MNRAS.446.3788B} {446, 3788}

\bibitem[\protect\citeauthoryear{{Brasser}, {Duncan}  \& {Levison}}{{Brasser}
  et~al.}{2006}]{Brasser2006}
{Brasser} R.,  {Duncan} M.~J.,   {Levison} H.~F.,  2006, \mn@doi [\icarus]
  {10.1016/j.icarus.2006.04.010}, \href
  {http://adsabs.harvard.edu/abs/2006Icar..184...59B} {184, 59}

\bibitem[\protect\citeauthoryear{{Brasser}, {Duncan}  \& {Levison}}{{Brasser}
  et~al.}{2007}]{Brasser2007}
{Brasser} R.,  {Duncan} M.~J.,   {Levison} H.~F.,  2007, \mn@doi [\icarus]
  {10.1016/j.icarus.2007.05.003}, \href
  {http://adsabs.harvard.edu/abs/2007Icar..191..413B} {191, 413}

\bibitem[\protect\citeauthoryear{{Brasser}, {Duncan}  \& {Levison}}{{Brasser}
  et~al.}{2008}]{Brasser2008}
{Brasser} R.,  {Duncan} M.~J.,   {Levison} H.~F.,  2008, \mn@doi [\icarus]
  {10.1016/j.icarus.2008.02.016}, \href
  {http://adsabs.harvard.edu/abs/2008Icar..196..274B} {196, 274}

\bibitem[\protect\citeauthoryear{{Broekhoven-Fiene} et~al.,}{{Broekhoven-Fiene}
  et~al.}{2013}]{Broekhoven-Fiene2013}
{Broekhoven-Fiene} H.,  et~al., 2013, \mn@doi [\apj]
  {10.1088/0004-637X/762/1/52}, \href
  {http://adsabs.harvard.edu/abs/2013ApJ...762...52B} {762, 52}

\bibitem[\protect\citeauthoryear{{Brown}, {Trujillo}  \& {Rabinowitz}}{{Brown}
  et~al.}{2004}]{Brown2004}
{Brown} M.~E.,  {Trujillo} C.,   {Rabinowitz} D.,  2004, \mn@doi [\apj]
  {10.1086/422095}, \href {http://adsabs.harvard.edu/abs/2004ApJ...617..645B}
  {617, 645}

\bibitem[\protect\citeauthoryear{{Bryan}, {Bowler}, {Knutson}, {Kraus},
  {Hinkley}, {Mawet}, {Nielsen}  \& {Blunt}}{{Bryan} et~al.}{2016}]{Bryan2016}
{Bryan} M.~L.,  {Bowler} B.~P.,  {Knutson} H.~A.,  {Kraus} A.~L.,  {Hinkley}
  S.,  {Mawet} D.,  {Nielsen} E.~L.,   {Blunt} S.~C.,  2016, preprint, \href
  {http://adsabs.harvard.edu/abs/2016arXiv160606744B} {} (\mn@eprint {arXiv}
  {1606.06744})

\bibitem[\protect\citeauthoryear{{Canup} \& {Asphaug}}{{Canup} \&
  {Asphaug}}{2001}]{Canup2001}
{Canup} R.~M.,  {Asphaug} E.,  2001, \nat, \href
  {http://adsabs.harvard.edu/abs/2001Natur.412..708C} {412, 708}

\bibitem[\protect\citeauthoryear{{Chambers}}{{Chambers}}{2001}]{Chambers2001}
{Chambers} J.~E.,  2001, \mn@doi [\icarus] {10.1006/icar.2001.6639}, \href
  {http://adsabs.harvard.edu/abs/2001Icar..152..205C} {152, 205}

\bibitem[\protect\citeauthoryear{{Chatterjee}, {Ford}, {Matsumura}  \&
  {Rasio}}{{Chatterjee} et~al.}{2008}]{Chatterjee2008}
{Chatterjee} S.,  {Ford} E.~B.,  {Matsumura} S.,   {Rasio} F.~A.,  2008,
  \mn@doi [\apj] {10.1086/590227}, \href
  {http://adsabs.harvard.edu/abs/2008ApJ...686..580C} {686, 580}

\bibitem[\protect\citeauthoryear{{Chen}, {Mittal}, {Kuchner}, {Forrest},
  {Lisse}, {Manoj}, {Sargent}  \& {Watson}}{{Chen} et~al.}{2014}]{Chen2014}
{Chen} C.~H.,  {Mittal} T.,  {Kuchner} M.,  {Forrest} W.~J.,  {Lisse} C.~M.,
  {Manoj} P.,  {Sargent} B.~A.,   {Watson} D.~M.,  2014, \mn@doi [\apjs]
  {10.1088/0067-0049/211/2/25}, \href
  {http://adsabs.harvard.edu/abs/2014ApJS..211...25C} {211, 25}

\bibitem[\protect\citeauthoryear{{Contro}, {Wittenmyer}, {Horner}  \&
  {Marshall}}{{Contro} et~al.}{2015}]{Contro2015}
{Contro} B.,  {Wittenmyer} R.~A.,  {Horner} J.,   {Marshall} J.~P.,  2015,
  preprint, \href {http://adsabs.harvard.edu/abs/2015arXiv150503198C} {}
  (\mn@eprint {arXiv} {1505.03198})

\bibitem[\protect\citeauthoryear{{Cotten} \& {Song}}{{Cotten} \&
  {Song}}{2016}]{Cotten2016}
{Cotten} T.~H.,  {Song} I.,  2016, preprint, \href
  {http://adsabs.harvard.edu/abs/2016arXiv160601134C} {} (\mn@eprint {arXiv}
  {1606.01134})

\bibitem[\protect\citeauthoryear{{Defr{\`e}re} et~al.,}{{Defr{\`e}re}
  et~al.}{2015}]{Defrere2015}
{Defr{\`e}re} D.,  et~al., 2015, \mn@doi [\apj] {10.1088/0004-637X/799/1/42},
  \href {http://adsabs.harvard.edu/abs/2015ApJ...799...42D} {799, 42}

\bibitem[\protect\citeauthoryear{{Defr{\`e}re} et~al.,}{{Defr{\`e}re}
  et~al.}{2016}]{Defrere2016}
{Defr{\`e}re} D.,  et~al., 2016, \mn@doi [\apj] {10.3847/0004-637X/824/2/66},
  \href {http://adsabs.harvard.edu/abs/2016ApJ...824...66D} {824, 66}

\bibitem[\protect\citeauthoryear{{Dent} et~al.,}{{Dent}
  et~al.}{2014}]{Dent2014}
{Dent} W.~R.~F.,  et~al., 2014, \mn@doi [Science] {10.1126/science.1248726},
  \href {http://adsabs.harvard.edu/abs/2014Sci...343.1490D} {343, 1490}

\bibitem[\protect\citeauthoryear{{Dones}, {Weissman}, {Levison}  \&
  {Duncan}}{{Dones} et~al.}{2004}]{Dones2004}
{Dones} L.,  {Weissman} P.~R.,  {Levison} H.~F.,   {Duncan} M.~J.,  2004, in
  {Johnstone} D.,  {Adams} F.~C.,  {Lin} D.~N.~C.,  {Neufeeld} D.~A.,
  {Ostriker} E.~C.,  eds,  Astronomical Society of the Pacific Conference
  Series Vol. 323, Star Formation in the Interstellar Medium: In Honor of David
  Hollenbach. p.~371

\bibitem[\protect\citeauthoryear{{Dones}, {Brasser}, {Kaib}  \&
  {Rickman}}{{Dones} et~al.}{2015}]{Dones2015}
{Dones} L.,  {Brasser} R.,  {Kaib} N.,   {Rickman} H.,  2015, \mn@doi [\ssr]
  {10.1007/s11214-015-0223-2}, \href
  {http://adsabs.harvard.edu/abs/2015SSRv..197..191D} {197, 191}

\bibitem[\protect\citeauthoryear{{Duch{\^e}ne} et~al.,}{{Duch{\^e}ne}
  et~al.}{2014}]{Duchene2014}
{Duch{\^e}ne} G.,  et~al., 2014, \mn@doi [\apj] {10.1088/0004-637X/784/2/148},
  \href {http://adsabs.harvard.edu/abs/2014ApJ...784..148D} {784, 148}

\bibitem[\protect\citeauthoryear{{Duncan} \& {Levison}}{{Duncan} \&
  {Levison}}{1997}]{Duncan1997}
{Duncan} M.~J.,  {Levison} H.~F.,  1997, \mn@doi [Science]
  {10.1126/science.276.5319.1670}, \href
  {http://adsabs.harvard.edu/abs/1997Sci...276.1670D} {276, 1670}

\bibitem[\protect\citeauthoryear{{Duncan}, {Quinn}  \& {Tremaine}}{{Duncan}
  et~al.}{1987}]{Duncan1987}
{Duncan} M.,  {Quinn} T.,   {Tremaine} S.,  1987, \mn@doi [\aj]
  {10.1086/114571}, \href {http://adsabs.harvard.edu/abs/1987AJ.....94.1330D}
  {94, 1330}

\bibitem[\protect\citeauthoryear{{Eiroa} et~al.,}{{Eiroa}
  et~al.}{2013}]{Eiroa2013}
{Eiroa} C.,  et~al., 2013, \mn@doi [\aap] {10.1051/0004-6361/201321050}, \href
  {http://adsabs.harvard.edu/abs/2013A%26A...555A..11E} {555, A11}

\bibitem[\protect\citeauthoryear{{Ertel} et~al.,}{{Ertel}
  et~al.}{2014}]{Ertel2014}
{Ertel} S.,  et~al., 2014, \mn@doi [\aap] {10.1051/0004-6361/201424438}, \href
  {http://adsabs.harvard.edu/abs/2014A%26A...570A.128E} {570, A128}

\bibitem[\protect\citeauthoryear{{Evans} \& {Tabachnik}}{{Evans} \&
  {Tabachnik}}{1999}]{Evans1999}
{Evans} N.~W.,  {Tabachnik} S.,  1999, \mn@doi [\nat] {10.1038/19919}, \href
  {http://adsabs.harvard.edu/abs/1999Natur.399...41E} {399, 41}

\bibitem[\protect\citeauthoryear{{Faber} \& {Quillen}}{{Faber} \&
  {Quillen}}{2007}]{Faber2007}
{Faber} P.,  {Quillen} A.~C.,  2007, \mn@doi [\mnras]
  {10.1111/j.1365-2966.2007.12490.x}, \href
  {http://adsabs.harvard.edu/abs/2007MNRAS.382.1823F} {382, 1823}

\bibitem[\protect\citeauthoryear{{Fabrycky} \& {Murray-Clay}}{{Fabrycky} \&
  {Murray-Clay}}{2010}]{Fabrycky&Murray-Clay2010}
{Fabrycky} D.~C.,  {Murray-Clay} R.~A.,  2010, \mn@doi [\apj]
  {10.1088/0004-637X/710/2/1408}, \href
  {http://adsabs.harvard.edu/abs/2010ApJ...710.1408F} {710, 1408}

\bibitem[\protect\citeauthoryear{{Ford} \& {Rasio}}{{Ford} \&
  {Rasio}}{2008}]{Ford2008}
{Ford} E.~B.,  {Rasio} F.~A.,  2008, \mn@doi [\apj] {10.1086/590926}, \href
  {http://adsabs.harvard.edu/abs/2008ApJ...686..621F} {686, 621}

\bibitem[\protect\citeauthoryear{{Fouchard}, {Rickman}, {Froeschl{\'e}}  \&
  {Valsecchi}}{{Fouchard} et~al.}{2011}]{Fouchard2011}
{Fouchard} M.,  {Rickman} H.,  {Froeschl{\'e}} C.,   {Valsecchi} G.~B.,  2011,
  \mn@doi [\aap] {10.1051/0004-6361/201116514}, \href
  {http://adsabs.harvard.edu/abs/2011A%26A...535A..86F} {535, A86}

\bibitem[\protect\citeauthoryear{{Frewen} \& {Hansen}}{{Frewen} \&
  {Hansen}}{2014}]{Frewen2014}
{Frewen} S.~F.~N.,  {Hansen} B.~M.~S.,  2014, \mn@doi [\mnras]
  {10.1093/mnras/stu097}, \href
  {http://adsabs.harvard.edu/abs/2014MNRAS.439.2442F} {439, 2442}

\bibitem[\protect\citeauthoryear{{Fujiwara}, {Onaka}, {Yamashita}, {Ishihara},
  {Kataza}, {Fukagawa}, {Takeda}  \& {Murakami}}{{Fujiwara}
  et~al.}{2012}]{Fujiwara2012}
{Fujiwara} H.,  {Onaka} T.,  {Yamashita} T.,  {Ishihara} D.,  {Kataza} H.,
  {Fukagawa} M.,  {Takeda} Y.,   {Murakami} H.,  2012, \mn@doi [\apjl]
  {10.1088/2041-8205/749/2/L29}, \href
  {http://adsabs.harvard.edu/abs/2012ApJ...749L..29F} {749, L29}

\bibitem[\protect\citeauthoryear{{Genda}, {Kobayashi}  \& {Kokubo}}{{Genda}
  et~al.}{2015}]{Genda2015}
{Genda} H.,  {Kobayashi} H.,   {Kokubo} E.,  2015, \mn@doi [\apj]
  {10.1088/0004-637X/810/2/136}, \href
  {http://adsabs.harvard.edu/abs/2015ApJ...810..136G} {810, 136}

\bibitem[\protect\citeauthoryear{{Gillon} et~al.,}{{Gillon}
  et~al.}{2016}]{Gillon2016}
{Gillon} M.,  et~al., 2016, \mn@doi [\nat] {10.1038/nature17448}, \href
  {http://adsabs.harvard.edu/abs/2016Natur.533..221G} {533, 221}

\bibitem[\protect\citeauthoryear{{Goldreich}, {Lithwick}  \&
  {Sari}}{{Goldreich} et~al.}{2004}]{Goldreich2004}
{Goldreich} P.,  {Lithwick} Y.,   {Sari} R.,  2004, \mn@doi [\apj]
  {10.1086/423612}, \href {http://adsabs.harvard.edu/abs/2004ApJ...614..497G}
  {614, 497}

\bibitem[\protect\citeauthoryear{{Gomes}, {Levison}, {Tsiganis}  \&
  {Morbidelli}}{{Gomes} et~al.}{2005}]{Gomes2005}
{Gomes} R.,  {Levison} H.~F.,  {Tsiganis} K.,   {Morbidelli} A.,  2005, \mn@doi
  [\nat] {10.1038/nature03676}, \href
  {http://adsabs.harvard.edu/abs/2005Natur.435..466G} {435, 466}

\bibitem[\protect\citeauthoryear{{Go{\'z}dziewski} \&
  {Migaszewski}}{{Go{\'z}dziewski} \& {Migaszewski}}{2014}]{Gozdziewski2014}
{Go{\'z}dziewski} K.,  {Migaszewski} C.,  2014, \mn@doi [\mnras]
  {10.1093/mnras/stu455}, \href
  {http://adsabs.harvard.edu/abs/2014MNRAS.440.3140G} {440, 3140}

\bibitem[\protect\citeauthoryear{{Heisler} \& {Tremaine}}{{Heisler} \&
  {Tremaine}}{1986}]{Heisler1986}
{Heisler} J.,  {Tremaine} S.,  1986, \mn@doi [\icarus]
  {10.1016/0019-1035(86)90060-6}, \href
  {http://adsabs.harvard.edu/abs/1986Icar...65...13H} {65, 13}

\bibitem[\protect\citeauthoryear{{Heng} \& {Tremaine}}{{Heng} \&
  {Tremaine}}{2010}]{Heng2010}
{Heng} K.,  {Tremaine} S.,  2010, \mn@doi [\mnras]
  {10.1111/j.1365-2966.2009.15739.x}, \href
  {http://adsabs.harvard.edu/abs/2010MNRAS.401..867H} {401, 867}

\bibitem[\protect\citeauthoryear{{Higuchi}, {Kokubo}  \& {Mukai}}{{Higuchi}
  et~al.}{2006}]{Higuchi2006}
{Higuchi} A.,  {Kokubo} E.,   {Mukai} T.,  2006, \mn@doi [\aj]
  {10.1086/498892}, \href {http://adsabs.harvard.edu/abs/2006AJ....131.1119H}
  {131, 1119}

\bibitem[\protect\citeauthoryear{{Hills}}{{Hills}}{1981}]{Hills1981}
{Hills} J.~G.,  1981, \mn@doi [\aj] {10.1086/113058}, \href
  {http://adsabs.harvard.edu/abs/1981AJ.....86.1730H} {86, 1730}

\bibitem[\protect\citeauthoryear{{Holmberg} \& {Flynn}}{{Holmberg} \&
  {Flynn}}{2000}]{Holmberg2000}
{Holmberg} J.,  {Flynn} C.,  2000, \mn@doi [\mnras]
  {10.1046/j.1365-8711.2000.02905.x}, \href
  {http://adsabs.harvard.edu/abs/2000MNRAS.313..209H} {313, 209}

\bibitem[\protect\citeauthoryear{{Horner} \& {Jones}}{{Horner} \&
  {Jones}}{2009}]{Horner2009}
{Horner} J.,  {Jones} B.~W.,  2009, \mn@doi [International Journal of
  Astrobiology] {10.1017/S1473550408004357}, \href
  {http://adsabs.harvard.edu/abs/2009IJAsB...8...75H} {8, 75}

\bibitem[\protect\citeauthoryear{{Horner}, {Jones}  \& {Chambers}}{{Horner}
  et~al.}{2010}]{Horner2010}
{Horner} J.,  {Jones} B.~W.,   {Chambers} J.,  2010, \mn@doi [International
  Journal of Astrobiology] {10.1017/S1473550409990346}, \href
  {http://adsabs.harvard.edu/abs/2010IJAsB...9....1H} {9, 1}

\bibitem[\protect\citeauthoryear{{Howard} et~al.,}{{Howard}
  et~al.}{2010}]{Howard2010}
{Howard} A.~W.,  et~al., 2010, \mn@doi [Science] {10.1126/science.1194854},
  \href {http://adsabs.harvard.edu/abs/2010Sci...330..653H} {330, 653}

\bibitem[\protect\citeauthoryear{{Howe} \& {Rafikov}}{{Howe} \&
  {Rafikov}}{2014}]{Howe2014}
{Howe} A.~R.,  {Rafikov} R.~R.,  2014, \mn@doi [\apj]
  {10.1088/0004-637X/781/1/52}, \href
  {http://adsabs.harvard.edu/abs/2014ApJ...781...52H} {781, 52}

\bibitem[\protect\citeauthoryear{{Ida} \& {Lin}}{{Ida} \&
  {Lin}}{2004}]{Ida2004}
{Ida} S.,  {Lin} D.~N.~C.,  2004, \mn@doi [\apj] {10.1086/424830}, \href
  {http://adsabs.harvard.edu/abs/2004ApJ...616..567I} {616, 567}

\bibitem[\protect\citeauthoryear{{Inamdar} \& {Schlichting}}{{Inamdar} \&
  {Schlichting}}{2016}]{Inamdar2016}
{Inamdar} N.~K.,  {Schlichting} H.~E.,  2016, \mn@doi [\apjl]
  {10.3847/2041-8205/817/2/L13}, \href
  {http://adsabs.harvard.edu/abs/2016ApJ...817L..13I} {817, L13}

\bibitem[\protect\citeauthoryear{{Jackson} \& {Wyatt}}{{Jackson} \&
  {Wyatt}}{2012}]{Jackson2012}
{Jackson} A.~P.,  {Wyatt} M.~C.,  2012, \mn@doi [\mnras]
  {10.1111/j.1365-2966.2012.21546.x}, \href
  {http://adsabs.harvard.edu/abs/2012MNRAS.425..657J} {425, 657}

\bibitem[\protect\citeauthoryear{{Jackson}, {Wyatt}, {Bonsor}  \&
  {Veras}}{{Jackson} et~al.}{2014}]{Jackson2014}
{Jackson} A.~P.,  {Wyatt} M.~C.,  {Bonsor} A.,   {Veras} D.,  2014, \mn@doi
  [\mnras] {10.1093/mnras/stu476}, \href
  {http://adsabs.harvard.edu/abs/2014MNRAS.440.3757J} {440, 3757}

\bibitem[\protect\citeauthoryear{{Johnson} \& {Melosh}}{{Johnson} \&
  {Melosh}}{2012}]{Johnson2012}
{Johnson} B.~C.,  {Melosh} H.~J.,  2012, \mn@doi [\icarus]
  {10.1016/j.icarus.2011.11.020}, \href
  {http://adsabs.harvard.edu/abs/2012Icar..217..416J} {217, 416}

\bibitem[\protect\citeauthoryear{{Jura}}{{Jura}}{2005}]{Jura2005}
{Jura} M.,  2005, \mn@doi [\apj] {10.1086/426902}, \href
  {http://adsabs.harvard.edu/abs/2005ApJ...620..487J} {620, 487}

\bibitem[\protect\citeauthoryear{{Jura}}{{Jura}}{2011}]{Jura2011}
{Jura} M.,  2011, \mn@doi [\aj] {10.1088/0004-6256/141/5/155}, \href
  {http://adsabs.harvard.edu/abs/2011AJ....141..155J} {141, 155}

\bibitem[\protect\citeauthoryear{{Juri{\'c}} \& {Tremaine}}{{Juri{\'c}} \&
  {Tremaine}}{2008}]{Juric2008}
{Juri{\'c}} M.,  {Tremaine} S.,  2008, \mn@doi [\apj] {10.1086/590047}, \href
  {http://adsabs.harvard.edu/abs/2008ApJ...686..603J} {686, 603}

\bibitem[\protect\citeauthoryear{{Kaib} \& {Quinn}}{{Kaib} \&
  {Quinn}}{2008}]{Kaib2008}
{Kaib} N.~A.,  {Quinn} T.,  2008, \mn@doi [\icarus]
  {10.1016/j.icarus.2008.03.020}, \href
  {http://adsabs.harvard.edu/abs/2008Icar..197..221K} {197, 221}

\bibitem[\protect\citeauthoryear{{Kaib}, {Ro{\v s}kar}  \& {Quinn}}{{Kaib}
  et~al.}{2011}]{Kaib2011}
{Kaib} N.~A.,  {Ro{\v s}kar} R.,   {Quinn} T.,  2011, \mn@doi [\icarus]
  {10.1016/j.icarus.2011.07.037}, \href
  {http://adsabs.harvard.edu/abs/2011Icar..215..491K} {215, 491}

\bibitem[\protect\citeauthoryear{{Kalas}, {Graham}  \& {Clampin}}{{Kalas}
  et~al.}{2005}]{Kalas2005}
{Kalas} P.,  {Graham} J.~R.,   {Clampin} M.,  2005, \mn@doi [\nat]
  {10.1038/nature03601}, \href
  {http://adsabs.harvard.edu/abs/2005Natur.435.1067K} {435, 1067}

\bibitem[\protect\citeauthoryear{{Kalas}, {Graham}, {Fitzgerald}  \&
  {Clampin}}{{Kalas} et~al.}{2013}]{Kalas2013}
{Kalas} P.,  {Graham} J.~R.,  {Fitzgerald} M.~P.,   {Clampin} M.,  2013,
  \mn@doi [\apj] {10.1088/0004-637X/775/1/56}, \href
  {http://adsabs.harvard.edu/abs/2013ApJ...775...56K} {775, 56}

\bibitem[\protect\citeauthoryear{{Kennedy} \& {Piette}}{{Kennedy} \&
  {Piette}}{2015}]{Kennedy2015prdrag}
{Kennedy} G.~M.,  {Piette} A.,  2015, \mn@doi [\mnras] {10.1093/mnras/stv453},
  \href {http://adsabs.harvard.edu/abs/2015MNRAS.449.2304K} {449, 2304}

\bibitem[\protect\citeauthoryear{{Kennedy} \& {Wyatt}}{{Kennedy} \&
  {Wyatt}}{2011}]{Kennedy2011}
{Kennedy} G.~M.,  {Wyatt} M.~C.,  2011, \mn@doi [\mnras]
  {10.1111/j.1365-2966.2010.18041.x}, \href
  {http://adsabs.harvard.edu/abs/2011MNRAS.412.2137K} {412, 2137}

\bibitem[\protect\citeauthoryear{{Kennedy} \& {Wyatt}}{{Kennedy} \&
  {Wyatt}}{2013}]{Kennedy2013}
{Kennedy} G.~M.,  {Wyatt} M.~C.,  2013, \mn@doi [\mnras]
  {10.1093/mnras/stt900}, \href
  {http://adsabs.harvard.edu/abs/2013MNRAS.433.2334K} {433, 2334}

\bibitem[\protect\citeauthoryear{{Kennedy} \& {Wyatt}}{{Kennedy} \&
  {Wyatt}}{2014}]{Kennedy2014}
{Kennedy} G.~M.,  {Wyatt} M.~C.,  2014, \mn@doi [\mnras]
  {10.1093/mnras/stu1665}, \href
  {http://adsabs.harvard.edu/abs/2014MNRAS.444.3164K} {444, 3164}

\bibitem[\protect\citeauthoryear{{Kennedy} et~al.,}{{Kennedy}
  et~al.}{2015}]{Kennedy2015lbti}
{Kennedy} G.~M.,  et~al., 2015, \mn@doi [\apjs] {10.1088/0067-0049/216/2/23},
  \href {http://adsabs.harvard.edu/abs/2015ApJS..216...23K} {216, 23}

\bibitem[\protect\citeauthoryear{{Kenyon} \& {Bromley}}{{Kenyon} \&
  {Bromley}}{2006}]{Kenyon2006}
{Kenyon} S.~J.,  {Bromley} B.~C.,  2006, \mn@doi [\aj] {10.1086/499807}, \href
  {http://adsabs.harvard.edu/abs/2006AJ....131.1837K} {131, 1837}

\bibitem[\protect\citeauthoryear{{Kiefer}, {Lecavelier des Etangs}, {Boissier},
  {Vidal-Madjar}, {Beust}, {Lagrange}, {H{\'e}brard}  \& {Ferlet}}{{Kiefer}
  et~al.}{2014}]{Kiefer2014b}
{Kiefer} F.,  {Lecavelier des Etangs} A.,  {Boissier} J.,  {Vidal-Madjar} A.,
  {Beust} H.,  {Lagrange} A.-M.,  {H{\'e}brard} G.,   {Ferlet} R.,  2014,
  \mn@doi [\nat] {10.1038/nature13849}, \href
  {http://adsabs.harvard.edu/abs/2014Natur.514..462K} {514, 462}

\bibitem[\protect\citeauthoryear{{Konopacky}, {Marois}, {Macintosh},
  {Galicher}, {Barman}, {Metchev}  \& {Zuckerman}}{{Konopacky}
  et~al.}{2016}]{Konopacky2016}
{Konopacky} Q.~M.,  {Marois} C.,  {Macintosh} B.~A.,  {Galicher} R.,  {Barman}
  T.~S.,  {Metchev} S.~A.,   {Zuckerman} B.,  2016, preprint, \href
  {http://adsabs.harvard.edu/abs/2016arXiv160408157K} {} (\mn@eprint {arXiv}
  {1604.08157})

\bibitem[\protect\citeauthoryear{{Krijt} \& {Kama}}{{Krijt} \&
  {Kama}}{2014}]{Krijt2014}
{Krijt} S.,  {Kama} M.,  2014, \mn@doi [\aap] {10.1051/0004-6361/201423862},
  \href {http://adsabs.harvard.edu/abs/2014A%26A...566L...2K} {566, L2}

\bibitem[\protect\citeauthoryear{{Lagrange} et~al.,}{{Lagrange}
  et~al.}{2010}]{Lagrange2010}
{Lagrange} A.-M.,  et~al., 2010, \mn@doi [Science] {10.1126/science.1187187},
  \href {http://adsabs.harvard.edu/abs/2010Sci...329...57L} {329, 57}

\bibitem[\protect\citeauthoryear{{Lawler}, {Greenstreet}  \&
  {Gladman}}{{Lawler} et~al.}{2015}]{Lawler2015}
{Lawler} S.~M.,  {Greenstreet} S.,   {Gladman} B.,  2015, \mn@doi [\apjl]
  {10.1088/2041-8205/802/2/L20}, \href
  {http://adsabs.harvard.edu/abs/2015ApJ...802L..20L} {802, L20}

\bibitem[\protect\citeauthoryear{{Lee} \& {Peale}}{{Lee} \&
  {Peale}}{2003}]{LeePeale2003}
{Lee} M.~H.,  {Peale} S.~J.,  2003, \mn@doi [\apj] {10.1086/375857}, \href
  {http://adsabs.harvard.edu/abs/2003ApJ...592.1201L} {592, 1201}

\bibitem[\protect\citeauthoryear{{Levison} \& {Agnor}}{{Levison} \&
  {Agnor}}{2003}]{Levison2003}
{Levison} H.~F.,  {Agnor} C.,  2003, \mn@doi [\aj] {10.1086/374625}, \href
  {http://adsabs.harvard.edu/abs/2003AJ....125.2692L} {125, 2692}

\bibitem[\protect\citeauthoryear{{Levison} \& {Stewart}}{{Levison} \&
  {Stewart}}{2001}]{Levison2001}
{Levison} H.~F.,  {Stewart} G.~R.,  2001, \mn@doi [\icarus]
  {10.1006/icar.2001.6672}, \href
  {http://adsabs.harvard.edu/abs/2001Icar..153..224L} {153, 224}

\bibitem[\protect\citeauthoryear{{Levison}, {Terrell}, {Wiegert}, {Dones}  \&
  {Duncan}}{{Levison} et~al.}{2006}]{Levison2006}
{Levison} H.~F.,  {Terrell} D.,  {Wiegert} P.~A.,  {Dones} L.,   {Duncan}
  M.~J.,  2006, \mn@doi [\icarus] {10.1016/j.icarus.2005.12.016}, \href
  {http://adsabs.harvard.edu/abs/2006Icar..182..161L} {182, 161}

\bibitem[\protect\citeauthoryear{{Levison}, {Duncan}, {Brasser}  \&
  {Kaufmann}}{{Levison} et~al.}{2010}]{Levison2010}
{Levison} H.~F.,  {Duncan} M.~J.,  {Brasser} R.,   {Kaufmann} D.~E.,  2010,
  \mn@doi [Science] {10.1126/science.1187535}, \href
  {http://adsabs.harvard.edu/abs/2010Sci...329..187L} {329, 187}

\bibitem[\protect\citeauthoryear{{Levison}, {Kretke}  \& {Duncan}}{{Levison}
  et~al.}{2015}]{Levison2015}
{Levison} H.~F.,  {Kretke} K.~A.,   {Duncan} M.~J.,  2015, \mn@doi [\nat]
  {10.1038/nature14675}, \href
  {http://adsabs.harvard.edu/abs/2015Natur.524..322L} {524, 322}

\bibitem[\protect\citeauthoryear{{Lewis}, {Quinn}  \& {Kaib}}{{Lewis}
  et~al.}{2013}]{Lewis2013}
{Lewis} A.~R.,  {Quinn} T.,   {Kaib} N.~A.,  2013, \mn@doi [\aj]
  {10.1088/0004-6256/146/1/16}, \href
  {http://adsabs.harvard.edu/abs/2013AJ....146...16L} {146, 16}

\bibitem[\protect\citeauthoryear{{Lisse}, {Chen}, {Wyatt}  \& {Morlok}}{{Lisse}
  et~al.}{2008}]{Lisse2008}
{Lisse} C.~M.,  {Chen} C.~H.,  {Wyatt} M.~C.,   {Morlok} A.,  2008, \mn@doi
  [\apj] {10.1086/523626}, \href
  {http://adsabs.harvard.edu/abs/2008ApJ...673.1106L} {673, 1106}

\bibitem[\protect\citeauthoryear{{Lisse}, {Chen}, {Wyatt}, {Morlok}, {Song},
  {Bryden}  \& {Sheehan}}{{Lisse} et~al.}{2009}]{Lisse2009}
{Lisse} C.~M.,  {Chen} C.~H.,  {Wyatt} M.~C.,  {Morlok} A.,  {Song} I.,
  {Bryden} G.,   {Sheehan} P.,  2009, \mn@doi [\apj]
  {10.1088/0004-637X/701/2/2019}, \href
  {http://adsabs.harvard.edu/abs/2009ApJ...701.2019L} {701, 2019}

\bibitem[\protect\citeauthoryear{{Lisse} et~al.,}{{Lisse}
  et~al.}{2012}]{Lisse2012}
{Lisse} C.~M.,  et~al., 2012, \mn@doi [\apj] {10.1088/0004-637X/747/2/93},
  \href {http://adsabs.harvard.edu/abs/2012ApJ...747...93L} {747, 93}

\bibitem[\protect\citeauthoryear{{L{\"o}hne} et~al.,}{{L{\"o}hne}
  et~al.}{2012}]{Lohne2012}
{L{\"o}hne} T.,  et~al., 2012, \mn@doi [\aap] {10.1051/0004-6361/201117731},
  \href {http://adsabs.harvard.edu/abs/2012A%26A...537A.110L} {537, A110}

\bibitem[\protect\citeauthoryear{{Marois}, {Macintosh}, {Barman}, {Zuckerman},
  {Song}, {Patience}, {Lafreni{\`e}re}  \& {Doyon}}{{Marois}
  et~al.}{2008}]{Marois2008}
{Marois} C.,  {Macintosh} B.,  {Barman} T.,  {Zuckerman} B.,  {Song} I.,
  {Patience} J.,  {Lafreni{\`e}re} D.,   {Doyon} R.,  2008, \mn@doi [Science]
  {10.1126/science.1166585}, \href
  {http://adsabs.harvard.edu/abs/2008Sci...322.1348M} {322, 1348}

\bibitem[\protect\citeauthoryear{{Marois}, {Zuckerman}, {Konopacky},
  {Macintosh}  \& {Barman}}{{Marois} et~al.}{2010}]{Marois2010}
{Marois} C.,  {Zuckerman} B.,  {Konopacky} Q.~M.,  {Macintosh} B.,   {Barman}
  T.,  2010, \mn@doi [\nat] {10.1038/nature09684}, \href
  {http://adsabs.harvard.edu/abs/2010Natur.468.1080M} {468, 1080}

\bibitem[\protect\citeauthoryear{{Martin} \& {Triaud}}{{Martin} \&
  {Triaud}}{2016}]{Martin2016}
{Martin} D.~V.,  {Triaud} A.~H.~M.~J.,  2016, \mn@doi [\mnras]
  {10.1093/mnrasl/slv139}, \href
  {http://adsabs.harvard.edu/abs/2016MNRAS.455L..46M} {455, L46}

\bibitem[\protect\citeauthoryear{{Matthews}, {Krivov}, {Wyatt}, {Bryden}  \&
  {Eiroa}}{{Matthews} et~al.}{2014a}]{Matthews2014pp6}
{Matthews} B.~C.,  {Krivov} A.~V.,  {Wyatt} M.~C.,  {Bryden} G.,   {Eiroa} C.,
  2014a, \mn@doi [Protostars and Planets VI]
  {10.2458/azu_uapress_9780816531240-ch023}, \href
  {http://adsabs.harvard.edu/abs/2014prpl.conf..521M} {pp 521--544}

\bibitem[\protect\citeauthoryear{{Matthews}, {Kennedy}, {Sibthorpe}, {Booth},
  {Wyatt}, {Broekhoven-Fiene}, {Macintosh}  \& {Marois}}{{Matthews}
  et~al.}{2014b}]{Matthews2014hr8799}
{Matthews} B.,  {Kennedy} G.,  {Sibthorpe} B.,  {Booth} M.,  {Wyatt} M.,
  {Broekhoven-Fiene} H.,  {Macintosh} B.,   {Marois} C.,  2014b, \mn@doi [\apj]
  {10.1088/0004-637X/780/1/97}, \href
  {http://adsabs.harvard.edu/abs/2014ApJ...780...97M} {780, 97}

\bibitem[\protect\citeauthoryear{{Melis}, {Zuckerman}, {Rhee}  \&
  {Song}}{{Melis} et~al.}{2010}]{Melis2010}
{Melis} C.,  {Zuckerman} B.,  {Rhee} J.~H.,   {Song} I.,  2010, \mn@doi [\apjl]
  {10.1088/2041-8205/717/1/L57}, \href
  {http://adsabs.harvard.edu/abs/2010ApJ...717L..57M} {717, L57}

\bibitem[\protect\citeauthoryear{{Melis}, {Zuckerman}, {Rhee}, {Song}, {Murphy}
   \& {Bessell}}{{Melis} et~al.}{2012}]{Melis2012}
{Melis} C.,  {Zuckerman} B.,  {Rhee} J.~H.,  {Song} I.,  {Murphy} S.~J.,
  {Bessell} M.~S.,  2012, \mn@doi [\nat] {10.1038/nature11210}, \href
  {http://adsabs.harvard.edu/abs/2012Natur.487...74M} {487, 74}

\bibitem[\protect\citeauthoryear{{Meng} et~al.,}{{Meng}
  et~al.}{2014}]{Meng2014}
{Meng} H.~Y.~A.,  et~al., 2014, \mn@doi [Science] {10.1126/science.1255153},
  \href {http://adsabs.harvard.edu/abs/2014Sci...345.1032M} {345, 1032}

\bibitem[\protect\citeauthoryear{{Mennesson} et~al.,}{{Mennesson}
  et~al.}{2014}]{Mennesson2014}
{Mennesson} B.,  et~al., 2014, \mn@doi [\apj] {10.1088/0004-637X/797/2/119},
  \href {http://adsabs.harvard.edu/abs/2014ApJ...797..119M} {797, 119}

\bibitem[\protect\citeauthoryear{{Morales} et~al.,}{{Morales}
  et~al.}{2009}]{Morales2009}
{Morales} F.~Y.,  et~al., 2009, \mn@doi [\apj] {10.1088/0004-637X/699/2/1067},
  \href {http://adsabs.harvard.edu/abs/2009ApJ...699.1067M} {699, 1067}

\bibitem[\protect\citeauthoryear{{Morales}, {Rieke}, {Werner}, {Bryden},
  {Stapelfeldt}  \& {Su}}{{Morales} et~al.}{2011}]{Morales2011}
{Morales} F.~Y.,  {Rieke} G.~H.,  {Werner} M.~W.,  {Bryden} G.,  {Stapelfeldt}
  K.~R.,   {Su} K.~Y.~L.,  2011, \mn@doi [\apjl] {10.1088/2041-8205/730/2/L29},
  \href {http://adsabs.harvard.edu/abs/2011ApJ...730L..29M} {730, L29}

\bibitem[\protect\citeauthoryear{{Morbidelli} \& {Wood}}{{Morbidelli} \&
  {Wood}}{2015}]{Morbidelli2015}
{Morbidelli} A.,  {Wood} B.~J.,  2015, {Late Accretion and the Late Veneer}.
John Wiley \& Sons, pp 71--82, \mn@doi{10.1002/9781118860359.ch4}

\bibitem[\protect\citeauthoryear{{Mordasini}, {Alibert}  \& {Benz}}{{Mordasini}
  et~al.}{2009}]{Mordasini2009}
{Mordasini} C.,  {Alibert} Y.,   {Benz} W.,  2009, \mn@doi [\aap]
  {10.1051/0004-6361/200810301}, \href
  {http://adsabs.harvard.edu/abs/2009A%26A...501.1139M} {501, 1139}

\bibitem[\protect\citeauthoryear{{Moro-Mart{\'{\i}}n}
  et~al.,}{{Moro-Mart{\'{\i}}n} et~al.}{2015}]{Moro-Martin2015}
{Moro-Mart{\'{\i}}n} A.,  et~al., 2015, \mn@doi [\apj]
  {10.1088/0004-637X/801/2/143}, \href
  {http://adsabs.harvard.edu/abs/2015ApJ...801..143M} {801, 143}

\bibitem[\protect\citeauthoryear{{Mu{\~n}oz-Guti{\'e}rrez}, {Pichardo},
  {Reyes-Ruiz}  \& {Peimbert}}{{Mu{\~n}oz-Guti{\'e}rrez}
  et~al.}{2015}]{Munoz-Gutierrez2015}
{Mu{\~n}oz-Guti{\'e}rrez} M.~A.,  {Pichardo} B.,  {Reyes-Ruiz} M.,   {Peimbert}
  A.,  2015, \mn@doi [\apjl] {10.1088/2041-8205/811/2/L21}, \href
  {http://adsabs.harvard.edu/abs/2015ApJ...811L..21M} {811, L21}

\bibitem[\protect\citeauthoryear{{Murray} \& {Dermott}}{{Murray} \&
  {Dermott}}{1999}]{MurrayDermott1999}
{Murray} C.~D.,  {Dermott} S.~F.,  1999, {Solar system dynamics}.
Cambridge University Press

\bibitem[\protect\citeauthoryear{{Murray} \& {Holman}}{{Murray} \&
  {Holman}}{1997}]{Murray1997}
{Murray} N.,  {Holman} M.,  1997, \mn@doi [\aj] {10.1086/118558}, \href
  {http://adsabs.harvard.edu/abs/1997AJ....114.1246M} {114, 1246}

\bibitem[\protect\citeauthoryear{{Mustill}, {Raymond}  \& {Davies}}{{Mustill}
  et~al.}{2016}]{Mustill2016}
{Mustill} A.~J.,  {Raymond} S.~N.,   {Davies} M.~B.,  2016, \mn@doi [\mnras]
  {10.1093/mnrasl/slw075}, \href
  {http://adsabs.harvard.edu/abs/2016MNRAS.tmpL..57M} {}

\bibitem[\protect\citeauthoryear{{Nesvorn{\'y}}, {Jenniskens}, {Levison},
  {Bottke}, {Vokrouhlick{\'y}}  \& {Gounelle}}{{Nesvorn{\'y}}
  et~al.}{2010}]{Nesvorny2010}
{Nesvorn{\'y}} D.,  {Jenniskens} P.,  {Levison} H.~F.,  {Bottke} W.~F.,
  {Vokrouhlick{\'y}} D.,   {Gounelle} M.,  2010, \mn@doi [\apj]
  {10.1088/0004-637X/713/2/816}, \href
  {http://adsabs.harvard.edu/abs/2010ApJ...713..816N} {713, 816}

\bibitem[\protect\citeauthoryear{{Pawellek} \& {Krivov}}{{Pawellek} \&
  {Krivov}}{2015}]{Pawellek2015}
{Pawellek} N.,  {Krivov} A.~V.,  2015, \mn@doi [\mnras]
  {10.1093/mnras/stv2142}, \href
  {http://adsabs.harvard.edu/abs/2015MNRAS.454.3207P} {454, 3207}

\bibitem[\protect\citeauthoryear{{Pawellek}, {Krivov}, {Marshall},
  {Montesinos}, {{\'A}brah{\'a}m}, {Mo{\'o}r}, {Bryden}  \& {Eiroa}}{{Pawellek}
  et~al.}{2014}]{Pawellek2014}
{Pawellek} N.,  {Krivov} A.~V.,  {Marshall} J.~P.,  {Montesinos} B.,
  {{\'A}brah{\'a}m} P.,  {Mo{\'o}r} A.,  {Bryden} G.,   {Eiroa} C.,  2014,
  \mn@doi [\apj] {10.1088/0004-637X/792/1/65}, \href
  {http://adsabs.harvard.edu/abs/2014ApJ...792...65P} {792, 65}

\bibitem[\protect\citeauthoryear{{Payne} \& {Lodato}}{{Payne} \&
  {Lodato}}{2007}]{Payne2007}
{Payne} M.~J.,  {Lodato} G.,  2007, \mn@doi [\mnras]
  {10.1111/j.1365-2966.2007.12362.x}, \href
  {http://adsabs.harvard.edu/abs/2007MNRAS.381.1597P} {381, 1597}

\bibitem[\protect\citeauthoryear{{Payne}, {Ford}, {Wyatt}  \& {Booth}}{{Payne}
  et~al.}{2009}]{Payne2009}
{Payne} M.~J.,  {Ford} E.~B.,  {Wyatt} M.~C.,   {Booth} M.,  2009, \mn@doi
  [\mnras] {10.1111/j.1365-2966.2008.14338.x}, \href
  {http://adsabs.harvard.edu/abs/2009MNRAS.393.1219P} {393, 1219}

\bibitem[\protect\citeauthoryear{{Pearce} \& {Wyatt}}{{Pearce} \&
  {Wyatt}}{2014}]{Pearce2014}
{Pearce} T.~D.,  {Wyatt} M.~C.,  2014, \mn@doi [\mnras]
  {10.1093/mnras/stu1302}, \href
  {http://adsabs.harvard.edu/abs/2014MNRAS.443.2541P} {443, 2541}

\bibitem[\protect\citeauthoryear{{Petrovich}, {Tremaine}  \&
  {Rafikov}}{{Petrovich} et~al.}{2014}]{Petrovich2014}
{Petrovich} C.,  {Tremaine} S.,   {Rafikov} R.,  2014, \mn@doi [\apj]
  {10.1088/0004-637X/786/2/101}, \href
  {http://adsabs.harvard.edu/abs/2014ApJ...786..101P} {786, 101}

\bibitem[\protect\citeauthoryear{{Rasio}, {Nicholson}, {Shapiro}  \&
  {Teukolsky}}{{Rasio} et~al.}{1992}]{Rasio1992}
{Rasio} F.~A.,  {Nicholson} P.~D.,  {Shapiro} S.~L.,   {Teukolsky} S.~A.,
  1992, \mn@doi [\nat] {10.1038/355325a0}, \href
  {http://adsabs.harvard.edu/abs/1992Natur.355..325R} {355, 325}

\bibitem[\protect\citeauthoryear{{Raymond} \& {Armitage}}{{Raymond} \&
  {Armitage}}{2013}]{Raymond2013}
{Raymond} S.~N.,  {Armitage} P.~J.,  2013, \mn@doi [\mnras]
  {10.1093/mnrasl/sls033}, \href
  {http://adsabs.harvard.edu/abs/2013MNRAS.429L..99R} {429, L99}

\bibitem[\protect\citeauthoryear{{Raymond} \& {Bonsor}}{{Raymond} \&
  {Bonsor}}{2014}]{Raymond2014}
{Raymond} S.~N.,  {Bonsor} A.,  2014, \mn@doi [\mnras] {10.1093/mnrasl/slu048},
  \href {http://adsabs.harvard.edu/abs/2014MNRAS.442L..18R} {442, L18}

\bibitem[\protect\citeauthoryear{{Raymond}, {Armitage}  \&
  {Gorelick}}{{Raymond} et~al.}{2010}]{Raymond2010}
{Raymond} S.~N.,  {Armitage} P.~J.,   {Gorelick} N.,  2010, \mn@doi [\apj]
  {10.1088/0004-637X/711/2/772}, \href
  {http://adsabs.harvard.edu/abs/2010ApJ...711..772R} {711, 772}

\bibitem[\protect\citeauthoryear{{Read} \& {Wyatt}}{{Read} \&
  {Wyatt}}{2016}]{Read2016}
{Read} M.~J.,  {Wyatt} M.~C.,  2016, \mn@doi [\mnras] {10.1093/mnras/stv2968},
  \href {http://adsabs.harvard.edu/abs/2016MNRAS.457..465R} {457, 465}

\bibitem[\protect\citeauthoryear{{Rhee}, {Song}  \& {Zuckerman}}{{Rhee}
  et~al.}{2007}]{Rhee2007}
{Rhee} J.~H.,  {Song} I.,   {Zuckerman} B.,  2007, \mn@doi [\apj]
  {10.1086/520760}, \href {http://adsabs.harvard.edu/abs/2007ApJ...671..616R}
  {671, 616}

\bibitem[\protect\citeauthoryear{{Rhee}, {Song}  \& {Zuckerman}}{{Rhee}
  et~al.}{2008}]{Rhee2008}
{Rhee} J.~H.,  {Song} I.,   {Zuckerman} B.,  2008, \mn@doi [\apj]
  {10.1086/524935}, \href
  {http://adsabs.harvard.edu/abs/2008ApJ...675..777Rhttp://adsabs.harvard.edu/abs/2008ApJ...675..777R}
  {675, 777}

\bibitem[\protect\citeauthoryear{{Rogers}}{{Rogers}}{2015}]{Rogers2015}
{Rogers} L.~A.,  2015, \mn@doi [\apj] {10.1088/0004-637X/801/1/41}, \href
  {http://adsabs.harvard.edu/abs/2015ApJ...801...41R} {801, 41}

\bibitem[\protect\citeauthoryear{{Schlichting}, {Warren}  \&
  {Yin}}{{Schlichting} et~al.}{2012}]{Schlichting2012}
{Schlichting} H.~E.,  {Warren} P.~H.,   {Yin} Q.-Z.,  2012, \mn@doi [\apj]
  {10.1088/0004-637X/752/1/8}, \href
  {http://adsabs.harvard.edu/abs/2012ApJ...752....8S} {752, 8}

\bibitem[\protect\citeauthoryear{{Schwamb}, {Brown}  \& {Rabinowitz}}{{Schwamb}
  et~al.}{2009}]{Schwamb2009}
{Schwamb} M.~E.,  {Brown} M.~E.,   {Rabinowitz} D.~L.,  2009, \mn@doi [\apjl]
  {10.1088/0004-637X/694/1/L45}, \href
  {http://adsabs.harvard.edu/abs/2009ApJ...694L..45S} {694, L45}

\bibitem[\protect\citeauthoryear{{Sibthorpe} et~al.,}{{Sibthorpe}
  et~al.}{2010}]{Sibthorpe2010}
{Sibthorpe} B.,  et~al., 2010, \mn@doi [\aap] {10.1051/0004-6361/201014574},
  \href {http://adsabs.harvard.edu/abs/2010A%26A...518L.130S} {518, L130}

\bibitem[\protect\citeauthoryear{{Smith}, {Churcher}, {Wyatt}, {Moerchen}  \&
  {Telesco}}{{Smith} et~al.}{2009a}]{Smith2009etatel}
{Smith} R.,  {Churcher} L.~J.,  {Wyatt} M.~C.,  {Moerchen} M.~M.,   {Telesco}
  C.~M.,  2009a, \mn@doi [\aap] {10.1051/0004-6361:200810706}, \href
  {http://adsabs.harvard.edu/abs/2009A%26A...493..299S} {493, 299}

\bibitem[\protect\citeauthoryear{{Smith}, {Wyatt}  \& {Haniff}}{{Smith}
  et~al.}{2009b}]{Smith2009etacorvi}
{Smith} R.,  {Wyatt} M.~C.,   {Haniff} C.~A.,  2009b, \mn@doi [\aap]
  {10.1051/0004-6361/200911626}, \href
  {http://adsabs.harvard.edu/abs/2009A%26A...503..265S} {503, 265}

\bibitem[\protect\citeauthoryear{{Smith}, {Wyatt}  \& {Haniff}}{{Smith}
  et~al.}{2012}]{Smith2012}
{Smith} R.,  {Wyatt} M.~C.,   {Haniff} C.~A.,  2012, \mn@doi [\mnras]
  {10.1111/j.1365-2966.2012.20816.x}, \href
  {http://adsabs.harvard.edu/abs/2012MNRAS.422.2560S} {422, 2560}

\bibitem[\protect\citeauthoryear{{Smullen}, {Kratter}  \& {Shannon}}{{Smullen}
  et~al.}{2016}]{Smullen2016}
{Smullen} R.~A.,  {Kratter} K.~M.,   {Shannon} A.,  2016, \mn@doi [\mnras]
  {10.1093/mnras/stw1347}, \href
  {http://adsabs.harvard.edu/abs/2016MNRAS.tmp.1002S} {}

\bibitem[\protect\citeauthoryear{{Steffl}, {Cunningham}, {Shinn}, {Durda}  \&
  {Stern}}{{Steffl} et~al.}{2013}]{Steffl2013}
{Steffl} A.~J.,  {Cunningham} N.~J.,  {Shinn} A.~B.,  {Durda} D.~D.,   {Stern}
  S.~A.,  2013, \mn@doi [\icarus] {10.1016/j.icarus.2012.11.031}, \href
  {http://adsabs.harvard.edu/abs/2013Icar..223...48S} {223, 48}

\bibitem[\protect\citeauthoryear{{Stern} \& {Durda}}{{Stern} \&
  {Durda}}{2000}]{Stern2000}
{Stern} S.~A.,  {Durda} D.~D.,  2000, \mn@doi [\icarus]
  {10.1006/icar.1999.6263}, \href
  {http://adsabs.harvard.edu/abs/2000Icar..143..360S} {143, 360}

\bibitem[\protect\citeauthoryear{{Su} et~al.,}{{Su} et~al.}{2009}]{Su2009}
{Su} K.~Y.~L.,  et~al., 2009, \mn@doi [\apj] {10.1088/0004-637X/705/1/314},
  \href {http://adsabs.harvard.edu/abs/2009ApJ...705..314S} {705, 314}

\bibitem[\protect\citeauthoryear{{Sumi} et~al.,}{{Sumi}
  et~al.}{2010}]{Sumi2010}
{Sumi} T.,  et~al., 2010, \mn@doi [\apj] {10.1088/0004-637X/710/2/1641}, \href
  {http://adsabs.harvard.edu/abs/2010ApJ...710.1641S} {710, 1641}

\bibitem[\protect\citeauthoryear{{Sumi} et~al.,}{{Sumi}
  et~al.}{2011}]{Sumi2011}
{Sumi} T.,  et~al., 2011, \mn@doi [\nat] {10.1038/nature10092}, \href
  {http://adsabs.harvard.edu/abs/2011Natur.473..349S} {473, 349}

\bibitem[\protect\citeauthoryear{{Tamayo}}{{Tamayo}}{2014}]{Tamayo2014}
{Tamayo} D.,  2014, \mn@doi [\mnras] {10.1093/mnras/stt2473}, \href
  {http://adsabs.harvard.edu/abs/2014MNRAS.438.3577T} {438, 3577}

\bibitem[\protect\citeauthoryear{{Thebault}}{{Thebault}}{2016}]{Thebault2016}
{Thebault} P.,  2016, \mn@doi [\aap] {10.1051/0004-6361/201527626}, \href
  {http://adsabs.harvard.edu/abs/2016A%26A...587A..88T} {587, A88}

\bibitem[\protect\citeauthoryear{{Th{\'e}bault} \& {Wu}}{{Th{\'e}bault} \&
  {Wu}}{2008}]{Thebault2008}
{Th{\'e}bault} P.,  {Wu} Y.,  2008, \mn@doi [\aap]
  {10.1051/0004-6361:20079133}, \href
  {http://adsabs.harvard.edu/abs/2008A%26A...481..713T} {481, 713}

\bibitem[\protect\citeauthoryear{{Thureau} et~al.,}{{Thureau}
  et~al.}{2014}]{Thureau2014}
{Thureau} N.~D.,  et~al., 2014, \mn@doi [\mnras] {10.1093/mnras/stu1864}, \href
  {http://adsabs.harvard.edu/abs/2014MNRAS.445.2558T} {445, 2558}

\bibitem[\protect\citeauthoryear{{Traub}}{{Traub}}{2012}]{traub2012}
{Traub} W.~A.,  2012, \mn@doi [\apj] {10.1088/0004-637X/745/1/20}, \href
  {http://adsabs.harvard.edu/abs/2012ApJ...745...20T} {745, 20}

\bibitem[\protect\citeauthoryear{{Tremaine}}{{Tremaine}}{1993}]{Tremaine1993}
{Tremaine} S.,  1993, in {Phillips} J.~A.,  {Thorsett} S.~E.,   {Kulkarni}
  S.~R.,  eds,  Astronomical Society of the Pacific Conference Series Vol. 36,
  Planets Around Pulsars. pp 335--344

\bibitem[\protect\citeauthoryear{{Trujillo} \& {Sheppard}}{{Trujillo} \&
  {Sheppard}}{2014}]{Trujillo2014}
{Trujillo} C.~A.,  {Sheppard} S.~S.,  2014, \mn@doi [\nat]
  {10.1038/nature13156}, \href
  {http://adsabs.harvard.edu/abs/2014Natur.507..471T} {507, 471}

\bibitem[\protect\citeauthoryear{{Tsiganis}, {Gomes}, {Morbidelli}  \&
  {Levison}}{{Tsiganis} et~al.}{2005}]{Tsiganis2005}
{Tsiganis} K.,  {Gomes} R.,  {Morbidelli} A.,   {Levison} H.~F.,  2005, \mn@doi
  [\nat] {10.1038/nature03539}, \href
  {http://adsabs.harvard.edu/abs/2005Natur.435..459T} {435, 459}

\bibitem[\protect\citeauthoryear{{Udry} \& {Santos}}{{Udry} \&
  {Santos}}{2007}]{Udry2007}
{Udry} S.,  {Santos} N.~C.,  2007, \mn@doi [\araa]
  {10.1146/annurev.astro.45.051806.110529}, \href
  {http://adsabs.harvard.edu/abs/2007ARA%26A..45..397U} {45, 397}

\bibitem[\protect\citeauthoryear{{Vanderburg} et~al.,}{{Vanderburg}
  et~al.}{2015}]{Vanderburg2015}
{Vanderburg} A.,  et~al., 2015, \mn@doi [\nat] {10.1038/nature15527}, \href
  {http://adsabs.harvard.edu/abs/2015Natur.526..546V} {526, 546}

\bibitem[\protect\citeauthoryear{{Veras}, {Crepp}  \& {Ford}}{{Veras}
  et~al.}{2009}]{Veras2009}
{Veras} D.,  {Crepp} J.~R.,   {Ford} E.~B.,  2009, \mn@doi [\apj]
  {10.1088/0004-637X/696/2/1600}, \href
  {http://adsabs.harvard.edu/abs/2009ApJ...696.1600V} {696, 1600}

\bibitem[\protect\citeauthoryear{{Veras}, {Wyatt}, {Mustill}, {Bonsor}  \&
  {Eldridge}}{{Veras} et~al.}{2011}]{Veras2011}
{Veras} D.,  {Wyatt} M.~C.,  {Mustill} A.~J.,  {Bonsor} A.,   {Eldridge} J.~J.,
   2011, \mn@doi [\mnras] {10.1111/j.1365-2966.2011.19393.x}, \href
  {http://adsabs.harvard.edu/abs/2011MNRAS.417.2104V} {417, 2104}

\bibitem[\protect\citeauthoryear{{Veras}, {Mustill}, {G{\"a}nsicke},
  {Redfield}, {Georgakarakos}, {Bowler}  \& {Lloyd}}{{Veras}
  et~al.}{2016}]{Veras2016}
{Veras} D.,  {Mustill} A.~J.,  {G{\"a}nsicke} B.~T.,  {Redfield} S.,
  {Georgakarakos} N.,  {Bowler} A.~B.,   {Lloyd} M.~J.~S.,  2016, \mn@doi
  [\mnras] {10.1093/mnras/stw476}, \href
  {http://adsabs.harvard.edu/abs/2016MNRAS.458.3942V} {458, 3942}

\bibitem[\protect\citeauthoryear{{Weinberg}, {Shapiro}  \&
  {Wasserman}}{{Weinberg} et~al.}{1987}]{Weinberg1987}
{Weinberg} M.~D.,  {Shapiro} S.~L.,   {Wasserman} I.,  1987, \mn@doi [\apj]
  {10.1086/164883}, \href {http://adsabs.harvard.edu/abs/1987ApJ...312..367W}
  {312, 367}

\bibitem[\protect\citeauthoryear{{Wetherill}}{{Wetherill}}{1994}]{Wetherill1994}
{Wetherill} G.~W.,  1994, \mn@doi [\apss] {10.1007/BF00984505}, \href
  {http://adsabs.harvard.edu/abs/1994Ap%26SS.212...23W} {212, 23}

\bibitem[\protect\citeauthoryear{{Winn} \& {Fabrycky}}{{Winn} \&
  {Fabrycky}}{2015}]{Winn2015}
{Winn} J.~N.,  {Fabrycky} D.~C.,  2015, \mn@doi [\araa]
  {10.1146/annurev-astro-082214-122246}, \href
  {http://adsabs.harvard.edu/abs/2015ARA%26A..53..409W} {53, 409}

\bibitem[\protect\citeauthoryear{{Wyatt}}{{Wyatt}}{2003}]{Wyatt2003}
{Wyatt} M.~C.,  2003, \mn@doi [\apj] {10.1086/379064}, \href
  {http://adsabs.harvard.edu/abs/2003ApJ...598.1321W} {598, 1321}

\bibitem[\protect\citeauthoryear{{Wyatt}}{{Wyatt}}{2008}]{Wyatt2008}
{Wyatt} M.~C.,  2008, \mn@doi [\araa] {10.1146/annurev.astro.45.051806.110525},
  \href {http://adsabs.harvard.edu/abs/2008ARA%26A..46..339W} {46, 339}

\bibitem[\protect\citeauthoryear{{Wyatt} \& {Jackson}}{{Wyatt} \&
  {Jackson}}{2016}]{Wyatt2016}
{Wyatt} M.~C.,  {Jackson} A.~P.,  2016, \mn@doi [\ssr]
  {10.1007/s11214-016-0248-1}, \href
  {http://adsabs.harvard.edu/abs/2016SSRv..tmp...12W} {}

\bibitem[\protect\citeauthoryear{{Wyatt}, {Dermott}, {Telesco}, {Fisher},
  {Grogan}, {Holmes}  \& {Pi{\~n}a}}{{Wyatt} et~al.}{1999}]{Wyatt1999}
{Wyatt} M.~C.,  {Dermott} S.~F.,  {Telesco} C.~M.,  {Fisher} R.~S.,  {Grogan}
  K.,  {Holmes} E.~K.,   {Pi{\~n}a} R.~K.,  1999, \mn@doi [\apj]
  {10.1086/308093}, \href {http://adsabs.harvard.edu/abs/1999ApJ...527..918W}
  {527, 918}

\bibitem[\protect\citeauthoryear{{Wyatt}, {Greaves}, {Dent}  \&
  {Coulson}}{{Wyatt} et~al.}{2005}]{Wyatt2005}
{Wyatt} M.~C.,  {Greaves} J.~S.,  {Dent} W.~R.~F.,   {Coulson} I.~M.,  2005,
  \mn@doi [\apj] {10.1086/426929}, \href
  {http://adsabs.harvard.edu/abs/2005ApJ...620..492W} {620, 492}

\bibitem[\protect\citeauthoryear{{Wyatt}, {Smith}, {Greaves}, {Beichman},
  {Bryden}  \& {Lisse}}{{Wyatt} et~al.}{2007a}]{Wyatt2007hotdust}
{Wyatt} M.~C.,  {Smith} R.,  {Greaves} J.~S.,  {Beichman} C.~A.,  {Bryden} G.,
   {Lisse} C.~M.,  2007a, \mn@doi [\apj] {10.1086/510999}, \href
  {http://adsabs.harvard.edu/abs/2007ApJ...658..569W} {658, 569}

\bibitem[\protect\citeauthoryear{{Wyatt}, {Smith}, {Su}, {Rieke}, {Greaves},
  {Beichman}  \& {Bryden}}{{Wyatt} et~al.}{2007b}]{Wyatt2007collisionalcascade}
{Wyatt} M.~C.,  {Smith} R.,  {Su} K.~Y.~L.,  {Rieke} G.~H.,  {Greaves} J.~S.,
  {Beichman} C.~A.,   {Bryden} G.,  2007b, \mn@doi [\apj] {10.1086/518404},
  \href {http://adsabs.harvard.edu/abs/2007ApJ...663..365W} {663, 365}

\bibitem[\protect\citeauthoryear{{Wyatt}, {Booth}, {Payne}  \&
  {Churcher}}{{Wyatt} et~al.}{2010}]{Wyatt2010}
{Wyatt} M.~C.,  {Booth} M.,  {Payne} M.~J.,   {Churcher} L.~J.,  2010, \mn@doi
  [\mnras] {10.1111/j.1365-2966.2009.15930.x}, \href
  {http://adsabs.harvard.edu/abs/2010MNRAS.402..657W} {402, 657}

\bibitem[\protect\citeauthoryear{{Wyatt}, {Clarke}  \& {Booth}}{{Wyatt}
  et~al.}{2011}]{Wyatt2011}
{Wyatt} M.~C.,  {Clarke} C.~J.,   {Booth} M.,  2011, \mn@doi [Celestial
  Mechanics and Dynamical Astronomy] {10.1007/s10569-011-9345-3}, \href
  {http://adsabs.harvard.edu/abs/2011CeMDA.111....1W} {111, 1}

\bibitem[\protect\citeauthoryear{{Wyatt} et~al.,}{{Wyatt}
  et~al.}{2012}]{Wyatt2012}
{Wyatt} M.~C.,  et~al., 2012, \mn@doi [\mnras]
  {10.1111/j.1365-2966.2012.21298.x}, \href
  {http://adsabs.harvard.edu/abs/2012MNRAS.424.1206W} {424, 1206}

\bibitem[\protect\citeauthoryear{{van Lieshout}, {Dominik}, {Kama}  \&
  {Min}}{{van Lieshout} et~al.}{2014}]{vanLieshout2014}
{van Lieshout} R.,  {Dominik} C.,  {Kama} M.,   {Min} M.,  2014, \mn@doi [\aap]
  {10.1051/0004-6361/201322090}, \href
  {http://adsabs.harvard.edu/abs/2014A%26A...571A..51V} {571, A51}

\makeatother
\end{thebibliography}

\bsp    
\label{lastpage}
\end{document}